\documentclass[10pt,english]{paper}
\usepackage[latin9]{inputenc}
\usepackage[letterpaper]{geometry}
\usepackage{varioref}
\usepackage{textcomp}
\usepackage{amsmath}
\usepackage{graphicx}
\usepackage{amssymb}
\usepackage{esint}

\makeatletter

\providecommand*{\perispomeni}{\char126}
\AtBeginDocument{\DeclareRobustCommand{\greektext}{%
  \fontencoding{LGR}\selectfont\def\encodingdefault{LGR}%
  \renewcommand{\~}{\perispomeni}%
}}
\DeclareRobustCommand{\textgreek}[1]{\leavevmode{\greektext #1}}
\DeclareFontEncoding{LGR}{}{}

\newcommand{\lyxmathsym}[1]{\ifmmode\begingroup\def\b@ld{bold}
  \text{\ifx\math@version\b@ld\bfseries\fi#1}\endgroup\else#1\fi}

\newcommand{\lyxaddress}[1]{
\par {\raggedright #1
\vspace{1.4em}
\noindent\par}
}
\newenvironment{lyxlist}[1]
{\begin{list}{}
{\settowidth{\labelwidth}{#1}
 \setlength{\leftmargin}{\labelwidth}
 \addtolength{\leftmargin}{\labelsep}
 }}
{\end{list}}

\usepackage{cite}
\usepackage{hyperref}

\AtBeginDocument{
  
}

\makeatother

\usepackage{babel}

\begin{document}

\title{\noindent {\Large Toward a theory of curvature-scaling gravity}}

\author{\textrm{\emph{\normalsize Hoang Ky Nguyen}}%
\thanks{\emph{Electronic address:} HoangNguyen7@hotmail.com%
}}

\maketitle

\lyxaddress{New York City, USA}

\lyxaddress{{\small (February 26, 2013)}}
\begin{abstract}
A salient feature of Ho\v{r}ava-Lifshitz gravity is the anisotropic
time variable \cite{Horava}. We propose an alternative construction
of the spacetime manifold which naturally enables anisotropy in time
scaling. Our approach promotes the role of curvature: the Ricci scalar
$\mathcal{R}$ at a given point on the manifold sets the length scales
for physical processes -- including gravity -- in the local inertial
frames enclosing that point. The manifold is viewed as a patchwork
of local regions; each region is Lorentz-invariant and adopts a variable
local scale which we call the Ricci length $a_{\mathcal{R}}$ defined
as $a_{\mathcal{R}}\triangleq\left|\mathcal{R}\right|^{-1/2}$. In
each local patch, the length scales of physical processes are measured
relatively to the local Ricci length $a_{\mathcal{R}}$, and only
their dimensionless ratios participate in the dynamics of the physical
processes. The time anisotropy arises from the requirement that the
form -- but not necessarily the parameters -- of physical laws be
unchanged under variations of the local Ricci length as one moves
along any path on the manifold. The time duration is found to scale
as $dt\propto a_{\mathcal{R}}^{3/2}$ whereas the spatial differential
scales as $dx\propto a_{\mathcal{R}}$.

We show how to conjoin the local patches of the manifold in a way
which respects causality and other requirements of special relativity,
as well as the equivalence principle and the general covariance principle.
In our approach, all of Einstein's insights are preserved but the
parameters of physical laws are only valid locally and become functions
of the prevailing Ricci scalar. As such, the parameters are allowed
to vary on the manifold, together with the Ricci scalar. This alternative
construction of the manifold permits a unique choice for the Lagrangian
of gravity coupled with matter. Curvature thus acquires a new privileged
status -- it is actively involved in the dynamics of physical processes
by setting the scale for them; hence the name curvature-scaling gravity.

In vacuo curvature-scaling gravity takes the form of quadratic Lagrangian
$\mathcal{R}^{2}$, which adopts a larger set of solutions superseding
all solutions to the field equations based on the Einstein-Hilbert
action $\mathcal{R}$. We provide two static spherically symmetric
solutions: one solution connects our theory to Mannheim-Kazanas's
conformal gravity-based solution \cite{MK1} which Mannheim argues
to account for the galactic rotation curves \cite{Mannheim1,MOB1,MOB2,MOB3};
the other solution leads to novel properties for Schwarzschild-type
black holes. Additionally, we apply curvature-scaling gravity to address
an array of problems encountered in cosmology, and discuss its implications
in the quantization of gravity.

\tableofcontents{}
\end{abstract}

\section{\label{sec:1}Postulate on the dependence of local length scales
on the scalar curvature of spacetime}

Einstein's vision of spacetime consists of a manifold with (pseudo-)Riemannian
geometry. The curvature of the manifold is determined by the distribution
of matter and is governed by the Einstein equations of the metric
components $g_{\mu\nu}$. At a given point on the manifold, the metric
can be temporarily and locally approximated by a Minkowskian one which
in effect eliminates the effect of curvature at the point. In the
local inertial coordinate system, test particles follow geodesics
and all physical laws of non-gravitational origin obey the Lorentz
symmetry as established in special relativity. Gravity is interpreted
as the effects of curvature acting non-locally on the geodesics; in
the local inertial frame the effects of gravity disappear. These aspects
of spacetime and gravity have been theoretically established in general
relativity and experimentally verified with great precision for solar
system. The post-Newtonian phenomenology of gravity includes the three
classic tests -- the precession of Mercury's perihelion, the gravitational
redshift, the bending of starlight across the Sun disc -- and later
added to the list of successes for Einstein's theory is the decaying
orbits of binary pulsars.

In this report, we aim to extend Einstein's insights further. At the
heart of our undertaking is the role of curvature which we theorize
to be more profound than so far being perceived in literature and
carried out in practice. We are guided by the following observation.

\subsubsection*{Our observation:}

The equivalence principle states that at a given point on the manifold
one can choose a coordinate system (a {}``tangent'' frame) such
that the effects of gravity are instantaneously and locally eliminated.
Einstein was inspired by the example of a free falling elevator to
arrive at this insightful conclusion. The equivalence of acceleration
and local gravity establishes the equality between the inertial mass
and the passive gravitational mass of an object. Non-gravitational
forces act as if they were in a flat spacetime, with physical phenomena
within the local inertial frame obeying special relativity together
with other laws with non-gravitational origin, such as quantum mechanics.
What puzzles us, however, is that whilst on the one hand, the presence
of a real gravitation field -- i.e., the curved spacetime -- in the
region of the elevator cannot be eliminated since curvature is characterized
by the $20$ degrees of freedom inherent in the Riemann tensor (in
particular, the Ricci scalar being an invariant and thus manifest
in all coordinate systems), on the other hand, curvature curiously
plays no direct role whatsoever in the dynamics of the non-gravitational
physics in the local region once it is moved to an inertial frame.
The non-gravitational processes and laws in any of these {}``tangent''
frames are oblivious to the value of the Ricci scalar at that point.
Whilst this is a proceeding customarily employed in general relativity
and, de facto, all gravitational theories that followed since, we
suggest that the curvature plays a greater role than traditionally
assumed: it explicitly sets the size of physical objects in the {}``tangent''
frame -- a role that has been overlooked in conventional theories
of gravity. That is to say, being of the unit $\left[\mbox{length}\right]^{-2}$,
the Ricci scalar should determine the length scales of physical processes
that take place in the tangent region.

Let us define for each given point on the manifold a new length --
and name it the Ricci length $a_{\mathcal{R}}$ -- taken to be the
inverse of the square root of the absolute value of the Ricci scalar
at the point: $a_{\mathcal{R}}\triangleq\left|\mathcal{R}\right|^{-1/2}$.
The Ricci length is itself an invariant, i.e., having the same value
for all choices of the coordinate system, and is obviously of dimension
of length. We shall assert that the Ricci length is of a more fundamental
status than all other length scales, and that all other length scales
are denominated in terms of the Ricci length. That is to say, the
length scale of any physical process in a local tangent region is
dynamically determined by the Ricci scalar which in turn can vary
from one point to the next on the manifold (with the dynamics of $\mathcal{R}$
being related to that of $g_{\mu\nu}$), and only the dimensionless
relative ratio between the length scale of the process and the Ricci
length is relevant for the dynamics of the process. (We stress that
our ideas are not related to the dilaton or Brans-Dicke theories.
First, the scale setter in our approach is the Ricci scalar itself,
instead of an auxiliary field with its own dynamics. Second, as will
be shown in Section \ref{sec:5}, our approach enables anisotropy
in time scaling -- an important feature absent in dilaton and Brans-Dicke.)

More concretely, with a given metric $g_{\mu\nu}$ of the signature
convention $\left(+,-,-,-\right)$, in each local patch, a test particle
of mass $m$ travels on a timelike geodesics ($ds^{2}\triangleq g_{\mu\nu}dx^{\mu}dx^{\nu}>0$):\begin{equation}
\frac{d^{2}x^{\mu}}{ds^{2}}+\Gamma_{\nu\lambda}^{\mu}\frac{dx^{\nu}}{ds}\frac{dx^{\lambda}}{ds}=0,\ \ \ \ \ \mathcal{R}_{\mu\nu\lambda\sigma}\neq0\label{eq:1.1}\end{equation}
for an arbitrary choice of coordinate $x^{\mu}$. The equation of
motion \eqref{eq:1.1} is derivable from the action for the test particle
used in the standard theory of relativity \begin{equation}
\mathcal{S}\simeq\int ds.\label{eq:1.2}\end{equation}
This action is built from an invariant, the proper distance $ds$,
thus yielding the equation of motion \eqref{eq:1.1} in the tensorial
form, as required by the general covariance principle. At a given
point $P$, if one chooses a coordinate such that the Christoffel
$\Gamma_{\nu\lambda}^{\mu}$'s vanish (i.e., the inertial frame),
the effects of curvature -- interpreted as the effects of gravity
-- are locally eliminated around $P$ since the trajectory of the
test particle is a geodesics, and all physical phenomena of non-gravitational
origin look exactly like ones in the Minkowskian metric, i.e., satisfying
all the elements of special relativity. This is the essence of the
equivalence principle. Yet, as we noted before, the curvature curiously
drops out of the geodesics equation \eqref{eq:1.1} when $\Gamma_{\nu\lambda}^{\mu}=0$
and is decidedly absent in action \eqref{eq:1.2}. This is not necessarily
so, however: we can fully retain the validity of the equivalence principle
even if the proper distance in action \eqref{eq:1.2} is divided by
the Ricci length, $a_{\mathcal{R}}$ -- itself an invariant; namely,
the integrand is replaced by $ds/a_{\mathcal{R}}$ -- a combination
of invariants. That is to say:\begin{equation}
\mathcal{S}\simeq\int\frac{ds}{a_{\mathcal{R}}}.\label{eq:1.2-1}\end{equation}
The equivalence principle in no way prevents this possibility; the
extra divisor $a_{\mathcal{R}}$ does not imperil the equivalence
principle which is local in nature %
\footnote{The preservation of the equivalence principle in this manner was previously
recognized by Mannheim in his review \cite{Mannheim1}. See Eq. (11)
and the comment before Eq. (145) of \cite{Mannheim1} in which the
test particle is coupled with a scalar field $S$: $\mathcal{S}\simeq\int ds\, S(x)$;
this action formally does not imperil the equivalence principle. In
our approach, we furnish the physical meaning for the scalar field
in which $S$ is set to be inverse of the Ricci length.%
}. Moreover, this new choice for the integrand, $ds/a_{\mathcal{R}}$,
fully respects all the requirements of special relativity in the local
inertial frame. In particular, thanks to the local Lorentz symmetry,
the order of precedence of causally-connected events is an invariant;
as such, causality is protected globally. In addition, with such an
integrand being built from invariants, the proper distance and the
Ricci scalar, the general covariance principle is automatically and
globally ensured. As such, all of Einstein's insights of relativity are respected.

We therefore conclude that the de facto adoption of $ds$ in action
\eqref{eq:1.2} is not the only choice allowable, with $ds/a_{\mathcal{R}}$
as in action \eqref{eq:1.2-1} being the other acceptable option as
well. We project that conventional wisdom has been under-appreciating
the role of curvature, and that the theory of general relativity can
be made tighter: not only does the scalar curvature determine the
action (such as action \eqref{eq:1.2-1} above, which equals $\int ds\,\left|\mathcal{R}\right|^{1/2}$,
and the Einstein-Hilbert action $\int d^{4}x\,\sqrt{-\det g}\,\mathcal{R}$),
the curvature can further partake in the dynamics of the physical
phenomena and processes in the local inertial frame by setting the
scale for them. We thus upgrade the curvature to an eminent role,
a role which we believe has been overlooked in the development of
gravitational theories: the Ricci scalar directly sets the length
scales for all physical phenomena that take place in the local inertial
frame (and indeed all non-inertial frames since the Ricci scalar is
an invariant.) In other words, all length scales %
\footnote{Including the Bohr radius which is governed by quantum mechanics.%
} of physical processes in a local pocket of spacetime are pegged to
the Ricci length prevailing in the pocket. As intuitive as our arguments
are, we nonetheless cast them in two following postulates.\bigskip{}

\textbf{Postulate I:} \label{Postulate1}\textbf{\emph{On the invariance
of the form of physical laws}}
\begin{quote}
\emph{Physical laws retain their forms -- but not necessarily their
parameters -- in every local region on the spacetime manifold.}
\end{quote}
\medskip{}

\textbf{Postulate II:} \textbf{\emph{On the role of the Ricci scalar
as a local scale-setter}}
\begin{quote}
\emph{The Ricci length, defined via the Ricci scalar $\mathcal{R}$
as $a_{\mathcal{R}}\triangleq\left|\mathcal{R}\right|^{-1/2}$, is
the common denominator for all length scales of physical processes
-- gravitational and non-gravitational alike -- that take place in
a local region. Physical laws are dependent only on the dimensionless
ratios of lengths normalized by the Ricci length.}\textbf{\emph{\vspace{7mm}
}}
\end{quote}
Postulate (I), at the first look, is nothing but a reinforcement of
special relativity and the equivalence principle. In each local region,
all physical processes satisfy special relativity (i.e., the Lorentz
symmetry) as well as all other established laws in the quantum realm.
For example, the Dirac equation must retain its form for every local
pocket. Our emphasis, however, is the allowance for the parameters
of the physical laws to vary from one local region to the next on
the manifold. Indeed, since the equivalence principle is local in
nature, special relativity is only required to hold locally. What
we further enforce here is that the parameters of physical laws be
valid locally too. What is left unaddressed in Postulate (I) is how
the parameters are specified for each local region.

Postulate (II) quantifies how the parameters are determined in each
local pocket. To do this, it requests that the length scale of every
physical process be pegged into the Ricci length in the local pocket.
If the Ricci length varies, the length scale of any physical process
must vary in exact proportion. As a result, physical laws must be
expressed in terms of dimensionless ratios of lengths denominated
by the Ricci length. Whereas the form of the physical law that governs
a process is unchanged -- per Postulate (I) -- as we move from one
pocket to another on the manifold, the length scale intrinsic in the
law must not; rather, it has to adapt to the prevailing value of the
Ricci length, per Postulate (II). The Ricci length -- holding a higher
status than all other length scales -- acquires a dynamics from the
evolution of the spacetime manifold, viz. the gravitational field
equations, which are to be derived in this report. As we shall see,
the power of these two postulates goes beyond non-gravitational laws;
they too embrace gravitational laws by prescribing the dynamical equations
for $g_{\mu\nu}$ as well, which is the ultimate objective of this
report.

We thus arrive at a new depiction of the spacetime manifold: the spacetime
manifold constitutes a patchwork of local regions, each of which strictly
obeys special relativity and established non-gravitational physics
(be it the standard quantum field theory with $SU(3)\times SU(2)\times U(1)$
gauge symmetries plus spontaneous symmetry breaking, the phonons and
quantum Hall effect in solids, or the tunneling effect in nuclear
decays.) There is no fundamental length scale applicable globally
to the whole manifold, but each individual region adopts a local length
scale $a_{\mathcal{R}}$ set by the Ricci scalar in the region. All
regions are indistinguishable, however; the observer cannot tell which
region he lives in by using only the measurement apparatus that are
available within his surroundings. It is permissible, as we shall
show in this report, to conjoin the local patches in a way which respects
(global) causality and the general covariance principle. Our new construction
of the spacetime manifold is a departure from Einstein's theory while
we preserve all of his insights -- the Lorentz symmetry, the Michelson-Morley
finding, the relativity principle, the equivalence principle, and
the general covariance principle -- in our construction. It is the
relinquishment of a global length scale -- but instead adopting a
dynamical length scale associated with the Ricci scalar -- that is
the only new crucial element in our approach.

In plain language, let us reconsider Einstein's gedanken free falling
elevator. Let an observer residing within the elevator throw a brick.
He will observe the brick moving on a straight line at constant speed,
using his own ruler and clock; i.e., the pulling effects of the Earth
under his feet is absent. This is the essence of the equivalence principle.
However, we go one step further. We assert that the scalar curvature
further determines the size of each object in the elevator (including
the brick and the observer's ruler) and the oscillatory rate of the
observer's wristwatch. %
\footnote{This latter point holds the key to the time anisotropy alluded to
in the Abstract, to be discussed in Section \ref{sec:5}.%
} In so doing, we strengthen the position of the curvature within Einstein's
theory by promoting the Ricci scalar to a prominent status: not only
does it play a crucial part in the underlying geometry, it actively
participates in the dynamical process of physical phenomena by setting
the scale for them. This new element -- encapsulated in Postulate
(II) -- is the centerpiece for all of our subsequent considerations.
This report is devoted to examine the ramifications of this idea.

\subsubsection*{The strategy:}

Our ultimate objective is to find the field equations for gravity.
Due to Postulate (II), the field equations will necessarily be different
from those in Einstein's gravity. To give the reader a sense of what
shall be expanded in this report, below is the outline of our program.
Only dimensionless combinations of lengths denominated by the Ricci
length, such as $ds/a_{\mathcal{R}}$ or $dx^{\mu}/a_{\mathcal{R}}$,
shall enter the action. Beside the conventional way that the metric
components enter the action (via the minimal coupling procedure which
allows $g_{\mu\nu}$ to enter the action via the Jacobian $\sqrt{-\det g}$,
the covariant derivatives $\nabla_{\mu}$, and the contrarian derivatives
$\nabla^{\mu}\triangleq g^{\mu\nu}\nabla_{\nu}$), the metric will
also accompany the Ricci length to enter the action via two additional
entrances:
\begin{itemize}
\item Via the Lagrangian of matter fields which is comprised of derivatives
of the matter fields. The (covariant) derivatives $\nabla_{\mu}$
will be normalized by the Ricci length; i.e., they are replaced as
$\nabla_{\mu}\rightarrow a_{\mathcal{R}}\nabla_{\mu}=\left|\mathcal{R}\right|^{-1/2}\nabla_{\mu}$;
\item Via the 4-volume $d^{4}x$ in the volume element $d^{4}x\,\sqrt{-\det g}$,
the first part of which is replaced as $d^{4}x\rightarrow\left(dx/a_{\mathcal{R}}\right)^{4}=d^{4}x\,\mathcal{R}^{2}$.
\end{itemize}
All these routes combined open the door for matter fields to couple
with gravity in an organic, natural, and unique way. Our procedure
outlined above is an important breakaway from the standard procedure
of minimal coupling, viz. $\int d^{4}x\,\sqrt{-\det g}\,\mathcal{L}_{m}$.\newpage{}

Mathematically, the global structure of the manifold remains to be
torsion-free pseudo-Riemannian geometry, with its curvature -- the
Riemann tensor $\mathcal{R}_{\nu\lambda\sigma}^{\mu}$ and its contracted
derivatives, the Ricci tensor $\mathcal{R}_{\mu\nu}$ and Ricci scalar
$\mathcal{R}$ -- obey the well-developed mathematics of Riemannian
geometry and diffeomorphism.

Conceptually speaking, unlike the equivalence principle which was
inspired by Einstein's gedanken elevator, the conventional choice
of $ds$ in action (\ref{eq:1.2}) did not appeal to any fundamental
principle, but rather has been a convenient choice %
\footnote{The Einstein-Hilbert action, being a second-order theory, has a well-posed
initial value problem for the metric, in which \textquotedblleft{}coordinates\textquotedblright{}
and \textquotedblleft{}momenta\textquotedblright{} specified at an
initial time can be used to predict future evolution. This is often
cited as an advantage, if not the major rationale for the theory.
A gravitational theory which involves higher-derivative terms would
require not only those data, but also some number of derivatives of
the momenta, thus becoming intractable both conceptually and practically.
We shall show in Section \ref{sec:3} that this is not the case for
curvature-scaling gravity which also only requires {}``coordinates''
and {}``momenta'' as its Cauchy data.%
}. By adopting the alternative $ds/a_{\mathcal{R}}$ as in action \eqref{eq:1.2-1}
we resort to a requisition that local scale no longer be an omnipresent
prefixed property but instead be dynamically determined by the very
structure of spacetime itself, a requisition which lends ontological
and aesthetical appeals to the theory.

\subsubsection*{The criteria:}

Our approach thus is an alternative construction of the spacetime
manifold which we shall explicitly build. The success or failure of
such a permissible construction needs be judged against the following
list of criteria:
\begin{enumerate}
\item Will it embrace the well-established principles, the most important
of which are the relativity principle, the Lorentz invariance (i.e.,
the Michelson-Morley experiment), the equivalence principle, and the
general covariance principle? Also being included in the list are
the laws of quantum physics (quantum mechanics and quantum field theories
of the electroweak and strong forces -- namely, the laws with non-gravitational
origin). %
\footnote{The laws of gravity are temporarily absent in this list precisely
because we are embarking on a search for the laws for gravity, away
from Einstein's gravity. Yet we still want to retain the equivalence
principle and the general covariance principle in the list of well-tested
and all-embracing principles to be satisfied.%
}
\item Will it protect causality? %
\footnote{We single out the causality principle as this is a very touchy issue
which merits special attention.%
}
\item Will it give rise to a form of gravity which guarantees to recover
the precise solar system tests as well as the decaying orbits of binary
pulsars?
\item Will it offer new interesting physics together with testable predictions?
Will it help -- in the Occam's Razor sense -- solve problems and difficulties
that plague cosmology, astrophysics, quantum gravity?
\item Will it offer new conceptual insights?
\item Last but not least, at what costs, mathematics-wise?
\item And concept-wise?
\end{enumerate}
The answers to the first five criteria are affirmative. Our approach
imposes no structural damage to the existing physical laws. Every
local pocket of spacetime, by construction, satisfies special relativity
(that is, including the Michelson-Morley experiment, the constancy
of light speed regardless of the state of the observer and/or the
light emitter) and satisfies the quantum laws. The laws retain exactly
the same form -- but not necessarily the parameters -- in all local
pockets. There is no local pocket that is more privileged than any
others. The parameters of the laws, being dynamically pegged to the
Ricci scalar, however, can inherit the dynamics of the Ricci scalar,
which itself is governed by the field equations of the metric $g_{\mu\nu}$.
But their variabilities in no way imperil the well-tested principles
of relativity. With the actions being constructed from invariants,
the general covariance principle is ensured as well.

Causality is protected since in every local region the Lorentz symmetry
remains intact. No coordinate transformation can frustrate the order
of precedence of any pair of causally-related events in the local
region, and thus causality globally holds for any pair of time-like
events on the manifold. Time-like trajectories and space-like trajectories
do not mix. Null geodesics remain null-geodesics, separating the set
of time-like trajectories from the set of space-like trajectories.

We shall show that the new Lagrangian of gravity duly recovers post-Newtonian
phenomenology. This is because in vacuo the new gravity to be built
in this report resemblances the quadratic Lagrangian $\mathcal{R}^{2}$
which is known to accept all solutions to the Einstein field equations
(based on the Einstein-Hilbert action $\mathcal{R}$) as its subset.
As long as the Ricci scalar, and thus the Ricci length, varies little
in the solar system, the effects of the Ricci length $a_{\mathcal{R}}$
on the three classic tests of Einstein's gravity as well as the decaying
orbits of binary pulsars should be negligible. 

There will be new interesting outcomes which we shall cover in this
report. These new outcomes will have significant impacts on a range
of fundamental issues encountered in astrophysics, cosmography, cosmology,
and the quantization of gravity.

Mathematically, the costs in formalism are virtually nil. We can fully
enjoy the machinery of Riemannian geometry and diffeomorphism developed
in the last 150 years or so. This is because the underlying geometry
remains to be torsion-free pseudo-Riemannian (not Weyl, for example).

Regarding Point (7), we must however note that there are conceptual
departures which we shall elaborate in due course. The conceptual
route we are undertaking logically leads to a number of substantial
adjustments in one's view regarding the underpinnings of physical
laws.

Before we embark on the program, we have three further remarks:
\begin{itemize}
\item Our approach is not related to the dilaton or Brans-Dicke theories.
The scale setter is the Ricci scalar the dynamics of which is that
of the metrics $g_{\mu\nu}$, instead of an auxiliary scalar field
that lives on the manifold and has its own dynamics. Futher fundamental
differences between our approach and Brans-Dicke/dilaton, such as
the time anisotropy which arises from our approach and is absent in
Brans-Dicke/dilaton, will be elucidated in Section \ref{sec:5}.
\item Comparison with conformal gravity: Our approach should not be mistaken
as any equivalent form as the well-studied conformal gravity. Indeed,
our approach is precisely the opposite: whereas in conformal gravity,
the curvature of spacetime is {}``gauged'' away by conformal transformations,
our approach upgrades the Ricci scalar to an eminent role: it sets
the length scales for all physical phenomena including gravity from
one spacetime point to the next.
\item An objection would be that in vacuo spacetime is Ricci-flat (i.e.,
$\mathcal{R}=0$) according to the Einstein field equations, thereby
rendering the Ricci length infinite. However, this is a premature
objection for two reasons: (i) The field equations for curvature-scaling
gravity in vacuo generally admits spacetime configurations that have
non-zero Ricci scalar. This is because the field equations in vacuo,
as we shall see, echo the field equations of the quadratic Lagrangian
$\mathcal{R}^{2}$ which is known to admit a richer set of solutions
than Einstein's gravity (based on $\mathcal{R}$) does. (ii) The Robertson-Walker
metric, considered to be applicable to the cosmos, generally has non-zero
Ricci scalar. In the Robertson-Walker metric, even if space is flat,
spacetime is not. Both of these reasons would confine the domain with
vanishing Ricci scalar to a set of zero-measure.
\end{itemize}
Our report is structured as follows. Section \ref{sec:2} presents
the Lagrangian and equation of motion for a test particle in curvature-scaling
gravitational field; we also show how causality is protected in curvature-scaling
gravity. Section \ref{sec:3} builds the Lagrangian of the curvature-scaling
gravitational field in vacuo. Section \ref{sec:4} shows our solution
to the curvature-scaling field equation in vacuo and its connection
with Manheim's phenomenological theory of galactic rotation curves
\cite{MK1,MOB1,MOB2,MOB3,Mannheim1}. Section \ref{sec:5} presents
a series of logical consequences of our two postulates declared \vpageref{Postulate1};
in particular, we show how the anisotropic time scaling would arise
from the two postulates by pure deduction. Section \ref{sec:6} presents
a nontrivial solution of our curvature-scaling field equations and,
combined with the logical deductions in Section \ref{sec:5}, makes
a prediction of a new type of black holes with novel properties. Section
\ref{sec:7} builds the Lagrangian of curvature-scaling gravity coupled
with matter. Section \ref{sec:8} details the implications of curvature-scaling
gravity in cosmography; we address the (over)estimation problem with
the Hubble constant (and related to which, the age problem), provide
a critical reassessment of Type Ia supernovae data, and work out an
alternative interpretation of these data bypassing the conventional
explanation based on accelerated expansion. Section \ref{sec:9} explores
the consequence of curvature-scaling gravity in cosmology; we show
how it helps resolve the cosmological constant problem, the horizon
problem, the flatness problem, among others while circumventing the
inflationary expansion scenario. Section \ref{sec:10} summarizes
the theory and offers an outlook how curvature-scaling gravity could
offer a reasonable starting point toward quantum gravity. All the
technical details are presented in the Appendices.\newpage{}

\section{\label{sec:2}Equation of motion of a point mass in a given curvature-scaling
gravitational field}

Consider a test point mass $m$ moving in a given metric $g_{\mu\nu}$.
In the standard theory of general relativity, its action is well-known
\begin{equation}
\mathcal{S}=-mc\int ds\label{eq:2.1}\end{equation}
in which $c$ is the speed of light and the length element $ds^{2}=g_{\mu\nu}dx^{\mu}dx^{\nu}>0$.
We prefer to rewrite is as \begin{equation}
\mathcal{S}\simeq\int ds\label{eq:2.2}\end{equation}
since only the integrand will be of relevance for our subsequent reasoning.
The quantity of interest $ds$ is of dimension of length. The particle
is expected to move on a geodesics per the equation of motion \eqref{eq:1.1}.
In curvature-scaling gravity, however, only the dimensionless ratio
$d\tilde{s}\triangleq ds/a_{\mathcal{R}}$, in which $a_{\mathcal{R}}=\left|\mathcal{R}\right|^{-1/2}$
is the local Ricci length, is the meaningful quantity to be used in
the action. Therefore, we make the following replacement in action
\eqref{eq:2.2}\begin{equation}
ds\rightarrow d\tilde{s}=\frac{ds}{a_{\mathcal{R}}}=\left|\mathcal{R}\right|^{1/2}ds.\label{eq:2.3}\end{equation}
The action now reads\begin{equation}
\mathcal{S}\simeq\int d\tilde{s}=\int ds\,\left|\mathcal{R}\right|^{1/2}.\label{eq:2.4}\end{equation}
Upon a functional variation of $x^{\mu}$ while holding $g_{\mu\nu}$
fixed, we obtain the equation of motion for the test point mass:\begin{equation}
\frac{d^{2}x^{\mu}}{ds^{2}}+\Gamma_{\nu\lambda}^{\mu}\frac{dx^{\nu}}{ds}\frac{dx^{\lambda}}{ds}+\frac{1}{2\mathcal{R}}\frac{\partial\mathcal{R}}{\partial x^{\nu}}\left(g^{\mu\nu}+\frac{dx^{\mu}}{ds}\frac{dx^{\nu}}{ds}\right)=0.\label{eq:2.5}\end{equation}
Formally, this type of action has been considered in a more general
form of action $\int ds\, f\left(\mathcal{R}\right)$. So at the first
look, action \eqref{eq:2.4} and the equation of motion \eqref{eq:2.5}
offer nothing new, and the fact that the last term in Eq. \eqref{eq:2.5}
renders the motion non-geodesics has been well-understood. This perception
is deceptive, though. The non-geodesics nature of the motion is the
least interesting aspect in our approach. Our action \eqref{eq:2.4}
and Eq. \eqref{eq:2.5} carry a very different physics from the formal
$\int ds\, f\left(\mathcal{R}\right)$ considerations: action \eqref{eq:2.4}
arises from the change in local scale reflected in $\left|\mathcal{R}\right|^{1/2}$
and Eq. \eqref{eq:2.5} describes the motion of a point mass which
adapts its scale dynamically as it moves on the manifold. At each
new point where it arrives, the point mass {}``sees'' its surrounding
region at a new scale -- the prevailing Ricci length $a_{\mathcal{R}}$.
The physics around the point mass still obeys exactly special relativity
as well as all other laws of non-gravitational origin (such as electromagnetism
and quantum mechanics). However, the length scale for the physics
in its surroundings is now a variable denominated by the prevailing
Ricci length. All lengths are pegged to the Ricci length, and only
their ratios with respect to the Ricci length appear in the action.
Locally, the observer riding along with the point mass will not detect
anything amiss; he will not be able to tell any difference in his
surroundings since all physical laws retain their forms and all objects
around him (including his ruler) adapt their scales accordingly to
the Ricci length. For example, let the observer carry along a Michelson-Morley
interferometer; at any point along his path, although the arms of
the interferometer resize accordingly to the Ricci length, he will
unambiguously replicate Michelson-Morley's null result regarding the
interference pattern in the interferometer. Therefore, the observer
will not be able to tell which region he currently resides in. Globally,
however, if he manages to exchange light signals with another observer
residing at a location with a different value of the Ricci scalar,
he would be able to detect the effects of the relative scale difference
between the two locations. This is the new feature in our action \eqref{eq:2.4}
and equation of motion \eqref{eq:2.5} %
\footnote{This feature absolutely does not mean that Gulliver and Lilliputians
can cohabitate. They cannot. If Gulliver and a Lilliputian ever meet,
each of them will adjust to the corresponding scale $a_{\mathcal{R}}$
at their meeting location, thus both creatures having the same size.%
}.

At any given point on it trajectory, the point mass and the observer
will not feel any effect of the curvature (apart from the non-local
tidal effect). Therefore, the equivalence principle, Lorentz invariance,
and the relativity principle are satisfied locally. Note that Lorentz
invariance and the relativity principle are only satisfied locally
since the equivalence principle itself limits their validity to each
individual local region as Einstein's insight from his gedanken elevator
requires. Since action \eqref{eq:2.4} is built from invariants, the
equation of motion \eqref{eq:2.5} is explicitly tensorial and the
general covariance principle is satisfied globally.

Causality deserves an examination on its own right, however. Note
that due to the new aspect of length scale adaptation, which is missing
in formal considerations (of $\int ds\, f\left(\mathcal{R}\right)$),
we must provide a thorough look at causality as we shall do below.

\subsubsection*{\label{sub:Causality}Causality:}

For a time-like path, the total integral $\int ds\,\left|\mathcal{R}\right|^{1/2}$
is an invariant since both terms $ds$ and $\mathcal{R}$ are invariants.
If two events are causally connected, the order of the events is strictly
an invariant regardless of the coordinate choice. Although the infinitesimal
$ds$ is {}``scaled'' by a variable factor $\left|\mathcal{R}\right|^{1/2}$,
in each local region, there cannot be any coordinate transformation
that can flip the temporal order of events on a timelike trajectory.

Consider a timelike trajectory of a massive object. Along the path,
consider a series of events $A\equiv A_{1},\ A_{2},\ A_{3},\dots,A_{n}\equiv B$
each being labeled by the time coordinate $t_{A}\equiv t_{1},\ t_{2},\ t_{3},\dots,\ t_{n}\equiv t_{B}$
in which event $A_{1}$ precedes event $A_{2}$ and so on. The order
of time precedence dictates that\begin{equation}
\begin{array}{c}
t_{1}<t_{2}\\
t_{2}<t_{3}\\
\dots\\
t_{n-1}<t_{n}\end{array}\label{eq:2.6}\end{equation}
At each time point $t_{k}$ along the trajectory, the tangent frame
is Lorentz-invariant. Therefore, no choice of the local coordinate
at $t_{k}$ can alter the time order in the pair $\left\{ t_{k},\ t_{k+1}\right\} $.
Even though the differences $t_{2}-t_{1},\ t_{3}-t_{2},\dots,t_{n}-t_{n-1}$
are dependent on the coordinate choice, the signs of them are not.
The order of precedence for the series of events is strictly invariant
regardless of the coordinate system. Moreover, since no coordinate
transform can frustrate the order of precedence\begin{equation}
t_{A}=t_{1}<t_{2}<t_{3}<\dots<t_{n-1}<t_{n}=t_{B},\label{eq:2.7}\end{equation}
causality holds globally for any pair of events, $A$ and $B$, which
are arbitrarily far apart on the timelike trajectory\[
t_{A}<t_{B}.\]
Causality is protected since the timelike trajectories $(ds^{2}>0)$
and the spacelike trajectories $(ds^{2}<0)$ form two disjoint sets.
Null geodesics connecting two points $C$ and $D$ on the {}``lightcone'',
for which the total $\int_{C}^{D}ds=0$ remains vanishing for all
coordinate choices, are separatrix, strictly disconnecting the set
of timelike trajectories from the set of spacelike trajectories.

There is another way to see why causality is protected. In each local
region, since the Lorentz symmetry holds, the light speed is the upper
limit of speed for all objects in the region. No massive particle
can overcome the light speed barrier to jeopardize causality. Superluminosity
is strictly forbidden and causality is strictly enforced.

Note that the proof of causality relies solely on the requirement
that each individual local frame be Lorentz-invariant. It does not
make any resort to whether or not the intrinsic scale (in this case,
the Ricci length) is fixed or is allowed to vary from one location
to the next on the manifold. Also, it is very important to note that
the proof of causality presented above does not require the light
speed to be universsal in all local regions. The conclusion regarding
the preservation of causality even in the adaptation of length scales
to the local Ricci length has far-reaching consequences, which we
shall elaborate in Section \ref{sec:5}.

\subsubsection*{Effects on post-Newtonian phenomenology:}

If the Ricci scalar $\mathcal{R}$ varies on the manifold, the motion
of a test point mass is globally non-geodesic, although the motion
is locally geodesic. If the Ricci scalar varies little within a region
under consideration, the last term in \eqref{eq:2.5} should have
minor effects. Since we do not expect the curvature to vary much within
the solar system, corrections to solar system phenomenology are negligible,
ensuring the recovery of post-Newtonian phenomenology. The task remains
to show that the metric in solar system is almost the same as the
Schwarzschild metric \cite{Schwarzschild}: \begin{equation}
ds^{2}=\left(1-\frac{r_{s}}{r}\right)\left(c\, dt\right)^{2}-\frac{dr^{2}}{1-\frac{r_{s}}{r}}-r^{2}\left(d\theta^{2}+\sin^{2}\theta\, d\phi^{2}\right)\label{eq:2.8}\end{equation}
with $r_{s}=\frac{GM}{2c^{2}}$ being the Schwarzschild radius represented
in terms of the Newton gravitational constant $G$, the Sun mass $M$
and the speed of light $c$. This task will be addressed in Section
\ref{sec:4}.\newpage{}

\section{\label{sec:3}Lagrangian and the field equations of curvature-scaling
gravity in vacuo}

Following our previous derivation, let us consider a Lagrangian of
a uniform medium with density $\rho$. As usual, the action is\begin{equation}
\mathcal{S}\simeq\int d^{4}x\,\sqrt{g}\,\rho\label{eq:3.1}\end{equation}
with $g\triangleq-\det g$ and $d^{4}x\,\sqrt{g}$ being the (invariant)
volume element. Following our previous procedure, we replace $dx^{\mu}$
by its dimensionless ratio $d\tilde{x}^{\mu}$ normalized by the Ricci
length\[
dx^{\mu}\rightarrow d\tilde{x}^{\mu}=\frac{dx^{\mu}}{a_{\mathcal{R}}}=\left|\mathcal{R}\right|^{1/2}\, dx^{\mu},\]
or\begin{equation}
d^{4}x\rightarrow d^{4}\tilde{x}=d^{4}x\,\mathcal{R}^{2}.\label{eq:3.1-1}\end{equation}
The action becomes\begin{equation}
\mathcal{S}\simeq\int d^{4}\tilde{x}^{\mu}\,\sqrt{g}\,\rho=\int d^{4}x^{\mu}\,\mathcal{R}^{2}\,\sqrt{g}\,\rho.\label{eq:3.2}\end{equation}
Formally sending $\rho$ to zero, we obtain the Lagrangian for gravitational
field in vacuo\begin{equation}
\mathcal{S}_{vacuo}=\int d^{4}x\,\sqrt{g}\,\mathcal{R}^{2}.\label{eq:3.3}\end{equation}
The procedure of setting the uniform matter density $\rho$ to zero
is of formality. We emphasize that, in our approach, gravitational
field is not allowed to exist by itself, i.e., in isolation. It must
always couple with matter, albeit this amount of matter might be negligible.
The matter which plays the role of vacuo could be tentatively taken
to be the zero-point energy background. We shall derive the full Lagrangian
of gravity coupled with matter in Section \ref{sec:7} then show how
to obtain $\mathcal{S}_{vacuo}$ via a formality there.

The form of $\mathcal{S}_{vacuo}$ is strongly restrained. It explicitly
forbids the cosmological term, the Einstein-Hilbert term $\mathcal{R}$,
as well as all other terms, except the quadratic term. Unlike the
formal quadratic Lagrangian, though, the $\mathcal{R}^{2}$ terms
in our approach arose organically via the volume element. Each of
the two terms $\mathcal{R}^{2}$ and $\sqrt{g}$ participates in action
\eqref{eq:3.3} for a legitimate reason, rather than an arbitrary
fashion. Also, the resemblance between $\mathcal{S}_{vacuo}$ and
the quadratic Lagrangian is misleading; our full Lagrangian is not
$\mathcal{R}^{2}$ gravity when coupled with matter as we shall show
in Section \ref{sec:7}. 

For the time being, it suffices to comment that the purely quadratic
Lagrangian has been studied quite extensively in the past, see e.g.
\cite{Buchdahl1,Buchdahl2,Sotiriou1,HighOrderGrav1,HighOrderGrav2,Schmidt1,Schmidt2}.
For the sake of completeness, we nonetheless cite the results here.
Upon the functional variation of $g_{\mu\nu}$, the field equations
are\begin{equation}
\mathcal{R}\left(\mathcal{R}_{\mu\nu}-\frac{1}{4}g_{\mu\nu}\mathcal{R}\right)+\left(g_{\mu\nu}\square-\nabla_{\mu}\nabla_{\nu}\right)\mathcal{R}=0,\label{eq:3.4}\end{equation}
with $\square\triangleq\nabla^{\mu}\nabla_{\mu}$. Upon taking the
trace\begin{equation}
\square\mathcal{R}=0,\label{eq:3.5}\end{equation}
the field equations in vacuo are simplified to \begin{equation}
\mathcal{R}\left(\mathcal{R}_{\mu\nu}-\frac{1}{4}g_{\mu\nu}\mathcal{R}\right)=\nabla_{\mu}\nabla_{\nu}\mathcal{R}\label{eq:3.6}\end{equation}
which is a set of fourth-order PDEs as compared with the second-order
Einstein field equations in vacuo\begin{equation}
\mathcal{R}_{\mu\nu}-\frac{1}{2}g_{\mu\nu}\mathcal{R}=0\label{eq:3.7}\end{equation}

\subsubsection*{The well-posed Cauchy problem:}

For a theory to have predictive power, it must yield a well-behaved
evolution. This entails two issues: (i) The evolution be unique from
a set of initial conditions, and (ii) The amount of information needed
in the initial condition be manageable.

The well-posedness of the initial value problem for $\mathcal{R}^{2}$
Lagrangian has been examined in literature \cite{Noakes,Teyssandier}.We
only review its main features here. The purely quadratic Lagrangian
$\mathcal{R}^{2}$ is equivalent to a second-order tensor-scalar Lagrangian
\begin{equation}
\int d^{4}x\,\sqrt{g}\,\left(\phi\mathcal{R}-\frac{1}{2}\phi^{2}\right)\label{eq:3.8}\end{equation}
with the unknown fields being the ten metric components $g_{\mu\nu}$
and an auxiliary scalar field $\phi$. Upon functional variation of
$g_{\mu\nu}$ and $\phi$, the field equations are\begin{equation}
\begin{cases}
\ \phi\left(\mathcal{R}_{\mu\nu}-\frac{1}{4}g_{\mu\nu}\phi\right)+\left(g_{\mu\nu}\square-\nabla_{\mu}\nabla_{\nu}\right)\phi=0\\
\ \ \ \ \ \phi=\mathcal{R}\end{cases}\label{eq:3.9}\end{equation}
The scalar field $\phi$ happens to coincide with the Ricci scalar
as a consequence of the field equations \eqref{eq:3.9}. The Cauchy
problem is thus well-posed, with the Cauchy data consisting of the
values of $g_{\mu\nu},\ \partial_{0}g_{\mu\nu},\ \mathcal{R},\ \partial_{0}\mathcal{R}$
on the initial 3-hypersurface $\Sigma$. Normally one would expect
a fourth-order theory to involve $g_{\mu\nu},\ \partial_{0}g_{\mu\nu},\ \partial_{0}^{2}g_{\mu\nu},\ \partial_{0}^{3}g_{\mu\nu}$
but the $\mathcal{R}^{2}$ theory is special: although $\mathcal{R}$
involves $\partial_{0}^{2}g_{\mu\nu}$, and $\partial_{0}\mathcal{R}$
involves $\partial_{0}^{3}g_{\mu\nu}$, its Cauchy data do not require
full information of $\partial_{0}^{2}g_{\mu\nu}$ and $\partial_{0}^{3}g_{\mu\nu}$
but just a portion of it, the amount of information which is parsimoniously
and neatly encoded in $\mathcal{R}$ and $\partial_{0}\mathcal{R}$.
Being cast in this way, the evolution of the gravitational field involves
two {}``positions'' ($g$ and $\mathcal{R}$) and their {}``momenta''
($\partial_{0}g$ and $\partial_{0}\mathcal{R}$) rather than one
{}``position'' ($g$) and its three lowest time-derivatives. The
utility of $\mathcal{R}$ as a {}``position'' also nicely dovetails
with its very meaning as the scale-setter for the Ricci length which
is needed to specify the length scale for every point on the initial
hypersurface $\Sigma$.

Regarding the amount of information needed in the initial condition,
one major reason often cited in favor of the Einstein-Hilbert action
$\mathcal{R}$ is that it is a second-order theory and thus its Cauchy
data advantageously avoid second- or higher-order derivatives w.r.t.
time. In light of our discussion above, $\mathcal{R}^{2}$ theory
is not much more expensive than Einstein's gravity; beside $g$ and
$\partial_{0}g$, it only requires information regarding the values
of the Ricci length and the first-order time-derivative of the Ricci
length on the initial hypersurface $\Sigma$.

\subsubsection*{Recovery of solutions to Einstein's gravity:}

As is well-understood, the field equation for $\mathcal{R}^{2}$ gravity
in vacuo accepts all solutions to the Einstein's field equation in
vacuo including the cosmological constant \begin{equation}
\mathcal{R}_{\mu\nu}-\frac{1}{2}g_{\mu\nu}\mathcal{R}+\Lambda g_{\mu\nu}=0\label{eq:3.10}\end{equation}
(which is equivalent to, upon taking the trace, $\mathcal{R}_{\mu\nu}=\Lambda g_{\mu\nu}$).
Thus the post-Newtonian phenomenology would be recovered if the corrections
prove to be small and $\Lambda$ takes on a small value at cosmic
distances. Also note that the $\mathcal{R}^{2}$ field equations \eqref{eq:3.6}
generally adopt solutions with $\mathcal{R}\neq0$, thereby justifying
the utility of the Ricci length $a_{\mathcal{R}}$ in our theory.

The $\mathcal{R}^{2}$ field equations in vacuo are sufficiently tractable;
we shall provide two static spherically symmetric solutions in Sections
\eqref{sec:4} and \eqref{sec:6}. The $\mathcal{R}^{2}$ Lagrangian
admits a richer phenomenology. In particular it allows solutions with
non-constant $\mathcal{R}$, a fertile ground for the scale adaptation
to manifest. We shall present one such solution in Section \ref{sec:6}
which has interesting implications in the physics of black holes.
Yet even for constant-$\mathcal{R}$ solutions in vacuo, thanks to
its extra boundary conditions, the solutions carry different meaning
and application from Schwarzschild-de Sitter.%
\footnote{Solutions that deviate from the Einstein field equations with cosmological
constant can be categorized in two groups. One group of solutions
entails $\mathcal{R}\neq0$. The other group of solutions still corresponds
to $\mathcal{R}=\mbox{const}$ but the extra boundary conditions introduce
additional dimensional parameters (in terms of length). The latter
group of solutions in general cannot be transformed into a constant-$\mathcal{R}$
solutions to the Einstein field equations with cosmological constant
by an everywhere-differentiable coordinate transform. This point is
important in our consideration of the dark matter problem in Section
\ref{sec:4}.%
} The rationale to study $\mathcal{R}^{2}$ did not stem from its richness
in phenomenology, however. Rather, the $\mathcal{R}^{2}$ term arises
organically via our replacement procedure. We note that in writing
down his field equations, Einstein did not appeal to a principle but
aimed at recovering Newtonian gravity. The Einstein-Hilbert action
is simplest one that yields nontrivial phenomenology (with the action
$\int d^{4}x\,\sqrt{g}$ yielding no dynamics, for example). The main
rationale to use the Einstein-Hilbert action lies with its second-order
nature with its Cauchy data for Einstein gravity consists only $g_{\mu\nu}$
and $\partial_{0}g_{\mu\nu}$. This simplicity is not exclusive for
Einstein gravity, however, but for $\mathcal{R}^{2}$ gravity as well
as we showed above.

With Einstein's gravity began to show limitations to capture several
recent problems, such as the dark matter problem, it is a usual practice
to add new form of matter to the stress-energy tensor. Curvature-scaling
gravity, on the other hand, provides a natural modification of the
geometrical sector of the field equation. This could open a pathway
to solve the dark matter problem without resorting to some hypothetical
non-luminous matter. We shall explore this possibility in Section
\eqref{sec:4}.

\subsubsection*{Buchdahl's solution for static spherically symmetric setup:}

A fourth-order theory, $\mathcal{R}^{2}$ gravity generally requires
more boundary conditions than the second-order Einstein's field equation.
The static spherically symmetric case was examined in details over
$50$ years ago by Buchdahl \cite{Buchdahl1}. We shall revisit his
derivation, extend it and recast his solution in a more transparent
way which helps illuminate the meaning of the terms in his solution.
Our detailed derivation is presented in Appendix \ref{sec:B}. Here,
we only summarize our finding.

We deliberately cast the line element in the following form, where
$x$ is a radial coordinate variable ($dx^{0}\triangleq c\, dt$ ):
\begin{equation}
ds^{2}=\frac{4}{\mathcal{R}(x)}\left\{ \frac{p(x)}{4}\left[-\frac{q(x)}{4x}\left(dx^{0}\right)^{2}+\frac{4x}{q(x)}dx^{2}\right]+x^{2}d\Omega^{2}\right\} \label{eq:3.11}\end{equation}
in which $\mathcal{R}$ is the Ricci scalar and $p,\ q$ two auxiliary
functions, satisfying the {}``evolution'' equations: \begin{eqnarray}
\mathcal{R} & = & \pm\Lambda e^{k\int\frac{dx}{x\, q}}\nonumber \\
q_{x} & = & \left(1\mp x^{2}\right)p\label{eq:3.12}\\
p_{x} & = & \frac{3k^{2}}{4x}\frac{p}{q}\nonumber \end{eqnarray}
The most important conclusion from Buchdahl's result is that spherically
symmetric solutions to $\mathcal{R}^{2}$ gravity in vacuo require
four parameters. By virtue of the solution (\ref{eq:3.11},\ref{eq:3.12})
above, the parameters are:
\begin{itemize}
\item $\Lambda$, which measures the large-distance curvature. It is nothing
but the de Sitter parameter. In Einstein gravity, $\Lambda=0$ unless
one expressly introduces a cosmological constant. In $\mathcal{R}^{2}$
gravity, a non-zero $\Lambda$ can freely arise. It should be viewed
as the boundary condition at $x\rightarrow\infty$, thus merging with
the cosmological background at cosmic distances.
\item $k$, which allows the Ricci scalar to deviate from constant $\Lambda$.
We shall call the deviation the anomalous curvature.
\item $p\left(x_{0}\right)$ and $q\left(x_{0}\right)$ at some $x_{0}$
of user's convenience. In the formulation presented above, the meaning
of these two parameters is obscure.
\end{itemize}
In Sections \ref{sec:4} and \ref{sec:6}, we shall utilize a more
conventional choice of coordinates. As such, the parameters are no
longer the same as shown in here, with their meanings becoming more
self-evident, but the number of degrees of freedom remains $4$.\newpage{}

\section{\label{sec:4}Connection of curvature-scaling gravity to Mannheim's
conformal gravity-based treatment of galactic rotation curves}

In this section, we provide a specific solution in closed form for
the static spherically symmetric setup. We next find a connection
of our solution to a solution which Mannheim and Kazanas obtained
in the theory of conformal gravity \cite{MK1}. Based on their solution,
Mannheim has advocated a detailed phenomelogical theory for galactic
rotation curves bypassing the need of dark matter \cite{Mannheim1,MOB1,MOB2,MOB3}.
As such, his theory opens a promising possibility to resolve the long-standing
dark matter problem. From the connection of our solution to Mannheim-Kazanas's
solution, we discuss the potential transition from curvature-scaling
gravity to Mannheim's theory of galactic rotation curves.

\subsubsection*{Our solution:}

In general, the full solution for the static spherically symmetric
setup is captured in \eqref{eq:3.11} and \eqref{eq:3.12}. We are
nonetheless interested in more explicit and closed-form solutions.
Let us start with the metric in the more conventional coordinate setup\begin{equation}
ds^{2}=e^{\alpha}\left[\Psi\left(dx^{0}\right){}^{2}-\frac{dr^{2}}{\Psi}-r^{2}d\Omega^{2}\right]\label{eq:4.1}\end{equation}
in which $r\in(0,\infty)$ is the radial coordinate and $dx^{0}=c\, dt$.
The $\mathcal{R}^{2}$ gravitational field equations for the $tt$-
and $\theta\theta$-components are\begin{eqnarray}
\left(\mathcal{R}_{tt}-\frac{1}{4}g_{tt}\mathcal{R}\right)\mathcal{R} & = & -\Gamma_{tt}^{r}\mathcal{R}'\label{eq:c.4-1}\\
\left(\mathcal{R}_{\theta\theta}-\frac{1}{4}g_{\theta\theta}\mathcal{R}\right)\mathcal{R} & = & -\Gamma_{\theta\theta}^{r}\mathcal{R}'\label{eq:c.5-1}\end{eqnarray}
which will be sufficient for the two unknowns $\Psi$ and $\alpha$.
We are able to find an analytical solution; the details of our calculation
are presented in Appendix \ref{sec:C}. There are three equivalent
representations for the solution. With the length element given in
\eqref{eq:4.1}, the two unknowns $\Psi$ and $\alpha$ are: 
\begin{itemize}
\item Representation I:

\begin{equation}
\begin{cases}
\ \Psi & =\left(1-3ar_{s}\right)-\frac{r_{s}}{r}-\Lambda r^{2}+a\left(2-3ar_{s}\right)r\\
\ e^{\alpha} & =\left(1+ar\right)^{-2}\end{cases}\label{eq:4.2}\end{equation}
There are three parameters: $\Lambda$, specifying the large-distance
curvature (the de Sitter parameter); $r_{s}$, the Schwarzschild radius
taken as a free parameter (instead of $\frac{GM}{2c^{2}}$, a combination
of $G,\ M$ and $c$ separately); and $a$ an additional parameter
with unit of $\left[\mbox{length}\right]^{-1}$.

\item Representation II:

\begin{equation}
\begin{cases}
\ \Psi & =\left(1-3\beta\gamma\right)-\frac{\beta\left(2-3\beta\gamma\right)}{r}-\Lambda r^{2}+\gamma r\\
\ e^{\alpha} & =\left(1+\frac{\gamma}{2-3\beta\gamma}r\right)^{-2}\end{cases}\label{eq:4.3}\end{equation}
This representation is a resemblance of Mannheim-Kazanas's solution
in conformal gravity \cite{MK1}. The $\Psi$ term here is precisely
the same term they obtained. Our solution is thus conformal to theirs
via a conformal transformation to {}``gauge'' away the $e^{\alpha}$
term.

\item Representation III:

\begin{equation}
\begin{cases}
\ \Psi & =\sqrt{1+\kappa}-\frac{r_{s}}{r}-\Lambda r^{2}-\frac{\kappa}{3r_{s}}r\\
\ e^{\alpha} & =\left(1+\frac{1-\sqrt{1+\kappa}}{3r_{s}}r\right)^{-2}\end{cases}\label{eq:4.4}\end{equation}
$\kappa$ is an additional parameter in place of $a$ in Representation
I.

\end{itemize}
\noindent In all three representations, there appears a new term linear
in $r$ corresponding to a linear potential in addition to the Newton
potential $\frac{r_{s}}{r}$. The main conclusion from our solution
is that $\mathcal{R}^{2}$ gravity gives rise to a modification of
Newton's gravitational potential: This linear term arises from the
geometrical structure (i.e., the metric) of spacetime itself, instead
of an extra source of matter which one would usually add to the stress-energy
tensor.

\subsubsection*{Connection of curvature-scaling gravity to Mannheim's phenomenological
theory for galactic rotation curves:}

Over the last twenty years or so, Mannheim advocated a phenomenological
theory to account for galactic rotation curves. The starting point
of his approach is the static spherically symmetric solution Mannheim
and Kazanas found in conformal gravity \cite{MK1}. The metric they
obtain, up to a conformal phase factor, is\begin{equation}
ds^{2}=\Psi dt^{2}-\frac{dr^{2}}{\Psi}-r^{2}d\Omega^{2}\label{eq:4.5}\end{equation}
in which\begin{equation}
\Psi=\left(1-3\beta\gamma\right)-\frac{\beta\left(2-3\beta\gamma\right)}{r}-\Lambda r^{2}+\gamma r.\label{eq:4.6}\end{equation}
This solution is conformal to our solution presented above (see Representation
II.) Beside the usual Schwarzschild term and the de Sitter term, their
solution contains a linear term with regard to the radial coordinate.
Mannheim subsequently argued that this new term arises from the boundary
condition and the distribution details of the mass source. For sufficiently
small $\gamma$, the metric recovers the well-known Schwarzschild-de
Sitter metric. The $\gamma r$ term presents an intriguing possibility
to alter the Newtonian gravitational potential, thereby affecting
the phenomenology of gravitation at larger distances, e.g. the galactic
scales, while retaining the post-Newtonian predictions, i.e., the
three classic tests of standard Einstein gravity if $\gamma$ indeed
has a small value. We refer the reader to the body of writing by Mannheim
and collaborators \cite{MK1,Mannheim1,Mannheim2,MOB1,MOB2,MOB3} for
a comprehensive exposition of Mannheim's phenomenological treatment
of galactic rotation curves. Here our objective is to build a bridge
from curvature-scaling gravity to the Mannheim's theory.

To be concise, Mannheim's line of reasoning is that a fourth-order
gravity theory (such as conformal gravity and curvature-scaling gravity)
allows a richer set of solutions via its additional degrees of freedom.
Classically, the Newton gravitational potential can be cast in term
of the Laplace equation. In vacuo, it becomes the second-order Poisson
equation (analogous to the Gauss law in electromagnetism). The radial
part of the Laplacian operator in spherical coordinate\begin{equation}
\nabla^{2}V=\frac{1}{r^{2}}\left(r^{2}V'\right)'\label{eq:4.7}\end{equation}
which, interestingly, happens to equal to\[
\nabla^{2}V=\frac{1}{r}\left(rV\right)'',\]
where prime denotes $\partial_{r}$. The Poisson equation\begin{equation}
\nabla^{2}V=0\label{eq:4.8}\end{equation}
thus has solution:\begin{equation}
V=a-\frac{b}{r}\label{eq:4.9}\end{equation}
with $a$ and \textbf{$b$} two integration constants. This static
potential mimics the $g_{00}$ component of the Schwarzschild metric,
$g_{00}=1-\frac{r_{s}}{r}$. The radial force is thus\begin{equation}
F=-\partial_{r}V=-\frac{b}{r^{2}}.\label{eq:4.10}\end{equation}
By the same token, the fourth-order {}``Laplacian'' operator in
spherical coordinate is\begin{equation}
\nabla^{4}V=\nabla^{2}\left(\nabla^{2}V\right)=\frac{1}{r}\left(r\left(\nabla^{2}V\right)\right)''=\frac{1}{r}\left(rV\right)''''.\label{eq:4.11}\end{equation}
The fourth-order Poisson equation\begin{equation}
\nabla^{4}V=0\label{eq:4.12}\end{equation}
thus adopts the solution\begin{equation}
V=a-\frac{b}{r}+cr+\frac{d}{2}r^{2}\label{eq:4.13}\end{equation}
with $4$ constants of integration $a,\ b,\ c,\ d$. This static potential
mimics the $\Psi$ term in Eq. \ref{eq:4.6}. The corresponding centripetal
force is\begin{equation}
F=-\partial_{r}V=-\frac{b}{r^{2}}+c+dr.\label{eq:4.14}\end{equation}
The constant and linear forces, arising from the linear and quadratic
terms respectively, bring in additional centripetal accelerations.
That is to say, the $\Psi$ term is a natural output of a fourth-order
classical theory of gravitation.

Mannheim and collaborators proposed the following formula for the
centripetal acceleration of a galaxy \begin{equation}
\frac{v_{TOT}^{2}\left(R\right)}{R}=\frac{v_{LOC}^{2}\left(R\right)}{R}+\frac{\gamma_{0}c^{2}}{2}-\kappa c^{2}R\label{eq:4.15}\end{equation}
with $R$ being the distance from the galaxy's center. The local galactic
potential is computed in \cite{Mannheim1} to be \begin{equation}
v_{LOC}^{2}\left(R\right)=\frac{N^{*}\beta^{*}c^{2}R^{2}}{2R_{0}^{3}}\left[I_{0}\left(\frac{R}{2R_{0}}\right)K_{0}\left(\frac{R}{2R_{0}}\right)-I_{1}\left(\frac{R}{2R_{0}}\right)K_{1}\left(\frac{R}{2R_{0}}\right)\right]+\frac{N^{*}\gamma^{*}c^{2}R^{2}}{2R_{0}}I_{1}\left(\frac{R}{2R_{0}}\right)K_{1}\left(\frac{R}{2R_{0}}\right)\label{eq:4.16}\end{equation}
with $R_{0}$ being the scale and $N^{*}M_{\odot}=M$ the total mass
of the galaxy. Formula \eqref{eq:4.15} yields the asymptotic limit\begin{equation}
v_{TOT}^{2}\left(R\right)\rightarrow\frac{N^{*}\beta^{*}c^{2}}{R}+\frac{N^{*}\gamma^{*}+\gamma_{0}}{2}c^{2}R-\kappa c^{2}R^{2}.\label{eq:4.17}\end{equation}
Beside the mass to light ratio being the only free parameter to be
adjusted for each galaxy, these formulae involve 3 parameters $\gamma^{*},\ \gamma_{0},\ \kappa$,
taken to be universal for all galaxies. In Formula \eqref{eq:4.17},
beside the usual falling off term (the $1/R$ term), the linear and
quadratic terms are expected to alter the behavior of the velocity
curves at large value of the distance $R$. In \cite{MOB1,MOB2,MOB3}
Mannheim and colleagues applied Formula \eqref{eq:4.15} to a comprehensive
set of 138 galactic rotation curves, a wide range of galaxies which
cover different types of galaxies. They established good fit to the
rotation curves and determined the values for the three universal
parameters

\begin{equation}
\gamma^{*}=5.42\times10^{-41}\mbox{cm}^{-1},\ \ \gamma_{0}=3.06\times10^{-30}\mbox{cm}^{-1},\ \ \kappa=9.54\times10^{-54}\mbox{cm}^{-2}.\label{eq:4.18}\end{equation}
In their fits, the need for dark matter was avoided. The linear potential
gives rise to a departure from Newton-Einstein at large distances,
precisely in the range wherever the dark matter problem is encountered.
With their values obtained in \eqref{eq:4.18}, the effects of $N^{\lyxmathsym{\textasteriskcentered}}\gamma^{*},\ \gamma_{0}$
and $\kappa$ only become as big as the Newtonian contribution at
galactic scales. Mannheim thus concludes that the phenomenology overcomes
the fine-tuning shortcomings of the dark matter-based framework. Since
then, his theory has become an active field of research \cite{CGother1,CGother2,CGother3,CGother4,CGother5,CGother6,CGother7,CGother8,CGother9,CGother10,CGother11}.

Given the impressive success of Mannheim's phenomenological treatment
for galactic rotation curves with parsimonious assumptions, it would
be interesting to see if curvature-scaling gravity can provide an
alternative logical foundation -- beside conformal gravity -- for
his theory. Like conformal gravity, curvature-scaling gravity is a
fourth-order theory; therefore, the classical fourth-order Poisson
equation argument repeated above is also applicable to it. Indeed,
curvature-scaling gravity does accept a solution (\ref{eq:4.1}, \ref{eq:4.2},
\ref{eq:4.3}, \ref{eq:4.4}) which is conformal to Mannheim-Kazanas's
solution (\ref{eq:4.5}, \ref{eq:4.6}) in conformal gravity. The
particular potential function that Mannheim employs is not specific
to the conformal gravity theory but rather is a generic feature, valid
in curvature-scaling gravity as well. As such, Mannheim's theory of
galactic rotation curves may not necessarily be an exclusive product
of conformal gravity but it can arise from curvature-scaling gravity,
or any fourth-order gravity in general. If this is the case, then
Mannheim's phenomenological theory and results would be quite robust. 

In recognition of Mannheim's endeavoring emphases on the importance
of the linear $\gamma r$ term in the behavior of galactic rotations,
we shall call this term (in curvature-scaling gravity, that is) the
Mannheim-Kazanas term, and $\gamma$ the Mannheim-Kazanas parameter.%
\footnote{We must note that the spherical solution to conformal gravity had
previously been obtained in \cite{Schimming}.%
} It is noteworthy that the $\gamma r$ term will also appear in another
solution that we shall derive in Section \ref{sec:6}. The $\gamma$
parameter will play an important role in the physics of black holes
in the context of our curvature-scaling gravity in that section.

In passing, the extension of Mannheim's idea, essentially the linear
Mannheim-Kazanas term playing a crucial role, to address the Pioneer
anomalies has been studied recently \cite{Varieschi1,Varieschi2,Varieschi3}.
Given that such a linear term is also present in our solution, we
hope that curvature-scaling gravity could be of relevance to these
studies as well.

\subsubsection*{On the inapplicability of Birkhoff's theorem for conformal gravity
and $\mathcal{R}^{2}$ gravity:}

The metric we found in (\ref{eq:4.1}, \ref{eq:4.2}, \ref{eq:4.3},
\ref{eq:4.4}) has a constant Ricci scalar. That is the reason why
only 3 parameters are explicitly present in the metric. The fourth
parameter which specifies the deviation of curvature from constancy
is zero. As such, we have to address whether our metric is equivalent
to the Schwarzschild-de Sitter metric which also has a constant Ricci
scalar, according to Birkhoff's theorem.

It is well-understood that Birkhoff's theorem does not hold for high-order
gravity; see \cite{Clifton1}, for example. Explicit solutions that
violate Birkhoff's theorem for $\mathcal{R}^{1+\delta}$ gravity have
also been found in \cite{Clifton2}. These solutions have non-constant
Ricci scalar.

But even for metrics with a constant Ricci scalar, there are technical
subtleties that prevent Birkhoff's theorem to be applicable for higher-order
gravity. This is our focus in the rest of this section. To be concise,
it is the extra boundary conditions present in a higher-order theory
that require additional dimensional parameters (in terms of length)
to specify them. These dimensional parameters cannot be transformed
away by an everywhere-differentiable coordinate transformation.

More concretely, consider the metric (\ref{eq:4.1}, \ref{eq:4.2})
in which $r\in(0,\infty)$ and $a>0$:\begin{eqnarray}
ds^{2} & = & e^{\alpha}\left[\Psi dt{}^{2}-\frac{dr^{2}}{\Psi}-r^{2}d\Omega^{2}\right]\nonumber \\
\Psi & = & \left(1-3ar_{s}\right)-\frac{r_{s}}{r}-\Lambda r^{2}+a\left(2-3ar_{s}\right)r\nonumber \\
e^{\alpha} & = & \left(1+ar\right)^{-2}\nonumber \\
\Lambda & = & \frac{\mathcal{R}_{0}}{12}+a^{2}\left(ar_{s}-1\right).\label{eq:4.18-1}\end{eqnarray}
Let us tentatively make the coordinate change\begin{equation}
r=\frac{r'}{1-ar'}.\label{eq:4.19}\end{equation}
It is straightforward to verify that, in the new coordinate $r'$,
the line element reads\begin{equation}
ds^{2}=\left(1-ar'\right)^{2}\Psi dt^{2}-\frac{d\rho^{2}}{\left(1-ar'\right)^{2}\Psi}-r'^{2}d\Omega\label{eq:4.19-1}\end{equation}
and that\begin{equation}
\left(1-ar'\right)^{2}\Psi=1-\frac{r_{s}}{r'}-\frac{\mathcal{R}_{0}}{12}r'^{2}\label{eq:4.19-2}\end{equation}
Thus\begin{equation}
ds^{2}=\left(1-\frac{r_{s}}{r'}-\frac{\mathcal{R}_{0}}{12}r'^{2}\right)dt^{2}-\frac{dr'^{2}}{1-\frac{r_{s}}{r'}-\frac{\mathcal{R}_{0}}{12}r'^{2}}-r'^{2}d\Omega^{2}\label{eq:4.20}\end{equation}
which is formally the Schwarzschild-de Sitter metric. However, this
formality fails (thus rendering Birkhoff's theorem inapplicable for
conformal gravity and curvature-scaling gravity) for four following
reasons:
\begin{enumerate}
\item The coordinate transform is not differentiable everywhere. It is not
analytic at $r'=\frac{1}{a}>0$.
\item Related to Point (1), the mapping is not entire. To cover the range
$r\in\left(0,\infty\right)$, the new coordinate only needs to cover
$r'\in\left(0,\frac{1}{a}\right)$; namely, only a finite range $\left(0,\frac{1}{a}\right)$
is needed to label the space. The metric above would be seen as Schwarzschild-de
Sitter if it held for the entire half axis. Since only a finite interval
$\left(0,\frac{1}{a}\right)$ is mappable to the physical region $r\in\left(0,\infty\right)$,
it does not approach the required limit as $r'\rightarrow\frac{1}{a}$
from below. Say, for $\Lambda'=0$, it is not asymptotically flat
as $r\rightarrow\infty$ (i.e., $r'\rightarrow\frac{1}{a}$ from below).
\footnote{We thank Mannheim for his affirmation of these technical points, which
himself and O'Brien utilized in \cite{MOB1,MOB2,MOB3}.%
}
\item The proof of Birkhoff's theorem is prejudicial in its choice of coordinates.
It starts with the following choice:\begin{equation}
ds^{2}=Adt^{2}-Bdr^{2}-r^{2}d\Omega^{2}\label{eq:4.21}\end{equation}
from which it deduces from the Einstein field equations that $B=A^{-1}$.
This choice however expressly suppresses the freedom for the term
$a$ in metric \eqref{eq:4.18-1} from being anything but zero, to
start with. Therefore, it is not surprising that Birkhoff's conclusion
is obtained. In Einstein gravity, this restriction poses no problem
since Einstein gravity is a second-order theory. For fourth-order
gravity such as conformal gravity or $\mathcal{R}^{2}$ gravity, there
are two more degrees of freedom in their static spherically symmetric
solutions. In both theories, one of the additional parameters is $a$
(or equivalently, the Mannheim-Kazanas parameter $\gamma$). The other
remaining degree of freedom is manifest only in $\mathcal{R}^{2}$
gravity, however. It measures the deviation of curvature away from
constancy -- we shall call it the anomalous curvature, controlled
by a new parameter $\epsilon$. The topics of anomalous curvature
and $\epsilon$ will be covered in Section \ref{sec:6}.
\item The additional boundary conditions in higher-order gravity introduce
new length scale to the solutions. In Einstein gravity, the freedom
of scale is broken in the form of the Schwarzschild radius $r_{s}$
(which is of unit of length). In $\mathcal{R}^{2}$ gravity, two extra
length scales appear in the form of $a$ (or, equivalently, $\gamma$)
and $\epsilon$ which are of dimension of $\left[\mbox{length}\right]^{-1}$
and $\left[\mbox{length}\right]^{-2}$ respectively.
\end{enumerate}
In conclusion, it is not permissible to make a coordinate transformation
to remove the Mannheim-Kazanas linear term, $\gamma r$, in either
conformal gravity or curvature-scaling gravity. This term is not an
artifact of coordinate choices; rather, it is set by boundary conditions,
and has physical and detectable effects on the motion of matter in
gravitational field when it is present %
\footnote{It will be interesting to explore how this conclusion affects gravitational
waves since Birkhoff's theorem no longer applies for $\mathcal{R}^{2}$
gravity. This question has been raised in \cite{Clifton1}, for example.%
}.

\newpage{}

\section{\label{sec:5}Logical inferences from the postulate of Ricci scalar
as dynamical scale-setter}

\subsection{\label{sub:Tannisotropy}The anisotropy in time scaling}

In a recent proposal put forward by Ho\v{r}ava \cite{Horava}, time
is treated anisotropically. In Ho\v{r}ava-Lifshitz gravity the time
direction, unlike the spatial directions, acquires a dynamical exponent
$z=3$ rather than $1$ in its scaling in the Wilsonian renormalization
group process. The anisotropy is invoked to deal with the inherent
asymmetry of the role of time vs. space in the dynamics of physical
processes. The Lorentz symmetry is expressly broken in the ultraviolet
limit, and is expected to emerge in the infrared limit, thus restoring
the space-time isotropy in the classical limit.

Interestingly the anisotropy of time vs. space also arises quite naturally
in our approach. Consider the Schrödinger equation for the Hydrogen
atom\begin{equation}
i\hbar\frac{\partial}{\partial t}\Psi=-\frac{\hbar^{2}}{2m_{e}}\nabla^{2}\Psi-\frac{e^{2}}{r}\Psi.\label{eq:5.1}\end{equation}
Let us recall that in Section \ref{sec:1}, Postulate (I) mandates
that the form of Eq. \eqref{eq:5.1} be the same in all pockets of
spacetime, whereas Postulate (II) requires that the length scale in
Eq. \eqref{eq:5.1} be pegged to the local Ricci length, defined as
$a_{\mathcal{R}}\triangleq\left|\mathcal{R}\right|^{-1/2}$. Consequently,
these two conditions mean that Eq. \eqref{eq:5.1} is invariant when
expressed in terms of dimensionless ratios of space and time using
the Ricci length as denominator. More concretely, rewriting the coordinate
differentials as\begin{equation}
\begin{cases}
\ d\vec{x} & =\ a_{\mathcal{R}}\, d\vec{\tilde{x}}\\
\ dt & =\ a_{\mathcal{R}}^{\eta}\, d\tilde{t}\end{cases}\label{eq:5.2}\end{equation}
in which we introduce an anomalous scaling exponent $\eta$ for time,
the Schrödinger equation \eqref{eq:5.1} becomes\begin{equation}
i\frac{\hbar}{a_{\mathcal{R}}^{\eta}}\frac{\partial}{\partial\tilde{t}}\Psi=-\frac{\hbar^{2}}{2m_{e}a_{\mathcal{R}}^{2}}\tilde{\nabla}^{2}\Psi-\frac{e^{2}}{a_{\mathcal{R}}\tilde{r}}\Psi,\label{eq:5.3}\end{equation}
or\begin{equation}
i\left(\hbar\, a_{\mathcal{R}}^{1-\eta}\right)\frac{\partial}{\partial\tilde{t}}\Psi=-\frac{\left(\hbar\, a_{\mathcal{R}}^{-1/2}\right)^{2}}{2m_{e}}\tilde{\nabla}^{2}\Psi-\frac{e^{2}}{\tilde{r}}\Psi.\label{eq:5.4}\end{equation}
The invariance requirement alluded above then enforces two conditions
that\begin{equation}
\begin{cases}
\ \hbar\, a_{\mathcal{R}}^{1-\eta} & =\ \mbox{const}\\
\ \hbar\, a_{\mathcal{R}}^{-1/2} & =\ \mbox{const}\end{cases}\label{eq:5.5}\end{equation}
Equating the exponents of the two questions, we obtain\[
\eta=\frac{3}{2}\]
which means an anisotropy in the scaling for time interval\begin{equation}
dt\propto a_{\mathcal{R}}^{3/2}\propto\left|\mathcal{R}\right|^{-3/4}.\label{eq:scaleT}\end{equation}
This conclusion is not limited to the nonrelativistic Schrödinger
equation for the Hydrogen atom. Rather, it holds generically for any
physical system. In particular, it holds for the partition of QED
which encompasses the (relativistic) Dirac equation and the Maxwell
equations of electromagnetism, as shown in Appendix \ref{sec:A}.
The emergence of time anisotropy is a conceptual conclusion derived
solely from Postulates (I) and (II). The time anisotropy is inherent
in all physical laws established to date.

As the observer moves from one local region to the next on the manifold,
the Ricci length varies (together with the metric $g_{\mu\nu}$),
but the observer would not be able to detect anything amiss using
objects, rulers and clocks within his surroundings as reference. Everything
in his surroundings scales in exact proportion, and every clock in
his surroundings speeds up or slows down in sync. For example, if
he is to conduct the Michelson-Morley experiment, he will obtain a
unique value for the speed of light regardless of his state of motion
and/or the direction of the light beam. Wherever he sits on the manifold,
the laws of physics look the same; it is only his ruler and clock
and all objects around him (including his own body size and his own
heart rate) adjust to the prevailing value of the Ricci scalar. The
observer, however, in principle can detect the discrepancies between
his region and another region by exchanging light signals with an
observer residing in the other region. We thus conclude:\bigskip{}

\textbf{The scaling of time duration:}
\begin{quote}
\emph{The time duration of physical processes (i.e., the inverse of
the oscillatory rate of clocks) at a given point on the spacetime
manifold obeys the scaling rule $dt\propto a_{\mathcal{R}}^{\eta}$
with $\eta=\frac{3}{2}$ and $a_{\mathcal{R}}$ being the local Ricci
length, defined as $a_{\mathcal{R}}\triangleq\left|\mathcal{R}\right|^{-1/2}$.}
\end{quote}

\subsection{\label{sub:Hvariable}The first conceptual departure}

The utility of the Ricci scalar as the local scale-setter for physical
laws leads to several conceptual departures from the conventional
belief. By the very nature of the equivalence principle, physical
laws are valid locally; as such, it is conceivable that the parameters
of the physical laws are also valid only locally. One of the conceptual
casualties we shall encounter is the abandonment of $\hbar$ as a
universal constant. Whilst the laws of physics -- say, the Schrödinger
equation -- retain their forms in each individual local region, the
parameters -- in this case, $\hbar$ -- which keep track of the laws
need not be universal. This can be seen directly from the conditions
\eqref{eq:5.5} above that\begin{equation}
\hbar\propto a_{\mathcal{R}}^{1/2}=\left|\mathcal{R}\right|^{-1/4}\label{eq:scaleH}\end{equation}
namely, the value of the Planck constant depends on the prevailing
Ricci scalar. At first this result looks alarming enough. After all,
it is an undisputable fact that $\hbar$ is a fundamental constant
which ubiquitously governs different branches of physics -- from photons
and quarks in particle physics, to phonons, magnons, quantum Hall
effect in condensed matter physics, to the periodic table in quantum
chemistry, to the alpha decay and the shell model in nuclear physics,
to name a few. For all these diversified branches of physics, it has
been established beyond doubt that there is one single value of $\hbar$.

Our approach unequivocally preserves the all-embracing domain of applications
of quantum physics. The Planck constant retains its sovereignty over
all quantum physical processes that occur in each individual local
region. Yet it is in no contradiction with the conceivability that
its value is valid only locally to each region. Moreover, it is imperative
to realize that the ubiquity in the value for $\hbar$ in all these
branches of physics has been based on measurements conducted in the
surrounding region of the Earth %
\footnote{One might object that light from distant galaxies and supernovae appears
to support a universal value of $\hbar$ as well. We shall dispense
with this point momentarily in the next two subsections \eqref{sub:Cvariable}
and \eqref{sub:invitation}.%
}. Since $\hbar$ does govern wide-ranging physics in a region -- i.e.,
our region -- it was natural to generalize its value to everywhere
else on the spacetime manifold. This extrapolation in the value of
$\hbar$ is overreaching and unnecessary theoretically, and has not
been falsified experimentally.

The Planck constant thus should not be taken for granted to be an
omnipotent fundamental; rather, in each local pocket on the manifold,
$\hbar$ should be allowed to adapt to the prevailing value of the
Ricci scalar $\mathcal{R}$ in the precise relationship \eqref{eq:scaleH},
$\hbar\propto\left|\mathcal{R}\right|^{-1/4}$. Note that this is
not a triviality of choosing a set of measurement apparatus at the
observer's discretion -- an idea often circulated in literature %
\footnote{\label{fn:Ellis}This issue has been misinterpreted, if not misunderstood,
in various papers in literature. One favorite critique pretends that
a dimensional $\hbar$ can be made to take any value at the observer's
will, by changing the units of length and time; e.g., see \cite{Ellis}.
Although handy, this argument is naïve and false for at least three
reasons. First, in our approach, the rulers are prepared identical,
and the clocks synchronized. The ratio in the values of $\mathcal{R}$
at two different locations is an objective dimensionless number; so
is the ratio in the values of $\hbar$ at those said locations regardless
of the choice of units. Second, the same critique must also dismiss
the Lorentz contraction and time dilation as a matter of unit choice
since length and time are dimensional. This conclusion is wrong; time
dilation is a real effect detectable in the lifetime of muons created
in the upper stratosphere yet able to reach the Earth's sea level.
Third, when this sort of critique dismisses the non-universality for
$\hbar$ as a matter of units, it simultaneously rejects the very
meaning of the universality of $\hbar$ which these authors deem sacred
(Wouldn't the equality of $\hbar$ at different locations be a matter
of unit choice too?) As such, this self-defeating critique should
be exorcised from the scientific vocabulary.%
}. The ratio of the Ricci scalar at two different locations is an objective
measure, and thus the ratio of in values of $\hbar$ at two different
locations is also an objective measure without being subject to any
arbitrary choice of ruler and clock at the observer's disposal. The
abandonment of the universality in the value of $\hbar$ does no violence
to any physical principles of special relativity, the equivalence
principle, the general covariance principle, as well as quantum physics.

There is a more intuitive way to obtain the scaling rule \eqref{eq:scaleH}
for $\hbar$. At first the astute reader might have noticed that in
Einstein's gedanken elevator, the size of every object should be determined
by quantum mechanics via, say, the Schrödinger equation which governs
its atoms. For example, the Bohr radius $r_{Bohr}=\frac{\hbar^{2}}{m_{e}e}$
of the Hydrogen atom should be fully fixed by constants of Nature,
such as $\hbar$. Equipped with this mindset, the reader may object
that the size of an object is complied with $\hbar$ rather than the
Ricci scalar $\mathcal{R}$ per our Postulate (II). Our approach,
however, is a radical departure from the orthodox view: we hold that
the Ricci length is of a more fundamental status than is the Bohr
radius. The Schrödinger equation retains exactly the same form in
every local spacetime region that the elevator falls though. Yet,
the Bohr radius -- and equivalently $\hbar$ -- is pegged to the Ricci
length, and thus is allowed (indeed required) to vary together with
$\mathcal{R}$ from point to point on the manifold. With $m_{e}$
and $e$ taken as intrinsic quantities of the electron, given that
$r_{Bohr}$ is pegged to $a_{\mathcal{R}}$ (i.e., $r_{Bohr}=\frac{\hbar^{2}}{m_{e}e}\propto a_{\mathcal{R}}$)
one can immediately deduce that $\hbar\propto a_{\mathcal{R}}^{1/2}$
in agreement with the scaling rule above. The (re-)assignment of fundamental
role from $\hbar$ to $\mathcal{R}$ in setting the length scales
for physical laws also has a conceptual appeal: spacetime should provide
the length scale against which non-gravitational processes are measured,
rather than the other way around.

Although the size of physical objects and the beating rate of clocks
can vary from one local region to the next on the manifold, the observer
cannot simply bring identical objects from two different regions together
side-by-side to catch out their relative difference. If he manages
to do so, the objects would adjust their sizes to the new Ricci length
where they sit and no relative difference in size would manifest.
Likewise, he cannot simply bring synchronized clocks from two different
regions together side-by-side to catch out a mismatch in their oscillatory
rates. If he manages to do so, the clocks will adjust their vibrations
to the new local scale and click in sync. To detect objective discrepancies
between two regions, the observer must compare the wavelengths and
frequencies of light signals exchanged between the two regions. This
procedure is not unlike the standard practice astronomers use to detect
the redshift of light from distant galaxies. Once again, in our illustration
so far, the rulers are identical, and the clocks synchronized. The
objective and relative differences in rulers' size and clocks' rate
arises from the difference in the values of the Ricci scalar in the
two locations.

The anisotropic time scaling leads to another conceptual departure
which is our focus in the following section.

\subsection{\label{sub:Cvariable}The second conceptual departure and the preservation
of causality}

Let us first review the essence of the Michelson-Morley experiment.
The Michelson-Morley interferometer consists of two light beams, one
beam -- the {}``longitudinal'' arm -- traveling along and against
the Earth's direction of motion in the solar system, whereas the other
beam -- the {}``transverse'' arm -- traveling perpendicularly to
the Earth's motion. The two beams are then made to meet and interfere.
The two beams were initially expected by Michelson and Morley to accumulate
different travel times because of their different alignments with
regard to the Earth's motion. As such, they were expected to experience
a relative phase shift and thus should produce an interference pattern.
The result that Michelson and Morley obtained was null; no interference
was found. Since then, the Michelson-Morley finding has been reconfirmed
with increasing accuracy. Their finding is interpreted in special
relativity as solid evidence in support of a constancy in $c$, that
light travels at the same speed regardless of its direction, whether
along or against or transverse to the Earth's motion in the solar
system.

We hold the Michelson-Morley experimental finding and Einstein's theoretical
conclusion of the constancy of $c$ as well-established truths, and
we shall enforce them as venerable facts in our approach. Nonetheless
it is imperative to realize that the Michelson-Morley experiment only
confirms the common value of $c$ for light beams that travel at the
same location, viz. within their interferometer stationed in Cleveland.
If we are to bring their interferometer to, say, the city of New York
(NYC), and repeat their experiment at the new place, their null finding
and thus a common value of $c$ for the two light beams within the
interferometer now residing in NYC should not be in question. However,
it is perfectly legitimate to question the equality between the Cleveland-bound
value of $c$ and the NYC-bound value of $c$. The Michelson-Morley
interferometer has nothing to say as of whether or not the values
of $c$ encountered at Cleveland and at NYC are identical; it was
not designed to answer that question. The Michelson-Morley experiment
commands no authority whatsoever over the equality (or the lack thereof)
of the values of $c$ measured at different locations. Indeed there
has never been any experimental or observational evidence in support
of a universal value for $c$ at different locations in spacetime
\footnote{There have been repeats of the Michelson-Morley experiment using star-light
sources. But this detail is irrelevant since in this type of experiments
the two light beams still met at the same location on Earth and the
common value of $c$ was obtained in the Earth-based lab. This type
of experiments have nothing to say about the values of $c$ at the
respective locations of the two light sources.%
}. This is a subtle yet significant point which has been underappreciated
in the development of relativity.

The importance of this observation of ours cannot be overstated. Imagine
a series of observers $\mathcal{O}_{1},\ \mathcal{O}_{2},\ \dots,\ \mathcal{O}_{n}$
aligned on a light beam emitted from a light source. Let each observer
separately measure the speed of the light beam as it passes by his
location. Regardless of the motion of Observer $\mathcal{O}_{k}$
(whether he runs toward the light source, away from it, or sideway),
he must record a unique value $c_{k}$ for the light speed at his
location. This is the requirement of the constancy of the speed of
light which we unconditionally enforce in our theory. Yet, the Michelson-Morley
finding has nothing to say about the equality (or the lack thereof)
for the members in the serial $\left\{ c_{1},\, c_{2},\,\dots,\, c_{n}\right\} $.
Each element in the series can depend on the state of spacetime where
it is measured. In curvature-scaling gravity, the said state of spacetime
is the Ricci scalar.

Let us prepare two identical replicas of the Michelson-Morley interferometer
and a pair of synchonized clocks at our location, denoted by $A$,
then send one replica and one clock to a remote location $B$. Let
us assume that the two locations have different values for the Ricci
scalar, such that $\mathcal{R}\left(B\right)=\frac{1}{4}\,\mathcal{R}\left(A\right)$.
The Ricci lengths at the two locations are related as $a_{\mathcal{R}}\left(B\right)=2\, a_{\mathcal{R}}\left(A\right)$.
The $B$-interferometer adapts its size accordingly; i.e., the length
of its two arms scale up by a factor of $2$. At the same time, the
$B$-clock which measures the vibration rate of atoms and light beams
at $B$ scales anisotropically per the time scaling rule \eqref{eq:scaleT}:
it requires an increase of $2^{\,3/2}$-fold for the atoms at $B$
to make a full vibration as compared with the atoms residing at $A$
to complete a full cycle. That is to say, the $B$-clock beats at
a rate $2\sqrt{2}$-fold slower than the synchronized $A$-clock.
The net result is that light travels at $B$ at a slower speed compared
with the light speed at $A$. This is because it requires a $2\sqrt{2}$-fold
more amount of time for light at $B$ to cover twice as long a distance,
thus: $c\left(B\right)=\frac{1}{\sqrt{2}}\, c\left(A\right)$. So,
each interferometer separately still registers a common value for
$c$ at its respective location -- that is to say, each individual
interferometer replicates the Michelson-Morley result at its respective
location -- yet the interferometers are free to take different values
of $c$ relative to each other. This is the second conceptual casualty
being deduced from our Postulates (I) and (II). The value of $c$
depends on the prevailing value of the Ricci scalar where the light
beam is measured according to the following scaling rule:\begin{equation}
c\propto a_{\mathcal{R}}^{-1/2}=\left|\mathcal{R}\right|^{1/4}.\label{eq:scaleC}\end{equation}
For the observers mentioned above, the values of light speed they
obtain are thus dependent on the values of the Ricci scalar where
they are located, $\left\{ c_{1},\, c_{2},\,\dots,\, c_{n}\right\} \propto\left\{ \left|\mathcal{R}_{1}\right|^{1/4},\,\left|\mathcal{R}_{2}\right|^{1/4},\,\dots,\,\left|\mathcal{R}_{n}\right|^{1/4}\right\} $.
This is, again, not a triviality in the choice of rulers and clocks
at the observers' disposal %
\footnote{See the related comment in footnote \vref{fn:Ellis}. %
}. The ratio $\mathcal{R}_{1}/\mathcal{R}_{2}$, e.g., is an objective
measure of the spacetime manifold; as the result, the ratio $c_{1}/c_{2}$
is also an objective measure.

The locality nature of the Lorentz symmetry, the relativity principle,
and the value of $c$ is enshrined in the very content of the equivalence
principle. The equivalence principle mandates that special relativity
hold locally in the tangent frames of each given point on the manifold.
It is not meaningful whatsoever to talk about a global scope for Lorentz
invariance, the relativity principle, and -- as we now see -- the
value of $c$ in a curved spacetime, although it is meaningful to
do so for a Minkowski spacetime. (On the other hand, causality, i.e.
the order in precedence of causally-connected events, is global in
all configurations of spacetime as we shall show right below.) Together
with $\hbar$, $c$ continue to oversee the physics within each local
region: $\hbar$ ubiquitously measures the strength of quantum effects
in different branches of physics, and $c$ controls the Lorentz symmetry.
Yet, under no premises does this feature prevent them from acquiring
a new value from one location to the next. Forcing them to have a
universal value -- a custom taken for granted in conventional wisdom
-- is an overreaching and unnecessary practice, if not an unfulfillment
of the spirit of the equivalence principle.

The lack of experimental support for a universal $c$ notwithstanding,
the reader might insist: Could there still be an underlying theoretical
reason which might otherwise enforce the equality in the value of
$c$ for all locations? In particular: (i) Would the causality principle
be the justifying agent, perhaps? (ii) Or, how then would a variability
of $c$ manage to avoid the troublesome superluminosity? (iii) Last
but not least, wouldn't a variability of $c$ in spacetime directly
violate the Lorentz symmetry? Ubiquitously, the equality in the value
of $c$ for all locations has been prejudiciously rationalized out
of the fears of an endangerment of causality and, related to it, an
encounter of faster-than-light travel. These fears are unwarranted
as we shall show momentarily. Our theory strictly forbids superluminosity
and strictly protects causality.

The proof of causality was provided in section Causality \vpageref{sub:Causality}.
For a pair of events no matter how far apart they are on the manifold,
as long as they are causally connected, the order of their precedence
is strictly protected. The proof only relies on the Lorentz invariance
for each individual local frame; it does not depend on whether or
not the speed of light is uniform along the timelike trajectory connecting
the two events. In the series of events \begin{equation}
t_{A}=t_{1}<t_{2}<t_{3}<\dots<t_{n-1}<t_{n}=t_{B}\label{eq:5.9}\end{equation}
although the differences $t_{2}-t_{1},\ t_{3}-t_{2},\ \dots,\ t_{n}-t_{n-1}$
are dependent on the coordinate choice, their signs are strictly positive.
No coordinate transformation can frustrate the order of precedence
\eqref{eq:5.9}. Timelike- and spacelike- trajectories do not mix.
Null geodesics remain null geodesics in all coordinate choices. In
all global coordinate systems, null geodesics strictly separate the
set of timelike paths from the set of spacelike paths. No coordinate
change can transform a spacelike path into a timelike path or vice
versa. Another way to view the denouement is that since our theory
only employs invariants (including $c$ itself which -- albeit a variable
-- is a Lorentz invariant thanks to its one-to-one correspondence
with the (invariant) Ricci scalar, $c\propto\left|\mathcal{R}\right|^{1/4}$),
causality is automatically ensured.

Preceding the lack of universality in light speed %
\footnote{It would be apt to instead say {}``the loss of universality in light
speed'' from the historical perspective.%
}, the fears of violation of the Michelson-Morley result and of causality
have been misplaced:
\begin{enumerate}
\item The speed of light at a given point and a given instant is the same
for all light beams that pass through that given point at that given
instant regardless of the directions of the light beams, viz. whether
they are traveling along or against the Earth's motion, for example.
By the same token, regardless of the motion of the observer, whether
he travels along or against a light beam, he will record one common
value for $c$ for the light beam at his location. Point (1) is the
very content of Einstein's postulate of constant light speed and of
the Michelson-Morley's experimental conclusion. These results are
fully retained in our approach.
\item The speed of light at a given point and a given instant is the maximum
velocity that any physical object can attain when it travels across
that given point at that given instant. Point (2) is a direct outcome
of the Lorentz symmetry -- a combination of Einstein's postulate of
constant light speed and his relativity principle. This conclusion
is also fully preserved in our approach. No objects could surpass
light at any point in space and at any instant in time. At any given
point on the manifold, light provides the upper bound of speed for
all objects. The fact that the upper bound of speed is variable from
one point to the next is irrelevant. Superluminosity is strictly forbidden.
As such, causality is strictly preserved.
\end{enumerate}
Another misplaced objection raised above is that a variability of
$c$ would necessarily endanger Lorentz invariance. This objection
is false. The Lorentz symmetry would be in jeopardy only if $c$ is
an explicit function of spacetime coordinates. In our approach, $c$
is not a function directly of spacetime coordinates. Rather, it is
solely a function of the intrinsic Ricci scalar which acquires a dynamics
via that of $g_{\mu\nu}$. At each point on the manifold, via \eqref{eq:scaleC}
the Ricci scalar appoints the value for $c$ which then enters the
length element for the local region enclosing the point. The length
element possesses the Lorentz symmetry. As such, Lorentz invariance
is enforced locally. We stress that $c$ (and $\hbar$ too) are not
auxiliary fields that live on the manifold. They do not have a dynamics
on their own right. That is to say, there are no terms such as $\partial_{t}c$
or $\vec{\nabla}\hbar$ present in our theory.

We attribute the conundrum to confusion in nomenclature: {}``constancy''
versus {}``universality''. Whereas {}``constancy'' means a common
value for $c$ at a given location regardless of the state of observer
and/or emitter (for example, regardless of the direction of the light
beam along or against the Earth's motion and/or regardless of whether
the observer is running along or against the light beam), {}``universality''
on the other hand means a common value for $c$ measured at different
locations. The {}``constancy'' of $c$ by no mean implies its {}``universality''.
Often, though, the two concepts are automatically paired hand-in-hand
out of fear to safeguard causality and ensure the Michelson-Morley
result everywhere on the manifold. The fear and practice are unwarranted;
it is permissible to construct a manifold -- as is done in our approach
-- which fully respects causality while preserving the constancy for
$c$ everywhere (i.e., ensuring the Michelson-Morley result at every
individual location) yet disowning its universality. Moreover, from
the constancy of $c$, any generalization to a universality for $c$
has been an overreaching act. The statement often taken for granted
that $c$ takes the same value everywhere does not reflect an experimentally
verified fact, but a presumption\emph{. }To be upheld with scientific
value, the assertion that the speed of light is universal needs be
falsified\emph{.} There have not been any experiments that falsify
it or even attempt to falsify it. The Michelson-Morley setup did not.
Neither have any experiments to date. Whereas the constancy-to-universality
generalization is valid for special relativity since all points in
a Minkowski spacetime are equal (no point in a flat spacetime holding
a privileged position over the rest), the generalization fails for
general relativity in which spacetime is inherently curved with local
regions no longer being identical. Spacetime points are not created
equal -- the Ricci scalar in principle varies from point to point.
General relativity breaks spacetime manifold into pockets each with
its own scalar curvature and thus with -- per our postulate on the
role of the Ricci scalar -- its own physical parameters, with $c$
and $\hbar$ being the most notable. The overreaching generalization
then is a relic of special relativity, a relic that needs be repudiated
in a general theory of spacetime %
\footnote{In getting acquainted with special relativity, the students are ubiquitously
familiarized with popular examples of Einstein's light-years-long
trains and twins taking light-years-long journeys. These examples
have not helped; indeed they sowed the seed of a misperception that
the speed of light must be uniform along all the way from Earth to
distant stars light years away even in curved spacetime. We need to
exorcise this misconception and correct the unfortunate course of
history.%
}.

Considering the perplexity of the Michelson-Morley finding compared
with one's daily life's {}``common sense'', as soon as the Michelson-Morley
conclusion crystallized in special relativity, it has been a safe
practice to superficially extrapolate its (correct) conclusion of
{}``constancy'' for $c$ to an (overreaching) speculation of {}``universality''
for $c$ and subconsciously hold the (unverified) presumption as an
unassailable truth. And it has been an equally safe practice to disregard
any serious attempt to challenge such a misperception as unworthy
of healthy discussions. Historical accounts indicate that, as of 1912,
Einstein came to recognize the locality nature of special relativity
and of the equivalence principle. Only in his later development of
general relativity, in a frantic race against Hilbert, did he adopt
a pragmatist's route without a thorough pursuit or exploitation of
the local -- instead of global -- validity of these principles.

In summary, we have arrived at the following depiction of the spacetime
manifold: it constitutes a patchwork of local pockets of spacetime,
each pocket obeying the special relativity (and all laws of non-gravitational
origin such as quantum mechanics) yet each adopting its own length
scale. The Ricci scalar $\mathcal{R}$ determines the length scale
for each individual local pocket. As one moves from one pocket to
the next on the manifold, the form of physical laws remains unchanged
but the beating rate of clocks and the size of rulers and every object
(including the observer's own body size and heart rate) must adapt
to the prevailing value of $\mathcal{R}$. The fundamental constants
$\hbar$ and $c$ continue to oversee the established physics in each
local pocket but they too must adapt to the prevailing value of $\mathcal{R}$.

Compared with the logical conclusion of $\hbar$-variability, the
relegation of the speed of light from its once-sacred position in
physics is far more alarming. Yet, both of the variabilities in $c$
and $\hbar$ are natural ramifications of our Postulate (II) proposed
in Section \ref{sec:1} that the Ricci length is of a more fundamental
status than all other lengths. Mathematically, their variabilities
arise from the anisotropy of time scaling \eqref{eq:scaleT} which
takes place in $3+1$ dimensions. The Planck constant and speed of
light are not God-given prefixed fundamentals; rather, their values
are determined by the Ricci scalar in the spacetime pocket in which
they are measured in a one-to-one correspondence: $\hbar\propto\left|\mathcal{R}\right|^{-1/4},\ c\propto\left|\mathcal{R}\right|^{1/4}$.
We must also emphasize once again that despite their variabilities,
$\hbar$ and $c$ are not auxiliary fields that live on the manifold
\footnote{This conclusion of ours is in contrast to the approach Moffat, Magueijo
and others subscribed to in \cite{Moffat,Magueijo1,Magueijo2,Magueijo3}.
These authors attempted several mechanistic assignments of a dynamics
for $c$. In light of our approach, such assignments were unnecessary.
If the reader nonetheless insists on a dynamics for $c$ and $\hbar$,
the reader may view the dynamics of $\mathcal{R}$, i.e. the field
equations for $g_{\mu\nu}$, as the underlying dynamics for $c$ and
$\hbar$ since these parameters are one-to-one related to $\mathcal{R}$.%
}.

Lastly, let us compare our approach with Ho\v{r}ava's proposal \cite{Horava}.
In Ho\v{r}ava's approach, Lorentz invariance is assumed to emerge
in the coarse-graining procedure in which fast-mode fluctuations are
step-by-step integrated out. At the microscopic ultraviolet limit,
Lorentz invariance is absent. In our approach, Lorentz invariance
is fully preserved: the anisotropy in time scaling $dt\propto a_{\mathcal{R}}^{3/2}$
is compensated by the $c$-variability, $c\propto a_{\mathcal{R}}^{-1/2}$
such that $dx^{0}\triangleq c\, dt\propto a_{\mathcal{R}}$ exactly
as its spatial counterparts $d\vec{x}\propto a_{\mathcal{R}}$.

\subsection{\label{sub:invitation}Applications of the anisotropic time scaling
in cosmology: An invitation}

The logical conclusion of $c$-variability in the preceding section
has very far-reaching consequences. If we are to repeat the Michelson-Morley
experiment in Cleveland today, it is not in question that we shall
reproduce their null finding for the interference pattern, and that
we shall unequivocally concur on a common value of $c$ for the two
light beams in our interferometer that exist today. Yet nothing in
principle -- viz., nothing amongst the causality principle, the relativity
principle, Lorentz invariance, the equivalence principle, or the general
covariance principle -- compels an equality between the value of $c$\emph{
}that we encounter in Cleveland today and the value of $c$ that Michelson
and Morley experienced in Cleveland back in 1887. According to our
conclusions deduced in the preceding section, the value of $c$ in
a local spacetime pocket is pegged to the Ricci scalar in the pocket
per the scaling rule \eqref{eq:scaleC}: $c\propto\left|\mathcal{R}\right|^{1/4}$,
and thus if the Ricci scalar in the local region has undergone a change
in value, then a change in the value of $c$ must follow. Obviously
we do not expect Cleveland's surroundings to have expanded during
the last 126 years (since galaxies resist cosmic expansion), yet our
Cleveland metaphor helped illustrate our logic.

The cosmos, however, is an ideal laboratory in which the $c$-variability
would show its glory. With the universe undergoing an expansion for
several billions years (i.e., its Ricci scalar having been steadily
decreasing) \cite{Hubble}, light emitted from stars at a distant
past has been traveling through a succession of local spacetime pockets
with steadily decreasing value of $c$. At any given point on its
transit toward Earth, the photon lives in a local tangent frame with
special relativity in full effect but the value of its speed progressively
adapts to the prevailing value of $\mathcal{R}$ where the photon
passes by. Over the life of the universe, $c$ has been falling by
several orders in magnitude, rendering qualitatively detectable and
quantitatively measurable effects in observational cosmology. This
conceptual point holds the key to resolving several pressing problems
in standard cosmography and cosmology. To be concise, the standard
paradigm of cosmology, together with the Friedmann model as its foundation,
overlook the progressive decrease in light speed in its treatment
of the Hubble law, the redshift-distance relationship, the interpretation
of high-$z$ objects (Type Ia supernovae), and the theoretical difficulties
with regard to the horizon, flatness, and cosmic coincidences. A full-fledged
excursion on the consequences of the $c$-variability in cosmography
and cosmology shall be presented in Sections \ref{sec:8} and \ref{sec:9}.
In what follows, we shall give a brief outline as of how the scaling
rule \eqref{eq:scaleC} alone will help resolve four outstanding problems:
the age problem, the interpretation of Type Ia supernovae (while bypassing
the accelerating expansion), the horizon problem, and the flatness
problem (while bypassing the inflationary expansion.) The results
presented below are model-independent. They directly follow from the
scaling rule \eqref{eq:scaleC}, $c\propto\left|\mathcal{R}\right|^{1/4}$
without any resort to any specific model of matter distribution (such
as a modification to the Friedmann model, for example.)

Imagine a distant galaxy which emitted a beam of photons. On their
journey toward the Earth, the photons went through a succession of
spacetime pockets with increasingly larger cosmic scale factor $a$,
i.e., decreasing scalar curvature $\mathcal{R}$. (Also, note that
the cosmic scale factor $a$ is proportional to the Ricci length $a_{\mathcal{R}}$,
so we shall work with $a$ in what follows.) The photon wavelength
gets {}``stretched out'', thus the photons getting {}``redshifted'',
a fact that astronomers agree upon. The standard formula for the redshift
$z$ is (see Eq. \eqref{eq:8.10}): \begin{equation}
1+z=a^{-1}\label{eq:5.10}\end{equation}
with $a$ being the cosmic factor at the moment the photons were emitted
(with the current cosmic factor set equal $1$.) However, this formula
-- as well as all other formulae in standard cosmology -- neglect
the effect of a steadily falling light speed, an effect which must
be taken into account. This effect modifies the redshift formula to
(see Eqs. \eqref{eq:8.18} and \eqref{eq:g.5}): \begin{equation}
1+z_{observed}=a^{-3/2}\label{eq:5.11}\end{equation}
where $z_{observed}$ is the redshift value that appears in the astronomer's
telescope (see Section \ref{sub:Lemaitre modRW} and Appendix \ref{sec:G}
for derivation of this new formula.) The exponent of $\frac{3}{2}$
directly stems from the anisotropic exponent in time scaling $\eta=\frac{3}{2}$
\footnote{One way to interpret Formula \eqref{eq:5.11} is that the beating
rate of clocks in the regions the photons passed by slowed down in
disproportion with the cosmic scale factor, viz. $dt\propto a^{3/2}$
instead of $a$ (due to the anisotropic exponent for time scaling
$\eta=\frac{3}{2}$.) One should realize that in observational cosmology
it is the shift in the photon frequency -- instead of its wavelength
-- that is measured. Yet this technical detail is of secondary importance.
Even if the redshift were measured in terms of the photon wavelength,
the redshift formula is still: $1+z_{observed}=a^{-3/2}$; see Appendix
\ref{sec:G} for explanation.%
}.
\begin{enumerate}
\item The age problem:

Given that the cosmos was smaller in the past, $a<1$, the new redshift
formula \eqref{eq:5.11} means that the actual redshift is enhanced:
$z_{observed}>z$. That is to say, if the astronomer unknowingly uses
the standard redshift formula \eqref{eq:5.10} to back out the value
of the cosmic factor $a$ from $z_{observed}$, she would inadvertently
overestimate the value of $a$. The overestimate is consistent for
light sources at all distances. Therefore, when expanding the redshift
formula for low-$z$ objects, the Hubble law is inadvertently missing
an extra multiplicative factor of $\frac{3}{2}$ (see Section \ref{sub:Hubble}
for details.) That is to say, the estimated value obtained from the
regression of $z_{observed}$ versus the distance $d$ of the objects
is not the actual Hubble constant $H_{0}$ but instead $\frac{3}{2}H_{0}$.
The estimate of $H_{0}$ has thus been biased upward. It is $\frac{3}{2}H_{0}$
that has the reported value of 70. The true value of $H_{0}$ is thus
only $\frac{2}{3}\times70\approx47$. From the age formula $t_{0}=\frac{2}{3H_{0}}$,
the actual age of the Universe is about $\frac{3}{2}\times9bl\approx14$
Glys in agreement with WMAP. The age problem disappears. (An otherwise
accepted value of $H_{0}=70$ yields an age of 9 bn years, an absurd
result given that the oldest stars are known to be 12 bn years old.)
The need to resort to a scenario of accelerating phase following a
decelerating phase such as in \cite{Turner} disappears.

\item An alternative interpretation of Type Ia supernovae data:

Consider two supernovae $A$ and $B$ at distances $3\, bn$ and $6\, bn$
light years away from the Earth, viz. $d_{B}=2\, d_{A}$. Standard
cosmology dictates the redshift values of $z_{A}$ and $z_{B}\approx2\, z_{A}$
for them (to first-order approximation.) However, light travelled
faster in a distant past than it did in a more recent epoch. Thus,
the $B$-photon covered twice as long the distance in less than twice
the amount of time as compared with the $A$-photon. Having spent
less time in transit than expected, the $B$-photon experienced less
cosmic expansion than expected, and thus less redshift than standard
cosmology predicts. Namely: $z_{B}<2\, z_{A}$. This result means
an upward slopping in the curve as we put the two supernovae's data
on the $d$ vs. $z$ plot. Conversely, a supernova $C$ with $z_{C}=2\, z_{A}$
must correspond to a distance greater than $6\, bn$ light years,
viz. $d_{C}>d_{B}=2\, d_{A}$, and thus is a fainter object. This
is precisely what being observed in the data of Type Ia supernovae
\cite{Riess1,Perlmutter}. In Section \ref{sec:8} we shall show that
the same value of $H_{0}$ which resolves the age problem alluded
above also gives the fit to the Type Ia supernovae data without any
adjustable parameter whatsoever. No fudge factors, such as the amount
of dark energy, are needed. The cosmic expansion is not accelerating;
the observed discrepancy in Type Ia supernovae as compared to the
critical expansion mode stems from the shortcoming of the standard
cosmological formulation to take into account the effects of $c$-variation
as function of the cosmic scale factor.

\item The horizon problem:

\noindent Observational data shows a highly uniform distribution (to
the accuracy of $10^{-5}$) of cosmic radiation across the horizon.
This uniformity presents a serious challenge to the Friedmann cosmological
model -- the observed uniformity cannot reconcile with the fact that
different segments of the current horizon are not causally connected.
However, as we pointed out before, the Friedmann model does not take
into account the variability of $c$ as function of the cosmic scale
factor $a$, per \eqref{eq:scaleC}: $c\propto a^{-1/2}$, which means
a much higher value of $c$ in the baby universe and thus a larger
cosmological horizon than what the Friedmann model dictates. Quantitatively,
this scaling rule precisely produces an infinite value for the cosmological
horizon (see Eq. \eqref{eq:9.15}) as we shall explicitly show in
Section \ref{sub:horizon prob}. This result neatly explains the near
uniformity in our current horizon since the entire universe was in
causal contact before stretching out to its current state. The idea
of employing a larger value of $c$ in the past (about 60 orders of
magnitude higher than now) to account for the horizon paradox was
first advanced by Moffat \cite{Moffat} and subsequently by Albrecht
and Magueijo \cite{Magueijo1}. What is new in our approach is an
organic mechanism behind the variation of $c$ via the role of the
Ricci scalar as the scale setter per Postulate (II). Our resolution
of the horizon paradox invokes no ad hoc assumptions as was done in
the inflationary universe hypothesis.

\item The flatness problem:

Observational data further confirm that the universe is almost flat.
This fact represents another serious challenge to the Friedmann model.
Flatness is very difficult to achieve in the Friedmann model; it requires
a fine tuning to an extraordinary level. Curvature-scaling gravity
offers a natural explanation to the flatness problem as well. As the
early universe expanded, the speed at which light travels decreased,
according to the scaling rule \eqref{eq:scaleC}: $c\propto a^{-1/2}$.
The effective horizon thus shrank relatively to the universe's size.
By the time of the decoupling event, at which point light began to
travel freely, as the universe had grown several orders over in size
and simultaneously the light speed had dropped several orders over
in magnitude, the observable region within which light signals that
emitted from could reach us today constituted a pocket several orders
of magnitude smaller than the universe. This explains the flatness
problem. In our resolution, we do not need to resort to any ad hoc
assumptions as was done in the inflationary expansion hypothesis %
\footnote{Beside the flatness and horizon problems, the Friedmann model suffers
Dicke's instability problem in which a tiny deviation away from the
critical density would rapidly drive the universe away from its current
state -- this is also known as the oldness problem. Curvature-scaling
gravity resolves this problem too. The evolution of the universe is
governed by a generic scaling rule which strictly forces the universe
to expand in the critical mode regardless of the its shape and/or
its matter content and density. The robust evolution law prevents
the universe from collapsing or expanding supercritically. The anthropic
argument is not needed. See Section \ref{sec:Dicke prob} for our
explanation.%
}.

\end{enumerate}
\noindent Curvature-scaling gravity provides a natural, unified, coherent,
comprehensive explanation -- and parsimonious in the Occam Razor's
sense -- for the four most pressing problems in cosmology. Our explanation
is model-independent (without resorting to any specific model of matter
distribution in the universe); it is also derived from first principle,
viz. Postulate (II) on the role of the Ricci scalar as the dynamical
scale-setter. It is noteworthy that curvature-scaling gravity was
not initially designed to resolve these problems but was motivated
by our desire to understand the underlying agent that decides the
size of things in our surroundings. In and of itself, curvature-scaling
gravity is theory of gravity and spacetime, instead of a theory of
cosmology alone.

If we subcribe to the new depiction of the spacetime manifold, a manifold
which allows each local region to have its own length scale -- the
Ricci length -- which has been in steady decline as the universe expands,
then the Friedmann model misses out this very important property of
spacetime -- viz. the adaptation of $c$ to the cosmic scale factor
-- and cannot be used as a theoretical foundation for cosmology. Such
a problematic model would inevitably lead to theoretical predictions
irreconcilable with observational data. Foremost among its problems
are the difficulties to account for Type Ia supernovae data, the age
of the cosmos, the near uniform observable horizon, the near flatness
of space, to name a few. The reason behind the difficulties for standard
cosmology to account for observations is that the cosmos serves as
the largest laboratory imaginable in which the effects of the Ricci
scalar as a dynamical scale-setter strongly manifest themselves.

To reconcile its predictions with observations, standard cosmology
has had to invent several ad hoc addenda to the Friedmann model: the
cosmological constant (viz. the dark energy), the accelerated expansion,
and the inflationary expansion, which introduce a new set of unsettling
fine-tunings and difficulties. Given the unified resolution from curvature-scaling
gravity for these once-troubling problems, these ad hoc supplementaries
-- which have nonetheless integrated into the vocabulary of cosmology
-- are unnecessary and lack a logical basis.

\newpage{}

\section{\label{sec:6}A nontrivial solution to the curvature-scaling field
equation and its consequences in the physics of black holes}

In this section, we shall derive a non-trivial solution to the field
equations of curvature-scaling gravity in vacuo and explore its implications
in the physics of black holes. The solution is found for the static
spherically symmetric setup up to the first-order perturbative expansion
in terms of the Mannheim-Kazanas parameter $\gamma$ alluded in Section
\ref{sec:4}. The parameter $\gamma$ was found in Mannheim's work
\cite{Mannheim1,MOB1,MOB2,MOB3} to be small for astronomical objects.

The solution presented below has a non-constant Ricci scalar which
diverges on the event horizon. From the logical deductions regarding
the variabilities in $\hbar$ and $c$ derived in the preceding section,
the solution allows $c$ to rise indefinitely and $\hbar$ to drop
to zero as one approaches the event horizon of a massive object in
the de Sitter background. Right on the event horizon, as the Ricci
scalar diverges logarithmically, $c$ is infinite and $\hbar$ vanishes.
Quantum effects thus diminish on the event horizon. How this result
affects the radiation of Schwarzschild-type black holes is an open,
yet intriguing, issue which merits further investigations.

\subsubsection*{The solution with non-constant Ricci scalar:}

The full solution for the static spherically symmetric setup is given
in Buchdahl's metric (\ref{eq:3.11}, \ref{eq:3.12}) in Section \ref{sec:3}.
Generally, Buchdahl's solution admits spacetime configurations with
non-constant Ricci scalar. In Section \ref{sec:4} we also found an
exact solution (\ref{eq:4.1}, \ref{eq:4.3}) which introduced a new
term, the linear Mannheim-Kazanas $\gamma r$ term in $\Psi=\sqrt{1-3r_{s}\gamma}-\frac{r_{s}}{r}-\Lambda r^{2}+\gamma r$
in the line element. This solution has a constant Ricci scalar, however.
Our aim in this section is to find a solution with non-constant Ricci
scalar.

We shall find a solution which is asymptotic to the constant-$\mathcal{R}$
solution (\ref{eq:4.1}, \ref{eq:4.3}) which contains 3 parameters;
the deviation from constant Ricci scalar will be specified by a new
parameter, bringing the number of parameters to 4 as demanded by Buchdahl's
treatment (see end of Section \ref{sec:3}.) We shall express the
metric in the form\begin{equation}
ds^{2}=e^{\alpha}\left[-\Psi\left(dx^{0}\right)^{2}+\frac{dr^{2}}{\Psi}+r^{2}d\Omega^{2}\right]\label{eq:6.1}\end{equation}
with two unknown functions $\alpha$ and $\Psi$. Taking the cue from
the solution (\ref{eq:4.1}, \ref{eq:4.3}), we shall find the 2 functions
$\alpha$ and $\Psi$ as perturbative expansion in terms of $\gamma$:
\begin{equation}
\begin{cases}
\ \Psi & =\Psi_{0}+\gamma\Psi_{1}+\mathcal{O}\left(\gamma^{2}\right)\\
\ \alpha & =\gamma\Phi_{1}+\mathcal{O}\left(\gamma^{2}\right)\end{cases}\label{eq:6.2}\end{equation}
with\begin{equation}
\Psi_{0}=1-\frac{r_{s}}{r}-\Lambda r^{2}.\label{eq:6.3}\end{equation}
We shall continue to call $\gamma$ the Mannheim-Kazanas parameter,
although their original utility of this parameter was limited to the
context of dark matter (see Section \ref{sec:4}.) Our derivation
for $\Psi_{1}$ and $\Phi_{1}$ is detailed in Appendix \ref{sec:D}.
Here is the outline of our derivation:
\begin{itemize}
\item For the two unknowns $\Psi_{1}$ and $\Phi_{1}$, the two field equations
needed are the $tt$- and the $\theta\theta$-components, which read\begin{eqnarray}
\left(\mathcal{R}_{tt}-\frac{1}{4}g_{tt}\mathcal{R}\right)\mathcal{R} & = & -\Gamma_{tt}^{r}\mathcal{R}'\label{eq:6.4}\\
\left(\mathcal{R}_{\theta\theta}-\frac{1}{4}g_{\theta\theta}\mathcal{R}\right)\mathcal{R} & = & -\Gamma_{\theta\theta}^{r}\mathcal{R}'\label{eq:6.5}\end{eqnarray}
in which $\mathcal{R}_{tt},\ \mathcal{R}_{\theta\theta}$ and $\mathcal{R}$
are obviously involved $\Psi_{1}$ and $\Phi_{1}$.
\item Next, expand the two field equations in terms of $\gamma$ order-by-order.
The zero-order terms in $\gamma$ disappear since we already specified
$\Psi_{0}=1-\frac{r_{s}}{r}-\Lambda r^{2}$. The first-order $\gamma$
terms thus constitute two equations for $\Psi_{1}$ and $\Phi_{1}$.
We shall ignore terms with second- and higher-order in $\gamma$.
\item The two equations for $\Psi_{1}$ and $\Phi_{1}$ just obtained (see
Eqs. \eqref{eq:d.15} and \eqref{eq:d.16} in Appendix \ref{sec:D})
contain $\Psi_{1}'''$ and $\Phi_{1}'''$ with prime denoting derivative
w.r.t. the radial coordinate $r$. Furthermore, these equations couple
with each other. So, an outright elimination procedure would necessarily
yield even higher derivatives and the problem thus looks intractable.
\end{itemize}
Despite this deceptive complexity, after a thicket of calculations
detailed in Appendix \ref{sec:D}, we were able to gather complicated
terms together %
\footnote{We first tried deriving the solution in the form of Laurent series
w.r.t. $r$. The series we found appeared to possess a number of interesting
properties, which guided us to reverse-engineer the solution. The
final outcome is the analytical formulae for $\Psi_{1}$ and $\Phi_{1}$
as in (\ref{eq:6.6}, \ref{eq:6.7}).%
} and obtain the formulae:\begin{eqnarray}
\Psi_{1} & = & r-\frac{3}{2}r_{s}\label{eq:6.6}\\
\Phi_{1} & = & -r-\epsilon\int\frac{dr}{r^{2}\left(1-\frac{r_{s}}{r}-\Lambda r^{2}\right)}\label{eq:6.7}\end{eqnarray}
Despite the complexity of the equations, the expressions for $\Psi_{1}$
and $\Phi_{1}$ are remarkably neat . The metric up to first-order
in $\gamma$ is thus \begin{eqnarray}
ds^{2} & = & e^{\alpha}\left[-\Psi(dx^{0})^{2}+\frac{dr^{2}}{\Psi}+r^{2}d\Omega^{2}\right]\nonumber \\
\Psi & = & 1-\frac{r_{s}}{r}-\Lambda r^{2}+\gamma\left(r-\frac{3}{2}r_{s}\right)+\mathcal{O}(\gamma^{2})\label{eq:6.8}\\
\alpha & = & -\gamma\left[r+\epsilon\int\frac{dr}{r^{2}\left(1-\frac{r_{s}}{r}-\Lambda r^{2}\right)}\right]+\mathcal{O}(\gamma^{2})\nonumber \end{eqnarray}
and the Ricci scalar is\begin{equation}
\mathcal{R}=12\Lambda\left[1-\gamma\epsilon\int\frac{dr}{r^{2}\left(1-\frac{r_{s}}{r}-\Lambda r^{2}\right)}\right]+\mathcal{O}(\gamma^{2}).\label{eq:6.9}\end{equation}
The metric now involves 4 parameters:
\begin{itemize}
\item $\Lambda$, the de Sitter parameter, specifying the large-distance
curvature.
\item $r_{s}$, the Schwarzschild radius (taken as a free parameter as a
whole, instead of a combination of separate terms $\frac{GM}{2c^{2}}$).
It specifies the {}``strength'' of the gravitational field source,
as usual.
\item $\gamma$, the Mannheim-Kazanas parameter, specifying the linear potential
term which Mannheim exploited in his theory of galactic rotation curves
\cite{Mannheim1,MOB1,MOB2,MOB3}.
\item An additional parameter $\epsilon$, allowing the Ricci scalar to
deviate from constancy. We shall call it the anomalous curvature parameter.
\end{itemize}
The number of parameters is 4, precisely the number derived from Buchdahl's
general study of spherically symmetric systems. The last parameter,
$\epsilon$, is not present in conformal gravity, e.g., in Mannheim-Kazanas's
solution. This parameter is specific to our solution since it appears
in the phase factor which would have been conformally {}``gauged''
away in conformal gravity. The anomalous curvature is only manifest
if these three conditions are met:
\begin{itemize}
\item $\Lambda\neq0$, the metric must not be asymptotically flat.
\item $\gamma\neq0$, the constant Mannheim-Kazanas centripetal acceleration
must be in effect.
\item $\epsilon\neq0$, the anomalous parameter, self-evidently, must not
vanish.
\end{itemize}
The parameter $\epsilon$ plays a central role in our prediction below,
in which we predict new behaviors for the Schwarzschild-type black
holes. The four parameters are dimensional: $[\Lambda]=\left[\text{length}\right]^{-2},\ [r_{s}]=\left[\text{length}\right],\ [\gamma]=\left[\text{length}\right]^{-1},\ [\epsilon]=\left[\text{length}\right]^{2}$,
all helping set the scale for the black holes. They depend on the
boundary conditions.

In the limit of small $\gamma$ and $\epsilon=0$, solution \eqref{eq:6.8}
is compatible with the constant-$\mathcal{R}$ solution (\ref{eq:4.1},
\ref{eq:4.3}) in Section \ref{sec:4}. To see this, let us Taylor
expand \eqref{eq:4.4} with respect to $\gamma$ of the solution in
Section \ref{sec:4}:\begin{equation}
\begin{cases}
\ \gamma & \triangleq-\frac{\kappa}{3r_{s}}\\
\ \Psi & =\sqrt{1-3r_{s}\gamma}-\frac{r_{s}}{r}-\Lambda r^{2}+\gamma r\\
\ e^{\alpha} & =\left(1+\frac{1-\sqrt{1-3r_{s}\gamma}}{3r_{s}}r\right)^{-2}\end{cases}\label{eq:6.10}\end{equation}
to obtain\begin{equation}
\begin{cases}
\ \Psi & =1-\frac{r_{s}}{r}-\Lambda r^{2}+\gamma\left(r-\frac{3}{2}r_{s}\right)+\mathcal{O}\left(\gamma^{2}\right)\\
\ \alpha & =-2\ln\left(1+\frac{1-\sqrt{1-3r_{s}\gamma}}{3r_{s}}r\right)=-\gamma r+\mathcal{O}\left(\gamma^{2}\right).\end{cases}\label{eq:6.11}\end{equation}
These Taylor series are compatible with solution \eqref{eq:6.8} with
$\epsilon=0$.

\subsubsection*{Prediction -- new properties of Schwarzschild-type black holes:}

Solution \eqref{eq:6.8} is interesting in either limit $r\rightarrow0$
or $r\rightarrow r_{s}$. We shall focus on the limit of $r\rightarrow r_{s}$
in what follows. Let us consider a small $\Lambda$ and denote $r_{\star}$
to be the root of the algebraic equation $1-\frac{r_{s}}{r}-\Lambda r^{2}=0$
that is closest to $r_{s}$. When $\epsilon=0$, the event horizon
is $r_{\star}$ since $\Psi_{0}\rightarrow0$ as $r\rightarrow r_{\star}$
obviously.

When $\epsilon\neq0$ and as $r\rightarrow r_{\star}$, the Ricci
scalar \eqref{eq:6.9} takes the form

\begin{equation}
\mathcal{R}\simeq\frac{\Lambda\gamma\epsilon}{r_{\star}}\,\ln\left|1-\frac{r_{\star}}{r}\right|.\label{eq:6.12}\end{equation}
The Ricci scalar diverges logarithmically as one approaches the event
horizon. This leads to an intriguing phenomenon. As the observer free
falls into a very massive object, the ruler and all objects in his
surroundings would shrink, per $dx\propto a_{\mathcal{R}}=\left|\mathcal{R}\right|^{-1/2}$,
and his wristwatch speeds up disproportionally, per $dt\propto a_{\mathcal{R}}^{3/2}=\left|\mathcal{R}\right|^{-3/4}$.
Note that the effect of time flow speeding up is an opposite effect
to the standard gravitational redshift which is also in play %
\footnote{We do not, however, expect the correction to the standard gravitational
redshift to be material for the solar system because we do not expect
the Ricci scalar to vary to any substantial amount within the solar
system.%
}. As the result of his clock and ruler scaling, the observer free
falling in the black hole would experience an increasing value of
$c$ (once again, superluminosity is strictly forbidden) and a decreasing
value of $\hbar$, according to\begin{eqnarray}
c & \simeq & \left|\ln\left|1-\frac{r_{\star}}{r}\right|\right|^{1/4}\label{eq:6.13}\\
\hbar & \simeq & \left|\ln\left|1-\frac{r_{\star}}{r}\right|\right|^{-1/4}\label{eq:6.14}\end{eqnarray}
Right on the event horizon, $\hbar=0$ and $c\rightarrow\infty$.
The variations are only logarithmical. Nonetheless, whether and/or
how the diminishment of quantum effects alter the behavior of black
holes such as their radiative properties, quantitatively and/or qualitatively,
deserve a thorough examination in future research.

We thus predict the novel properties of Schwarzschild-type black holes
residing on a de Sitter background. Obviously the tests of our prediction
regarding the behavior of black holes lie beyond the reach of man's
current technology. Our initial enquiry in the arena of black holes
is thus purely a theoretical adventure, which could help elucidate
the structure of spacetime in the strong gravitational field limit.
In addition, our mathematical endeavors in deriving the solutions
(\ref{eq:4.1}, \ref{eq:4.3}) and (\ref{eq:6.8}, \ref{eq:6.9})
illustrate the tractability of curvature-scaling gravity.\newpage{}

\section{\label{sec:7}Coupling of curvature-scaling gravity with matter:
a departure from conventional construction of Lagrangian}

Let us define three parameters with their units indicated respectively:\begin{eqnarray}
\hat{\hbar} & \triangleq & \frac{\hbar}{\sqrt{a_{\mathcal{R}}}}\sim\frac{\left[\mbox{mass}\right].\left[\mbox{length}\right]^{3/2}}{\left[\mbox{time}\right]}\label{eq:7.1}\\
\hat{c} & \triangleq & c\,\sqrt{a_{\mathcal{R}}}\sim\frac{\left[\mbox{length}\right]^{3/2}}{\left[\mbox{time}\right]}\label{eq:7.2}\\
\zeta & \triangleq & \frac{\hat{\hbar}}{\hat{c}}\sim\left[\mbox{mass}\right]\label{eq:7.3}\end{eqnarray}
with $a_{\mathcal{R}}=\left|\mathcal{R}\right|^{-1/2}$ being the
Ricci length. Given the scaling rules \eqref{eq:scaleH} and \eqref{eq:scaleC},
each of these parameters has a fixed value at all points on the manifold.
Conversely, the Planck constant and speed of light are dependent on
the Ricci scalar, per\begin{eqnarray}
\hbar & = & \hat{\hbar}\, a_{\mathcal{R}}^{1/2}\label{eq:7.4}\\
c & = & \hat{c}\, a_{\mathcal{R}}^{-1/2}\label{eq:7.5}\end{eqnarray}
Note that beside $\zeta$ which is independent of $\mathcal{R}$,
the dimensionless fine-structure constant $\alpha\triangleq e^{2}/\left(\hbar\, c\right)=e^{2}/\left(\hat{\hbar}\,\hat{c}\right)$
is also independent of $\mathcal{R}$.

The standard algorithm of covariantising an action (also known as
{}``the principle of minimal coupling'') consists of the following
steps:
\begin{itemize}
\item Write down the Lorentz-invariant action in special relativity.
\item Replace the Minkowski metric $\eta_{\mu\nu}$ by $g_{\mu\nu}$ whenever
applicable.
\item Replace a partial derivative $\partial_{\mu}$ by the covariant derivative
$\nabla_{\lyxmathsym{\textgreek{m}}}$ whenever applicable.
\item Replace the volume element $d^{4}x$ by its (invariant) counterpart
$d^{4}x\,\sqrt{g}$.
\end{itemize}
By construction, these equations are tensorial and true in the absence
of gravity and hence satisfy the general covariance principle. What
needs to be done in curvature-scaling gravity are three additional
steps:
\begin{itemize}
\item When writing down the Lorentz-invariant action in special relativity,
cast it in the terms of dimensionless quantities. Then proceed to
replace $\eta_{\mu\nu}\rightarrow g_{\mu\nu},\ \partial_{\mu}\rightarrow\nabla_{\mu},\ d^{4}x\rightarrow d^{4}x\,\sqrt{g}$
as previously required.
\item Replace $\hbar$ by $\hat{\hbar}$ and $c$ by $\hat{c}$ whenever
applicable.
\item Finally, make two following replacements, whenever applicable:\begin{equation}
\begin{cases}
\ dx^{\mu}\rightarrow d\tilde{x}^{\mu}=\frac{dx^{\mu}}{a_{\mathcal{R}}}=\left|\mathcal{R}\right|^{1/2}dx^{\mu}\\
\ \nabla_{\mu}\rightarrow\tilde{\nabla}_{\mu}=a_{\mathcal{R}}\nabla_{\mu}=\left|\mathcal{R}\right|^{-1/2}\nabla_{\mu}\end{cases}\label{eq:7.6}\end{equation}
In particular, the volume element is replaced as\begin{equation}
d^{4}x\,\sqrt{g}\rightarrow d^{4}x\,\mathcal{R}^{2}\,\sqrt{g}\label{eq:7.7}\end{equation}

\end{itemize}
Let us start from the action of the massive scalar field theory\begin{equation}
\mathcal{S}\simeq\int d^{4}x\,\left[-\frac{1}{2}\hbar^{2}c^{2}\partial^{\mu}\phi\partial_{\mu}\phi-\frac{1}{2}m^{2}c^{4}\phi^{2}\right].\label{eq:7.8}\end{equation}
We deliberately cast in terms of dimensionless quantities\begin{equation}
\mathcal{S}\simeq\int d^{4}x\,\left[-\frac{1}{2}\left(\frac{\hbar}{mc}\partial^{\mu}\phi\right)\left(\frac{\hbar}{mc}\partial_{\mu}\phi\right)-\frac{1}{2}\phi^{2}\right].\label{eq:7.9}\end{equation}
Upon covariantising it, we get (for scalar fields, $\nabla_{\mu}=\partial_{\mu}$,
though)\begin{equation}
\mathcal{S}\simeq\int d^{4}x\,\sqrt{g}\,\left[-\frac{1}{2}\left(\frac{\hbar}{mc}\nabla^{\mu}\phi\right)\left(\frac{\hbar}{mc}\nabla_{\mu}\phi\right)-\frac{1}{2}\phi^{2}\right].\label{eq:7.9-1}\end{equation}
Making the replacements $\hbar\rightarrow\hat{\hbar},\ c\rightarrow\hat{c}$
and \eqref{eq:7.6}, we obtain the action\begin{equation}
\mathcal{S}_{CSG}=\int d^{4}x\,\mathcal{R}^{2}\,\sqrt{g}\,\left[-\frac{1}{2}\left(\frac{\hat{\hbar}}{m\hat{c}}\left|\mathcal{R}\right|^{-1/2}\nabla^{\mu}\phi\right)\left(\frac{\hat{\hbar}}{m\hat{c}}\left|\mathcal{R}\right|^{-1/2}\nabla_{\mu}\phi\right)-\frac{1}{2}\phi^{2}\right]\label{eq:7.10}\end{equation}
with CSG abbreviating curvature-scaling gravity. In terms of $\zeta$,
it is recast as\begin{equation}
\mathcal{S}_{CSG}=\int d^{4}x\,\mathcal{R}^{2}\,\sqrt{g}\,\left[-\frac{\zeta^{2}}{2m^{2}\left|\mathcal{R}\right|}\nabla^{\mu}\phi\nabla_{\mu}\phi-\frac{1}{2}\phi^{2}\right].\label{eq:7.11}\end{equation}
The right-hand-side is dimensionless. More generally, for a covariantized
action cast in dimensionless quantities \begin{equation}
\mathcal{S}\simeq\int d^{4}x\,\sqrt{g}\,\mathcal{L}_{m}\left(\phi,\nabla_{\mu}\phi\right),\label{eq:7.12}\end{equation}
the curvature-scaling gravity action is\begin{equation}
\mathcal{S}_{CSG}=\int d^{4}x\,\mathcal{R}^{2}\,\sqrt{g}\,\mathcal{L}_{m}\left(\phi,\left|\mathcal{R}\right|^{-1/2}\nabla_{\mu}\phi\right)\label{eq:7.13}\end{equation}
This action contains all we need: the matter field $\phi$ and the
gravitational field $g_{\mu\nu}$. Functionally varying $\phi$ while
holding $g_{\mu\nu}$ fixed produces the equation of motion for the
matter field. Functionally varying $g_{\mu\nu}$ while holding $\phi$
fixed produces the gravitational field equations. As can be seen from
\eqref{eq:7.13}, gravity -- i.e., the metric tensor $g_{\mu\nu}$--
organically arises from matter fields via five simultaneous routes:
\begin{itemize}
\item $\left|\mathcal{R}\right|^{-1/2}\,\nabla_{\mu}$ in the Lagrangian
of matter $\mathcal{L}_{m}$;
\item $\mathcal{R}^{2}\, d^{4}x$ of the volume element;
\item the Jacobian $\sqrt{g}$, as usual;
\item the covariant derivatives $\nabla_{\mu}$;
\item the contravariant derivatives $\nabla^{\mu}=g^{\mu\nu}\nabla_{\nu}$.
\end{itemize}
Each of these terms participates in the Lagrangian for a legitimate
reason, rather than an ad hoc formality. $\mathcal{S}_{CSG}$ explicitly
excludes the cosmological term as well as the Einstein-Hilbert term
$\mathcal{R}$.

We stress that, in our approach, there is no separate Lagrangian for
a {}``free'' gravitational field. Gravitational field always couples
with matter; it does not exist in isolation. Our approach is thus
a radical departure from the standard practice. Traditionally, one
would first write down the Lagrangian for a {}``free'' gravitational
field called $\mathcal{L}_{free}$; after that, one would augment
it with the Lagrangian for the matter part $\mathcal{L}_{m}$ which
is often let couple with the gravitational field via the minimal coupling.
There also needed be a factor -- viz., the coupling constant -- to
specify the relative weight for $\mathcal{L}_{m}$ against$\mathcal{L}_{free}$.
For example, the Einstein-Hilbert full action is ($c=1$)\begin{equation}
\mathcal{S}_{EH}=-\frac{1}{16\pi G}\int d^{4}x\,\sqrt{g}\,\mathcal{R}+\int d^{4}x\,\sqrt{g}\,\mathcal{L}_{m},\label{eq:7.14}\end{equation}
the Weyl action of conformal gravity \cite{Mannheim1} is\begin{equation}
\mathcal{S}_{W}=-\alpha_{g}\int d^{4}x\,\sqrt{g}\,\left[\mathcal{R}_{\mu\nu\lambda\sigma}\mathcal{R}^{\mu\nu\lambda\sigma}-2\mathcal{R}_{\mu\nu}\mathcal{R}^{\mu\nu}+\frac{1}{3}\mathcal{R}^{2}\right]+\int d^{4}x\,\sqrt{g}\,\mathcal{L}_{m},\label{eq:7.15}\end{equation}
and the quadratic gravity (QG) action is\begin{equation}
\mathcal{S}_{QG}=-\alpha\int d^{4}x\,\sqrt{g}\,\mathcal{R}^{2}+\int d^{4}x\,\sqrt{g}\,\mathcal{L}_{m}\label{eq:7.16}\end{equation}
All gravitational theories devoted their focus on the form of the
$\mathcal{L}_{free}$ piece of the {}``free'' gravitational field.
We take a radically different route: in our approach, it is the matter
part that receives the primary focus. There is no {}``free'' gravitational
field in our approach; gravity does not exist in isolation. The single
term in action \eqref{eq:7.13} contains both the matter field and
the gravitational field simultaneously.

\pagebreak{}We need to address two issues:
\begin{itemize}
\item Since there is no $\mathcal{L}_{free}$ for the {}``free'' gravitational
field, how would the Lagrangian of gravitation field in vacuo be defined?
We formally define the vacuo as the configuration in which matter
has no dynamics -- i.e., the derivatives of the matter field vanish
and the density of matter field is subsequently set equal zero. This
definition is justified if we interpret the uniform zero-point energy
background as vacuum. In so doing, action \eqref{eq:7.11} yields\begin{equation}
\mathcal{S}_{vacuo}=\int d^{4}x\,\sqrt{g}\,\mathcal{R}^{2}\label{eq:7.17}\end{equation}
which coincides with action \eqref{eq:3.3} considered in Section
\eqref{sec:3}. Action \eqref{eq:7.7} happens to imitate the quadratic
action \eqref{eq:7.16} in vacuo. Note however that, in the presence
of matter, our action \eqref{eq:7.13} is not the full quadratic action
\eqref{eq:7.16}. In addition, our action \eqref{eq:7.13} was inspired
from Postulate (II) in Section \ref{sec:1} which mandates that only
dimensionless quantities be of relevance for physical laws.
\item The Newton constant, viz. the gravitational constant $G$, is absent
in our action. There is no input parameter that specifies the strength
of coupling between matter and the gravitation field. This is a rather
surprising aspect of our approach. The effective coupling can arise
when we compare the Lagrangian at two different locations relative
to each other, that is to say, the relative ratio of matter dynamics
and/or density at the two locations. The vacuum corresponds to a region
in which the dynamics of matter, viz. $\partial^{\mu}\phi$, is negligible;
in such a situation, the Lagrangian of vacuo reads: $\mathcal{S}_{vacuo}=\int d^{4}x\,\sqrt{g}\,\mathcal{R}^{2}\rho$.
On the other hand, in the region where the dynamics and/or density
of matter are not negligible, gravity couples with matter in full
form: $\mathcal{S}_{CSG}=\int d^{4}x\,\sqrt{g}\,\mathcal{R}^{2}\mathcal{L}_{m}\left(\left|\mathcal{R}\right|^{-1/2}\partial_{\mu}\phi\right)$.
The {}``strength'' of the coupling of matter $\mathcal{L}_{m}$
to gravity is seen as a result of a relative comparison of matter's
density and/or dynamics in different regions. %
\footnote{The Newton constant can emerge in an approximation outlined below.
Set the vacuo density to be a small value $\rho$ and take the scalar
curvature to vary slowly around $\mathcal{R}_{0}$. Next, approximate
$\mathcal{R}$ in $\mathcal{L}_{m}$ by $\mathcal{R}_{0},$ and linearize
$\mathcal{R}^{2}$ around $\mathcal{R}_{0}$: $\mathcal{R}^{2}\approx2\mathcal{R}_{0}\mathcal{R}-\mathcal{R}_{0}^{2}$.
Then approximate\[
\mathcal{R}^{2}\mathcal{L}_{m}=\mathcal{R}^{2}\rho\left[1+\frac{1}{\rho}\left(\mathcal{L}_{m}-\rho\right)\right]\approx\rho\left[\mathcal{R}^{2}+\frac{\mathcal{R}_{0}^{2}}{\rho}\bar{\mathcal{L}}_{m}\right]\approx2\mathcal{R}_{0}\rho\left[\mathcal{R}-\frac{1}{2}\mathcal{R}_{0}+\left(\frac{1}{2\rho}\mathcal{R}_{0}\right)\bar{\mathcal{L}}_{m}\right]\]
in which $\bar{\mathcal{L}}_{m}\triangleq\mathcal{L}_{m}-\rho$ is
the remaining matter part after the vacuum density is subtracted out.
The terms in the last square bracket mimic the Einstein-Hilbert action
with coupling $G=\frac{c^{4}}{16\pi\rho}\mathcal{R}_{0}$ and a cosmological
term $\Lambda=-\frac{1}{2}\mathcal{R}_{0}$. These manipulations are
only meant to illustrate how the Newton constant $G$ could emerge
in an approximation to our Lagrangian in the configuration of slow
varying field.%
}
\end{itemize}

\subsubsection*{The field equations of the metric components:}

Upon functionally varying $g_{\mu\nu}$ while keeping the matter field
fixed in the action of curvature-scaling gravity\begin{equation}
\mathcal{S}_{CSG}=\int d^{4}x\,\sqrt{g}\,\mathcal{R}^{2}\,\mathcal{L}_{m}\left(\left|\mathcal{R}\right|^{-1/2}\nabla_{\mu}\right),\label{eq:7.18}\end{equation}
we obtain the field equations \begin{equation}
\left(\mathcal{R}_{\mu\nu}-\nabla_{\mu}\nabla_{\nu}+g_{\mu\nu}\square\right)\left[2\mathcal{R}\mathcal{L}_{m}+\mathcal{R}^{2}\frac{\partial\mathcal{L}_{m}}{\partial\mathcal{R}}\right]=\frac{1}{2}\mathcal{R}^{2}T_{\mu\nu},\label{eq:7.19}\end{equation}
or, equivalently \begin{equation}
\left(\mathcal{R}_{\mu\nu}-\nabla_{\mu}\nabla_{\nu}+g_{\mu\nu}\square\right)\left[\mathcal{R}\left(\mathcal{\mathcal{L}}_{m}+\frac{1}{2}\left|\mathcal{R}\right|\frac{\partial\mathcal{L}_{m}}{\partial\left|\mathcal{R}\right|}\right)\right]=\frac{1}{4}\mathcal{R}^{2}T_{\mu\nu}.\label{eq:7.20}\end{equation}
The stress-energy tensor is, as usual\begin{equation}
T_{\mu\nu}=-\frac{2}{\sqrt{g}}\frac{\partial\left(\sqrt{g}\mathcal{L}_{m}\right)}{\partial g^{\mu\nu}}=g_{\mu\nu}\mathcal{L}_{m}-2\frac{\partial\mathcal{L}_{m}}{\partial g^{\mu\nu}}\label{eq:7.21}\end{equation}
but $\mathcal{L}_{m}$ now contains an extra dependence on $\mathcal{R}$
via $\left|\mathcal{R}\right|^{-\frac{1}{2}}\nabla_{\mu}\phi$. Note
that the matter Lagrangian $\mathcal{L}_{m}$ participates on both
sides of the field equations \eqref{eq:7.20}. The coupling parameter
(viz. the Newton constant $G$) is absent from the field equations.

In vacumm, $\mathcal{L}_{m}$ does not contain any derivatives $\partial^{\mu}\phi$;
as such, it does not contains $\mathcal{R}$. The field equations
\eqref{eq:7.20} recovers the field equation in vacuo (compared with
Eq. \eqref{eq:3.6}):\[
\left(\mathcal{R}_{\mu\nu}-\nabla_{\mu}\nabla_{\nu}+g_{\mu\nu}\square\right)\mathcal{R}=\frac{1}{4}\mathcal{R}^{2}g_{\mu\nu},\]
or, upon taking the trace\[
\left(\mathcal{R}_{\mu\nu}-\frac{1}{4}g_{\mu\nu}\mathcal{R}\right)\mathcal{R}=\nabla_{\mu}\nabla_{\nu}\mathcal{R}.\]

\subsubsection*{The divergence of stress-energy tensor:}

Taking the divergence of the field equation of curvature-scaling gravity
\eqref{eq:7.20}, we get %
\footnote{We will need to utilize $\nabla^{\mu}\mathcal{R}_{\mu\nu}=\frac{1}{2}\nabla_{\nu}\mathcal{R}$
and $\nabla^{\mu}\nabla_{\mu}\nabla_{\nu}f-\nabla_{\nu}\nabla^{\mu}\nabla_{\mu}f=\mathcal{R}_{\mu\nu}\nabla^{\mu}f$
which holds for any scalar function $f$.%
}\begin{equation}
\nabla^{\mu}\left(\mathcal{R}^{2}T_{\mu\nu}\right)=\left(\mathcal{L}_{m}+\frac{1}{2}\left|\mathcal{R}\right|\frac{\partial\mathcal{L}_{m}}{\partial\left|\mathcal{R}\right|}\right)\nabla_{\nu}\mathcal{R}^{2}\label{eq:7.22}\end{equation}
or, equivalently\[
\mathcal{R}^{2}\nabla^{\mu}T_{\mu\nu}=\nabla^{\mu}\mathcal{R}^{2}\left(g_{\mu\nu}\mathcal{L}_{m}+\frac{1}{2}g_{\mu\nu}\left|\mathcal{R}\right|\frac{\partial\mathcal{L}_{m}}{\partial\left|\mathcal{R}\right|}-T_{\mu\nu}\right)=2\nabla^{\mu}\mathcal{R}^{2}\left(\frac{\partial\mathcal{L}_{m}}{\partial g^{\mu\nu}}+\frac{1}{2}g_{\mu\nu}\left|\mathcal{R}\right|\frac{\partial\mathcal{L}_{m}}{\partial\left|\mathcal{R}\right|}\right).\]
Finally, we arrive at\begin{equation}
\nabla^{\mu}T_{\mu\nu}=2\left(\frac{\partial\mathcal{L}_{m}}{\partial g^{\mu\nu}}+\frac{1}{2}g_{\mu\nu}\left|\mathcal{R}\right|\frac{\partial\mathcal{L}_{m}}{\partial\left|\mathcal{R}\right|}\right)\nabla^{\mu}\ln\mathcal{R}^{2}.\label{eq:7.23}\end{equation}
By virtue of \eqref{eq:7.23}, in curvature-scaling gravity, energy
and momentum are no longer conserved. Energy and momentum are measured
in each local spacetime pocket and are dependent on the Ricci scalar.
\footnote{As is discussed in Appendix \ref{sec:A}, energy of\emph{ }every system
(be it electromagnetic, or a relativistic energy level of the Hydrogen
atom, or a massive object) scales as $a_{\mathcal{R}}^{-1}$ (whereas
momentum, on the other hand, scales as $a_{\mathcal{R}}^{-1/2}$.)
This effect of {}``energy loss'' has already been known for radiation
in the expanding universe via the {}``Doppler theft''. Curiously,
standard cosmology exempts other forms of matter, such as the baryonic
matter, from the energy loss effect. This treatment is unwarranted
for two reasons: 
\begin{itemize}
\item The binding energy in atoms is also of electromagnetic nature. It
then cannot avoid the {}``Doppler theft''. The fact that galaxies
resists the cosmic expansion is irrelevant.
\item Radiation and matter were treated differently in standard cosmology.
This is in violation of Postulate (I) which posits that physical laws
retain their form in every local spacetime pocket. As the universe
expands, both radiation and matter must transform exactly the same;
i.e., they both have to experience the {}``Doppler theft''.
\end{itemize}
As we shall discuss in Section \ref{sec:9}, this practice in standard
cosmology arose from the problematic Friedmann model which does not
take into account the variation of $c$ as function of the cosmic
scale.%
}\newpage{}

\section{\label{sec:8}Implications of curvature-scaling gravity in cosmography}

\noindent Per the equivalence principle, physical laws are valid locally
in every pocket of spacetime; namely, physical laws retain their form
in all local regions on the manifold. Curvature-scaling gravity extends
the equivalence principle further by requiring that the parameters
specifying the physical laws are also valid only locally. The Ricci
scalar $\mathcal{R}$ serves as the local scale setter for all physical
processes that take place in any given local region. As a result,
the observer's clock rate scales anisotropically per \eqref{eq:scaleT},
$dt\propto a_{\mathcal{R}}^{3/2}$ (with the Ricci length being defined
as $a_{\mathcal{R}}\triangleq\left|\mathcal{R}\right|^{-1/2}$), and
the Planck constant and the speed of light are dependent on the Ricci
scalar per \eqref{eq:scaleH}, $\hbar\propto\left|\mathcal{R}\right|^{-1/4}$,
and \eqref{eq:scaleC}, $c\propto\left|\mathcal{R}\right|^{1/4}$.
We emphasize that -- in contrary to long-standing belief -- the variability
of $c$ does not violate causality or any principle of relativity
-- the Michelson-Morley finding, the Lorentz symmetry, the relativity
principle, the equivalence principle, and the general covariance principle.
The Michelson-Morley experiment only establishes the equality of light
speed at one location regardless of the light beam's direction. It
does not enforce the equality of light speed at different locations
or at different time points. (See Section \ref{sec:5} for our elaboration.)

The impacts of the $c$-variability are most manifest in cosmology.
As the universe expands, its Ricci scalar drops, leading to a progressive
decline in the value of light speed since $c\propto\left|\mathcal{R}\right|^{1/4}$.
Over billions years, the accumulation of $c$-variation shows up in
observational data. A proper treatment of cosmology must then take
into account the effects of $c$-variation as the universe expands.
Failure to do so leads to difficulties in explaining observational
data (unless one augments theoretical cosmology with ad hoc assumptions
and fudge agents, such as dark energy, accelerated expansion, and
inflationary expansion.)

Standard cosmology neglects to take into account the dependence of
light speed on the Ricci scalar (i.e., the cosmic scale factor). As
a result, it produced a series of flawed redshift formulae based on
which observational data were analyzed. In this section, we shall
revise these redshift formulae by incorporating the $c$-variability.
The outline of this section is:
\begin{itemize}
\item We first include the $c$-variation in the Robertson-Walker metric,
resulting in the modified Robertson-Walker metric.
\item We then revise the Lemaitre redshift formula. The exponent $\frac{3}{2}$
in the time anisotropy $dt\propto a_{\mathcal{R}}^{3/2}$ enters the
exponent of the scale factor in the modified Lemaitre redshift formula.
Due to this $\frac{3}{2}$ factor (which is missing in the standard
redshift formula), the Hubble law needs be revised and the Hubble
constant reduced. This is the key to resolving the age problem in
cosmology.
\item We next modify the distance-redshift relationship and the luminosity-redshift
relationship. Formula \eqref{eq:CSGformula} is the centerpiece for
our reassessment of Type Ia supernovae data obtained in \cite{Riess1,Perlmutter,Riess2}.
\item We offer an alternative interpretation of the Type Ia supernovae data
in \cite{Riess1,Perlmutter,Riess2}. Our interpretation is based on
the variation of $c$ as function of the cosmic scale factor, bypassing
the standard explanation of an accelerated expansion.
\end{itemize}
For the sake of comparison, the derivations below are presented both
in the traditional framework and in our approach. Correspondingly,
the calculations are based on the traditional RW metric and the modified
RW metric.

\subsection{\label{sub:modifiedRW}The modified Robert-Walker metric}

\noindent The Robertson-Walker (RW hereafter) metric starts with the
assumption of homogeneity and isotropy of space. It also assumes that
the spatial component of the metric can be time-dependent. All of
the time dependence is in the function $a(t)$, known as the {}``scale
factor''. The RW metric is the only one that is spatially homogeneous
and isotropic \cite{RW1,RW2,RW3,RW4}. This is a geometrical result
and is not tied to the equations of gravitation field.

The RW metric has been determined to be\begin{align}
ds^{2} & =c_{0}^{2}dt^{2}-a^{2}(t)\left[\frac{dr^{2}}{1-kr^{2}}+r^{2}d\Omega^{2}\right]\label{eq:8.1}\\
d\Omega^{2} & =d\theta^{2}+\sin^{2}\theta\, d\phi^{2}\nonumber \end{align}
where $a\left(t\right)$ is the global scale factor of the universe
and is a function of the cosmic time $t$ only (with a$\left(t_{0}\right)=1$
for at our current time $t_{0}$), $k$ the curvature determining
the shape of the universe (open/flat/closed for $k>0$, $k=0$, $k<0$
respectively.) Also recall that the RW metric assumes a constant speed
of light which we denoted as $c_{0}$ in \eqref{eq:8.1}.

When applying curvature-scaling gravity to cosmology, we retain the
homogeneity and isotropy of space. As such, the RW metric remains
applicable with the only modification in the dependence of $c$ on
the Ricci scalar (viz. the cosmic scale factor %
\footnote{We shall prove in Section \ref{sub:evolution} that the Ricci length
$a_{\mathcal{R}}$ is indeed proportional to the scale factor $a$.
Therefore, we use $a$ in place of $a_{\mathcal{R}}$ in this current
section.%
}) By virtue of the scaling rule \eqref{eq:scaleC}, $c=c_{0}\, a^{-1/2}$,
we arrive at the modified Robertson-Walker metric:\begin{equation}
ds^{2}=\frac{c_{0}^{2}}{a(t)}dt^{2}-a^{2}(t)\left[\frac{dr^{2}}{1-kr^{2}}+r^{2}d\Omega^{2}\right]\label{eq:8.2}\end{equation}
in which $c_{0}$ is the speed of light measured in the outer space
(subject to cosmic expansion) at our current time.

\subsection{\label{sub:Lemaitre}Modification to Lemaitre's redshift formula}

\subsubsection{For the traditional RW metric:}

\noindent Light wave travels in the null geodesics, $ds^{2}=0$. The
traditional RW metric \eqref{eq:8.1} can be recast as\begin{equation}
ds^{2}=a^{2}\left(t\right)\left[\frac{c_{0}^{2}}{a^{2}\left(t\right)}dt^{2}-\frac{dr^{2}}{1-kr^{2}}-r^{2}d\Omega^{2}\right].\label{eq:8.2-1}\end{equation}
The null geodesics for a light wave traveling from a galaxy toward
Earth (viz. $d\Omega=0$) is thus\begin{equation}
\frac{c_{0}\, dt}{a(t)}=\frac{dr}{\sqrt{1-kr^{2}}}.\label{eq:8.3}\end{equation}
Denote $t_{e}$ and $t_{o}$ the emission and observation timepoints
of the light wave, and $r_{e}$ the comoving distance of the galaxy
from Earth. From \eqref{eq:8.3} we have\begin{equation}
\int_{t_{e}}^{t_{o}}\frac{c\, dt}{a(t)}=\int_{r_{e}}^{0}\frac{dr}{\sqrt{1-kr^{2}}}.\label{eq:8.4}\end{equation}
The next wavecrest to leave the galaxy at $t_{e}+\delta t_{e}$ and
arrives at Earth at $t_{o}+\delta t_{o}$ satisfies\begin{equation}
\int_{t_{e}+\delta t_{e}}^{t_{o}+\delta t_{o}}\frac{c\, dt}{a(t)}=\int_{r_{e}}^{0}\frac{dr}{\sqrt{1-kr^{2}}}.\label{eq:8.5}\end{equation}
Subtracting the two equations yields\begin{equation}
\frac{\delta t_{o}}{a(t_{o})}=\frac{\delta t_{e}}{a(t_{e})}.\label{eq:8.6}\end{equation}
It is important to realize that in observational astronomy it is the
shift in frequency -- instead of wavelength -- in the light spectrum
that is used to determine the redshift of the distant light source.
In terms of frequency shift, the redshift is defined as\begin{equation}
z\triangleq\frac{\nu_{e}-\nu_{o}}{\nu_{o}}.\label{eq:8.8}\end{equation}

\noindent In standard cosmology, since $c$ is a constant, the conversion
from frequency to wavelength ($\lambda=c/\nu$) is trivial, yielding
an equivalent formula\begin{equation}
z=\frac{\lambda_{o}-\lambda_{e}}{\lambda_{e}}.\label{eq:8.11}\end{equation}
By virtue of \eqref{eq:8.6}, the observed frequency is related to
the emitted frequency by (since $\nu=1/\delta t$):\begin{equation}
\frac{\nu_{o}}{\nu_{e}}=\frac{\delta t_{e}}{\delta t_{o}}=\frac{a(t_{e})}{a(t_{o})}.\label{eq:8.7}\end{equation}
We thus have\begin{equation}
z=\frac{a(t_{o})}{a(t_{e})}-1.\label{eq:8.9}\end{equation}
With $t_{o}=t_{0}$ and $a(t_{o})=a(t_{0})=1$, the traditional Lemaitre
redshift formula is thus\begin{equation}
1+z=a^{-1}(t_{e}).\label{eq:8.10}\end{equation}

\subsubsection{\label{sub:Lemaitre modRW}For the modified RW metric:}

\noindent The modified RW metric \eqref{eq:8.2} can be recast as\begin{equation}
ds^{2}=a^{2}(t)\left[\frac{c_{0}^{2}}{a^{3}\left(t\right)}dt^{2}-\frac{dr^{2}}{1-kr^{2}}-r^{2}d\Omega^{2}\right].\label{eq:8.10-1}\end{equation}
The null geodesics for a light wave traveling radially is thus\begin{equation}
\frac{c_{0}\, dt}{a^{3/2}(t)}=\frac{dr}{\sqrt{1-kr^{2}}},\label{eq:8.12}\end{equation}
leading to\begin{equation}
\int_{t_{e}}^{t_{o}}\frac{c_{0}\, dt}{a^{3/2}(t)}=\int_{r_{e}}^{0}\frac{dr}{\sqrt{1-kr^{2}}}\label{eq:8.13-1}\end{equation}
and\begin{equation}
\int_{t_{e}+\delta t_{e}}^{t_{o}+\delta t_{o}}\frac{c_{0}\, dt}{a^{3/2}(t)}=\int_{r_{e}}^{0}\frac{dr}{\sqrt{1-kr^{2}}}.\label{eq:8.14-1}\end{equation}
Subtracting the two equations yields:\begin{equation}
\frac{\delta t_{o}}{a^{3/2}(t_{o})}=\frac{\delta t_{e}}{a^{3/2}(t_{e})}\label{eq:8.15}\end{equation}
from which, the observed frequency is related to the emitted frequency
by\begin{equation}
\frac{\nu_{o}}{\nu_{e}}=\frac{\delta t_{e}}{\delta t_{o}}=\frac{a^{3/2}(t_{e})}{a^{3/2}(t_{o})}.\label{eq:8.16}\end{equation}
The redshift parameter defined in \eqref{eq:8.8}, $z=\frac{\nu_{e}-\nu_{o}}{\nu_{o}}$,
satisfies the modified Lemaitre redshift formula \begin{equation}
1+z=a^{-3/2}(t_{e}).\label{eq:8.18}\end{equation}
The most striking feature of \eqref{eq:8.18} is the appearance of
the time anisotropy exponent $\eta=\frac{3}{2}$ in the scale factor
\footnote{\noindent As we noted before, in observational cosmology it is the
shift in frequency -- instead of wavelength -- that is used to determine
the redshift. This detail is often overlooked in theoretical treatments.
Notwithstanding this technical detail however, we shall show in Appendix
\ref{sec:G} that as long as the galaxy that emitted the photons and
the Earth-based telescope are immune from cosmic expansion, the shift
in the photon wavelength, defined as in \eqref{eq:8.11}, remains
precisely the modified Lemaitre redshift formula \eqref{eq:8.18}
in which $a(t_{e})$ is the cosmic scale factor at the outskirt of
the galaxy that emitted the photons and $a(t_{o})$ is the cosmic
scale factor at the outskirt of our Milky Way. Namely, the prefactor
of $3/2$ still participates. %
}. This feature drastically alters the Hubble law and the estimation
of the Hubble constant, which we shall show momentarily.

\subsection{\label{sub:Hubble}The modified Hubble law and the corrected value
for Hubble constant}

\noindent From the definition of the Hubble constant \begin{equation}
H_{0}\triangleq\left.\frac{\dot{a}}{a}\right|_{t=t_{0}}\label{eq:8.18-1}\end{equation}
and for a small time difference $t_{o}-t_{e}$, we get\begin{equation}
a\left(t_{e}\right)=1-H_{0}\left(t_{0}-t_{e}\right)+\dots\label{eq:8.18-2}\end{equation}
Denote $d=c_{0}.(t_{0}-t_{e})$ as the distance from the Earth to
a galaxy. For small $z$ and $d$, the Taylor expansion of the modified
Lemaitre redshift formula \eqref{eq:8.18} combined with \eqref{eq:8.18-2}
yields\begin{equation}
\begin{array}{ccl}
1+z & = & a^{-3/2}\left(t_{e}\right)\\
 & = & \left(1-H_{0}.\left(t_{o}-t_{e}\right)+\dots\right)^{-3/2}\\
 & = & 1+\frac{3}{2}H_{0}.\left(t_{o}-t_{e}\right)+\dots\\
 & = & 1+\frac{3}{2}H_{0}\,\frac{d}{c_{0}}+\dots\end{array}\label{eq:8.19}\end{equation}
We thus obtain the modified Hubble law:\begin{equation}
z=\frac{3}{2}H_{0}\,\frac{d}{c_{0}}.\label{eq:8.20}\end{equation}
Compared with the conventional Hubble law (which can be obtained from
the Taylor expansion of the traditional Lemaitre redshift formula
\eqref{eq:8.10}), where the speed of light is explicitly restored:\begin{equation}
z=H_{0}\,\frac{d}{c_{0}},\label{eq:8.21}\end{equation}
the modified Hubble law \eqref{eq:8.20} acquires an additional prefactor
of $\frac{3}{2}$. That means the redshift $z$ detected in the astronomer's
telescope was enhanced by a factor of $\frac{3}{2}$, the exponent
in the anisotropic time scaling per \eqref{eq:scaleT}, $dt\propto a_{\mathcal{R}}^{3/2}$.
As such, to back out the actual scale factor at the distance galaxy,
the astronomer has to reduce his reading of $z$ by a factor of $\frac{3}{2}$.
Without doing this, she would unknowingly overestimate the Hubble
constant. This overestimation is precisely what happened in standard
cosmology.

To put it another way, in the plot of $z$ versus $\frac{d}{c_{0}}$,
it is the $\frac{3}{2}H_{0}$ what is the slope of the line. If the
astronomer mistook the slope to be $H_{0}$, he would inadvertently
overestimate the Hubble constant. With the reported value for $H_{0}$
being $71$, in the light of our reasoning, the actual value for $H_{0}$
should -- per Eq. \eqref{eq:8.20} -- be only $\frac{2}{3}\times71\approx47$.\newpage{}

The reduced Hubble constant has two remarkable consequences:
\begin{itemize}
\item It increases the age of the universe by a factor of $\frac{3}{2}$
to about $13.8$ Glys. This is the key to resolving the age problem
bypassing the need for an accelerating phase following a deceleration
phase as advocated in \cite{Turner}.
\item It leads to a lower critical density (since $\rho_{c}\propto H_{0}^{2}$)
than previous thought by a factor of $\left(\frac{3}{2}\right)^{2}$.
This is the key to resolve the budgetary shortfall problem without
the need of dark energy.
\end{itemize}
The reduction of $\frac{3}{2}$ for the Hubble constant is not the
end of the story yet, however. We shall shortly show that there are
two additional sources of correction: (i) The first is a new distance-redshift
formula in our model as compared to that in the \textgreek{L}CDM model;
(ii) The other is an extra modification factor when one converts the
luminosity distance into the proper distance, a practice of importance
for high-$z$ objects, such as Type Ia supernovae.

\subsection{\label{sub:distance-redshift}The modified distance-redshift relationship}

\subsubsection{For the Lambda-CDM model:}

\noindent Let us first re-derive the distance-redshift relationship
within the \textgreek{L}CDM model for a flat universe. With $\rho_{M}$
and $\rho_{\Lambda}$ being the density of matter (ordinary and non-luminous)
and of dark energy, the traditional Friedmann equation is being recast
as \cite{Gorbunov}\begin{equation}
\frac{\dot{a}^{2}}{a^{2}}=\frac{8\pi G}{3}\left(\rho_{M}+\rho_{\Lambda}\right)=H_{0}^{2}\left[\Omega_{M}\frac{1}{a^{3}}+\Omega_{\Lambda}\right]\label{eq:8.22}\end{equation}
in which \begin{align}
\Omega_{M} & \triangleq\frac{\rho_{M}(t_{0})}{\rho_{c}(t_{0})}=\frac{8\pi G}{3H_{0}^{2}}\rho_{M}(t_{0})\label{eq:8.23}\\
\Omega_{\Lambda} & \triangleq\frac{\rho_{\Lambda}(t_{0})}{\rho_{c}(t_{0})}=\frac{8\pi G}{3H_{0}^{2}}\rho_{\Lambda}\label{eq:8.24}\\
\Omega_{M}+\Omega_{\Lambda} & =1\label{eq:8.25}\end{align}
Note that the density of ordinary matter and non-luminous matter scales
inversely with volume, $\rho_{M}\propto a^{-3}$, whereas the density
of dark energy is a constant.

For the almost flat space, setting $k=0$ in Eq. \eqref{eq:8.4} and
denote $r$ for $r_{e}$, we obtain the proper distance from the galaxy
and the Earth\begin{equation}
r\approx c_{0}\int_{t_{e}}^{t_{o}}\frac{dt}{a(t)}.\label{eq:8.26}\end{equation}
Utilizing the traditional Lemaitre redshift formula \eqref{eq:8.10}:\begin{equation}
1+z=a^{-1}(t)\ \Rightarrow\ dz=-\frac{\dot{a}}{a^{2}}dt\ \Rightarrow\ \frac{dt}{a}=-\frac{dz}{\dot{a}/a}\label{eq:8.26-1}\end{equation}
we obtain\begin{equation}
r=c_{0}\int_{0}^{z}\frac{dz'}{(\dot{a}/a)(z')}.\label{eq:8.27}\end{equation}
By virtue of Eqs. \eqref{eq:8.22} and \eqref{eq:8.10}, we then get
the traditional distance-redshift relationship\begin{equation}
\frac{r}{c_{0}}=\frac{1}{H_{0}}\int_{0}^{z}\frac{dz'}{\sqrt{\Omega_{M}(1+z')^{3}+\Omega_{\Lambda}}}\label{eq:8.28}\end{equation}
For low $z$, it recovers the Hubble law \eqref{eq:8.21} as expected:\[
\frac{r}{c_{0}}\approx\frac{z}{H_{0}}.\]
Two special cases of interest: 
\begin{itemize}
\item The Einstein-de Sitter universe $(\Omega_{M}=1,\,\Omega_{\Lambda}=0)$
corresponds to

\begin{equation}
\frac{r}{c}=\frac{2}{H_{0}}\left(1-\frac{1}{\sqrt{1+z}}\right)\label{eq:8.29}\end{equation}

\item The de Sitter universe $(\Omega_{M}=0,\,\Omega_{\Lambda}=1)$ corresponds
to

\begin{equation}
\frac{r}{c}=\frac{z}{H_{0}}\label{eq:8.30}\end{equation}

\end{itemize}

\subsubsection{For curvature-scaling gravity:}

\noindent Let us derive the distance-redshift relation in our approach.
To proceed, we shall only need the following result, which is to be
derived in Section \ref{sec:9}. The evolution of the cosmic scale
factor, derived solely from the scaling rule of time duration, is
found to be (see Eq. \eqref{eq:9.6})

\noindent \begin{equation}
a\left(t\right)=\left(\frac{t}{t_{0}}\right)^{2/3}\label{eq:8.33}\end{equation}
with $a\left(t_{0}\right)$ being set equal 1. Taking derivative of
the above equation, we obtain the Hubble constant\begin{equation}
H\left(t\right)\triangleq\frac{\dot{a}\left(t\right)}{a\left(t\right)}=\frac{2}{3t}\label{eq:8.33-1}\end{equation}
from which, the age formula is thus\begin{equation}
t_{0}=\frac{2}{3H_{0}}.\label{eq:8.34}\end{equation}
We further have

\noindent \begin{equation}
\frac{\dot{a}}{a}=H=\frac{2}{3t}=\left(\frac{2}{3t_{0}}\right)\left(\frac{t_{0}}{t}\right)=H_{0}\frac{1}{a^{3/2}}.\label{eq:8.35}\end{equation}

\noindent With Eq. \eqref{eq:8.13-1} for $k=0$, the proper distance
in the almost flat space is \begin{equation}
r\approx\int_{t_{e}}^{t_{o}}\frac{c_{0}\, dt}{a^{3/2}(t)}.\label{eq:8.36}\end{equation}
Utilizing the modified Lemaitre redshift formula \eqref{eq:8.18}:
\begin{equation}
1+z=a^{-3/2}(t)\ \Rightarrow\ dz=-\frac{3}{2}\frac{\dot{a}}{a^{5/2}}dt\ \Rightarrow\ \frac{dt}{a^{3/2}}=-\frac{2}{3}\frac{dz}{a^{5/2}}\label{eq:8.36-1}\end{equation}
we obtain \begin{equation}
r=\frac{2}{3}c_{0}\int_{0}^{z}\frac{dz'}{(\dot{a}/a)(z')}\label{eq:8.36-2}\end{equation}
By virtue of Eqs. \eqref{eq:8.35} and \eqref{eq:8.18}, we then get
the modified distance-redshift relationship \begin{equation}
\frac{r}{c_{0}}=\frac{2}{3H_{0}}\int_{0}^{z}\frac{dz'}{1+z'}=\frac{2}{3H_{0}}\,\ln\left(1+z\right)\label{eq:8.37}\end{equation}
Two important features to note:
\begin{itemize}
\item Only $H_{0}$ participates in \eqref{eq:8.37}. The mass density and
the spatial curvature are not involved. This helps make the fitting
to observational data parsimonious and universal.
\item The correction factor of $\frac{3}{2}$ for $H_{0}$ is again manifest
in the right-hand-side of \eqref{eq:8.37}. For low $z$, it recovers
the modified Hubble law \eqref{eq:8.20} as expected, $\frac{r}{c_{0}}\approx\frac{2}{3H_{0}}z$. 
\end{itemize}

\subsection{\label{sub:luminosity-redshift}The modified luminosity-redshift
relationship}

\noindent So far we focused on the proper distance for cosmological
objects. However, proper distance is not directly measurable but must
be deduced from angular distance or luminosity distance. We shall
work out the conversion from the proper distance to the luminosity
distance in our approach.

\subsubsection{For the Lambda-CDM model:}

\noindent Consider a source located at the comoving coordinate $\chi_{e}$
with total luminosity $L$. The energy output at emission time $t_{e}$
within the window $\delta t_{e}$ is given by \begin{equation}
\Delta E_{e}=L\,\delta t_{e}.\label{eq:8.39}\end{equation}
As the photons traversed the distance, their energy gets {}``Doppler
thieved'' by the scale factor $a(t_{o})/a(t_{e})$. Therefore, at
the moment of observation, the observed energy will be \begin{equation}
\Delta E_{o}=\Delta E_{e}\, a(t_{e}).\label{eq:8.40}\end{equation}
The physical area of the sphere centered at $r_{e}$ and radius $r(z)$
to be crossed by photons today is\begin{equation}
S(z)=4\pi r^{2}(z).\label{eq:8.41}\end{equation}
The visible brightness (energy flux at observer's position) equals\begin{equation}
J=\frac{\Delta E_{o}}{S(z)\,\Delta t_{o}}=\frac{\Delta E_{e}\, a(t_{e})}{S(z)\,\Delta t_{o}}=\frac{(L\Delta t_{e})\, a(t_{e})}{S(z)\,\Delta t_{o}}=\frac{L}{S(z)}a^{2}(t_{e})\label{eq:8.42}\end{equation}
in which we have made use of \eqref{eq:8.6} \[
\frac{\delta t_{o}}{a(t_{o})}=\frac{\delta t_{e}}{a(t_{e})}.\]
Utilizing the traditional Lemaitre redshift formula \eqref{eq:8.10},
we obtain\begin{equation}
J=\frac{L}{4\pi r^{2}(z)}(1+z)^{-2}.\label{eq:8.44}\end{equation}
The photometric distance $d_{L}$ is defined via\begin{equation}
J=\frac{L}{4\pi d_{L}^{2}(z)}\label{eq:8.45}\end{equation}
which leads to\begin{equation}
d_{L}=(1+z)\, r(z).\label{eq:8.46}\end{equation}
Coupling with \eqref{eq:8.28}\begin{equation}
\frac{r}{c_{0}}=\frac{1}{H_{0}}\int_{0}^{z}\frac{dz'}{\sqrt{\Omega_{M}(1+z')^{3}+\Omega_{\Lambda}}},\label{eq:8.47}\end{equation}
we obtain the traditional luminosity-redshift relationship \begin{equation}
\frac{d_{L}}{c_{0}}=\frac{1+z}{H_{0}}\,\int_{0}^{z}\frac{dz'}{\sqrt{\Omega_{M}(1+z')^{3}+\Omega_{\Lambda}}}\label{eq:8.48}\end{equation}
For the sake of later comparison, the Einstein-de Sitter universe
$\left(\Omega_{M}=1,\ \Omega_{\Lambda}=0\right)$ has\begin{equation}
\frac{d_{L}}{c_{0}}=2\frac{1+z}{H_{0}}\,\left(1-\frac{1}{\sqrt{1+z}}\right)\label{eq:8.49}\end{equation}

\subsubsection{For curvature-scaling gravity:}

\noindent In our approach, although the speed at which photons arrive
at the observer gets modified, the Planck constant also gets modified.
With $c\propto a^{-1/2}$ and $\hbar\propto a^{1/2}$, the photon
energy continues to scale as $E=\hbar c/\lambda\propto a^{-1}$. Therefore,
the photon energy still gets {}``Doppler thieved'' by the scale
factor $a(t_{o})/a(t_{e})$ as in the traditional approached.

The visible brightness (energy flux at observer's position) equals\begin{equation}
J=\frac{\Delta E_{o}}{S(z)\,\Delta t_{o}}=\frac{\Delta E_{e}\, a(t_{e})}{S(z)\,\Delta t_{o}}=\frac{(L\Delta t_{e})\, a(t_{e})}{S(z)\,\Delta t_{o}}=\frac{L}{S(z)}a^{5/2}(t_{e})\label{eq:8.50}\end{equation}
in which we have made use of \eqref{eq:8.15}\[
\frac{\delta t_{o}}{a^{3/2}(t_{o})}=\frac{\delta t_{e}}{a^{3/2}(t_{e})}.\]
Utilizing the modified Lemaitre redshift formula \eqref{eq:8.18},
we obtain\begin{equation}
J=\frac{L}{4\pi r^{2}(z)}\left(1+z\right)^{-5/3}.\label{eq:8.51}\end{equation}
The photometric distance, defined in \eqref{eq:8.45}, now becomes\begin{equation}
d_{L}=\left(1+z\right)^{5/6}\, r(z).\label{eq:8.52}\end{equation}
Coupling with \eqref{eq:8.37} \begin{equation}
\frac{r}{c_{0}}=\frac{2}{3H_{0}}\,\ln\left(1+z\right)\label{eq:8.53}\end{equation}
we obtain the modified photometric distance-redshift relationship\begin{equation}
\frac{d_{L}}{c_{0}}=\frac{2}{3}\,\frac{\left(1+z\right)^{5/6}}{H_{0}}\,\ln(1+z)\label{eq:CSGformula}\end{equation}
This is the central formula of our curvature-scaling gravity approach
to re-assess the Type Ia supernovae data. It is universally applicable
to all shapes of the universe (flat/open/closed).

\subsection{\label{sec:SN1a}A critical analysis of Type Ia supernovae data}

\noindent In 1998, it was announced that Type Ia supernova of high-redshift
reveal a surprising behavior \cite{Riess1,Perlmutter}. The supernovae
appear fainter -- and thus farther -- than would have been expected
for cosmological objects of these redshifts based on the Friedmann
model. Type Ia supernovae are objects with stable luminosity, hence
the name standard candles. Astronomers exploit this property to deduce
the luminosity distance $d_{L}$ from Earth of these objects. Their
redshift values $z$, as usual, are independently obtained from the
Doppler effect. For Type Ia supernovae, the curve $d_{L}$ against
$z$ bends upward toward the high-$z$ end. This striking behavior
of Type Ia supernovae is said to be the evidence for dark energy,
which is modeled as the cosmological constant term $\Omega_{\Lambda}$
in the \textgreek{L}CDM model.

Standard cosmology, however, does not take into account the variation
of light speed as the universe expands. We shall show in this subsection
and the next that it is this very neglect in standard cosmology that
is responsible for behavior of the Type Ia supernovae. We must reexamine
the observational data using the revised redshift formulae derived
in this section so far. The centerpiece in our analysis below is the
modified photometric distance-redshift relationship \eqref{eq:CSGformula}:\[
\frac{d_{L}}{c_{0}}=\frac{2}{3}\,\frac{\left(1+z\right)^{5/6}}{H_{0}}\,\ln(1+z).\]
Observational data are cited in the form of distance modulus $m-M$
which is related to photometric distance $d_{L}$ as:\begin{equation}
m-M=5\log_{10}(d_{L}/Mpc)+25\label{eq:8.55}\end{equation}
We retrieve the observational data of 41 high-$z$ objects from Table
6 of \cite{Riess2}. The values of $z$ and of $m-M$ are listed in
columns 2 and 3 in the Table respectively. We then extract the photometric
distances $d_{L}$ via \eqref{eq:8.55}.

\begin{figure}
\centering{}\includegraphics[scale=0.75]{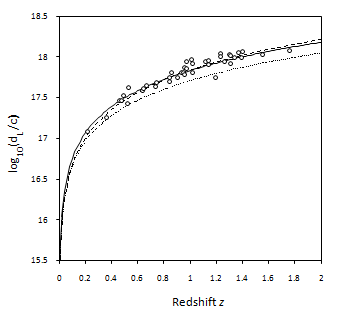} \caption{Comparison of various luminosity distance-redshift formulae to Type
Ia supernovae data. Open circles: 41 data points of high-$z$ objects
listed in Table 4 of Ref\cite{Riess2}. Long-dashed line: \textgreek{L}CDM
model's formula, Eq. \eqref{eq:8.48}, with $H_{0}=70.5,\,\Omega_{M}=0.27,\,\Omega_{\Lambda}=0.73$.
Dotted line: Einstein-de Sitter model's formula, Eq. \eqref{eq:8.49},
with $H_{0}=70.5$ ($\Omega_{M}=1,\,\Omega_{\Lambda}=0$). Solid line:
our formula, Eq. \eqref{eq:CSGformula}, using $H_{0}=37.4$.}

\end{figure}

Figure 1 is the main result of this section. The vertical axis measures
$\log_{10}\left(d_{L}/c\right)$ where $c$ is the value of speed
of light measured on Earth (i.e., 300,000$\,\mbox{km.sec}^{-1}$).
The set of 41 Type Ia supernovae data are shown in open circles. The
solid line shows our model's Formula \eqref{eq:CSGformula} corresponding
to the only parameter $H_{0}=37.4$ that best fits to the data. The
long-dashed line corresponds to the \textgreek{L}CDM model's Formula
\eqref{eq:8.48} with 3 parameters $H_{0}=70.5,\,\Omega_{M}=0.27,\,\Omega_{\Lambda}=0.73$
(besides $\Omega_{curv}=1-\Omega_{M}-\Omega_{\Lambda}=0$). The dotted
line shows Einstein-de Sitter universe's Formula \eqref{eq:8.49}
using $H_{0}=70.5$.

The mean average error (MAE) between a model's prediction and the
41 observational data is:\begin{equation}
\frac{1}{41}\sum_{i=1}^{41}\left|\log_{10}\frac{d_{L}^{model}(i)}{c}-\log_{10}\frac{d_{L}^{observed}(i)}{c}\right|\label{eq:8.56}\end{equation}
The MAE of our Formula \eqref{eq:CSGformula} is 0.0469 which is comparable
to the MAE of the \textgreek{L}CDM model \eqref{eq:8.48}, 0.0476.

The most striking result is a new -- and much lower -- value for Hubble
constant $H_{0}=37.4$. The reduction in value of $H_{0}$ (from the
conventionally accepted 70.5) leads to two important consequences: 
\begin{enumerate}
\item A reduction in value of the critical density: $\rho_{c}=3H_{0}^{2}/(8\pi G)$
is reduced by the factor $(70.5/37.4)^{2}\approx3.54$. That means
the actual $\rho_{c}$ should be only $28\%$ of the long-believed
value. Interestingly, the new value of $\rho_{c}$ approximately equals
the total amount of ordinary matter and {}``dark matter'' found
in the universe. %
\footnote{We must note that although the total 28\% leftover is said to be consisted
of ordinary matter and {}``dark matter'', in light of our consideration
in Section \ref{sec:4}, the {}``dark matter'' component is not
necessarily a hypothetical invisible form of matter. It could be nothing
but the linear Mannheim-Kazanas potential term (for ordinary matter)
which, after adding up ordinary matter in the universe altogether,
acts as surrogate for the {}``dark matter'' content.%
} The budgetary shortfall in density (to maintain a near-flat space)
disappears. {}``Dark energy'' is not needed.
\item The universe age: Our universal age formula \eqref{eq:8.34}, $t_{0}=\frac{2}{3H_{0}}$,
yields $17.4$ Glys. The overestimated value of 70.5 for $H_{0}$,
on the other hand, led to a much younger age at 9.3 Gly if one assumes
the Einstein-de Sitter model. Such a young universe would be at odd
with the established age of the oldest stars. %
\footnote{A standard resolution for the age problem is said to have been found
in \cite{Turner} via the \textgreek{L}CDM model's age formula: $t_{0}=(2\tanh^{-1}\sqrt{\Omega_{\Lambda}})/(3H_{0}\sqrt{\Omega_{\Lambda}})$
which yields $13.8$ Glys using $H_{0}=70.5,\ \Omega_{M}=0.27,\ \Omega_{\Lambda}=0.73.$
For completeness, the \textgreek{L}CDM model's age formula is derived
below. In a flat space, using Eq. \eqref{eq:8.22}, $\frac{\dot{a}}{a}=H_{0}\left(\Omega_{M}a^{-3}+\Omega_{\Lambda}\right)^{\frac{1}{2}}$,
we get

\begin{equation}
t_{0}=\frac{1}{H_{0}}\int_{0}^{1}\frac{da}{a\,\left(\Omega_{M}a^{-3}+\Omega_{\Lambda}\right)^{1/2}}=\frac{2}{3H_{0}}\frac{\tanh^{-1}\sqrt{\Omega_{\Lambda}}}{\sqrt{\Omega_{\Lambda}}}.\label{eq:LCDMage}\end{equation}
The \textgreek{L}CDM model assumes an acceleration phase accompanying
a deceleration phase. With $\Omega_{\Lambda}=0.73$, Eq. \eqref{eq:LCDMage}
yields $t_{0}\approx H_{0}^{-1}$. In light of the overestimation
of the Hubble constant in standard cosmology, the age problem is resolved
in our approach. The \textgreek{L}CDM answer \cite{Turner} based
on dark energy is no longer necessary.%
} 
\end{enumerate}
These two results will throw important light onto the orthodox interpretation
of an {}``accelerating'' universe discussed in the following section.

\subsection{\label{sec:alternative}An alternative interpretation to Type Ia
supernovae data based on curvature-scaling gravity}

\noindent A widely accepted view in standard cosmology is that the
universe expansion has been accelerating. The conclusion was inferred
from the data of Type Ia supernovae\cite{Riess1,Perlmutter}. In these
ground-breaking discoveries, distant supernovae appear substantially
fainter than what would have been expected for objects at the same
redshift (if the cosmological term is absent). This in turn indicates
that the supernovae were farther than what the standard model would
have predicted for their redshift. An explanation is that the space
expansion at the time was slower than expected from the standard Friedmann
model, or, in other words, the expansion has been speeding up; hence
the name {}``accelerating'' expansion. This interpretation has fueled
both observational and theoretical searches for a form of {}``dark
energy'' that supposedly speeds up the cosmic expansion.

As we pointed out in this section so far however, the Friedmann model
fails to take into account the dependence of $c$ on the cosmic scale
$a$. As such, the redshift formula \eqref{eq:8.48} is incorrect
and thus cannot be used to analyze the Type Ia supernovae data. The
conclusion regarding {}``acceleration expansion'' is based on this
flawed formula. Instead, the correct redshift formula is \eqref{eq:CSGformula}
which involves only one parameter $H_{0}$. The reported value of
$H_{0}\approx70.5$ was overestimated inadvertently from the flawed
redshift formula. Using the correct formula \eqref{eq:CSGformula},
the reduced value for $H_{0}\approx37.4$ simultaneously accounts
for three things: (i) The universe's correct age of 17.4 Glys, (ii)
The budgetary shortfall (the correct critical density being only 28\%
of what previously thought), and (iii) The fit to the Type Ia supernovae
data -- a fit as good as the \textgreek{L}CDM model produced (see
Fig. 1). Again, our resolution involves only $H_{0}$ and does not
resort to any fudge factors whatsoever, such as the amount of {}``dark
energy''.

Qualitatively, there are two intuitive interpretations for the observational
data of Type Ia supernovae:
\begin{enumerate}
\item Consider two supernovae $A$ and $B$ at distances $3\, bn$ and $6\, bn$
light years away from the Earth, viz. $d_{B}=2\, d_{A}$. Standard
cosmology dictates the redshift values of $z_{A}$ and $z_{B}\approx2\, z_{A}$
for them (to first-order approximation.) However, light travelled
faster in a distant past than it did in a more recent epoch. Thus,
the $B$-photon covered twice as long the distance in less than twice
the amount of time as compared with the $A$-photon. Having spent
less time in transit than expected, the $B$-photon experienced less
cosmic expansion than expected, and thus less redshift than standard
cosmology predicts. Namely: $z_{B}<2\, z_{A}$. This result means
an upward slopping in the curve as we put the two supernovae's data
on the $d$ vs. $z$ plot. Conversely, a supernova $C$ with $z_{C}=2\, z_{A}$
must correspond to a distance greater than $6\, bn$ light years,
viz. $d_{C}>d_{B}=2\, d_{A}$, and thus is a fainter object. This
is precisely what is observed in Type Ia supernovae \cite{Riess1,Perlmutter}.
\item Consider the Hubble law which is a good first-order approximation
for the distance-redshift relationship. (Although supernovae billions
light years away from the Earth do not exactly obey the Hubble law,
the Hubble law is used in this paragraph as a baseline to illustrate
the logic behind the deviation from the traditional distance-redshift
formula.) The farther the object, the higher its redshift. Per Hubble
law, its redshift $z$ and distance to Earth $d$ are in linear proportion
\footnote{As we clarified by now, there is an additional factor of $\frac{3}{2}$
in the Hubble law above, but this factor is not relevant for our reasoning
here.%
}:\begin{equation}
z=\frac{H}{c}\, d+\text{higher-order terms}\label{eq:8.57}\end{equation}
or\begin{equation}
d\approx z\,\frac{c}{H}\label{eq:8.58}\end{equation}
where $H$ is the rate of expansion at redshift $z$. The value of
$z$ is deduced from the photon's redshift. The value of $d$ is deduced
from the luminosity of the object. Prior to 1998, the supernova at
a given $z$ had been expected to be at a distance $d$ dictated by
Eq. \eqref{eq:8.58}. The fainter-than-expected supernovae observed
in 1998 \cite{Riess1} were thus at a greater distance than given
in \eqref{eq:8.58}. The conventional explanation then holds that
for a given $z$, the higher $d$ than usual implies a value of $H$
lower than usual. So, the supernovae must have experienced a lower
rate of expansion $H$ than now. One then concluded that the universe
has been expanding faster and faster.

Curvature-scaling gravity offers an alternative interpretation: for
a given $z$, the higher $d$ than usual implies a value of $c$ higher
than usual. So, light in the past must have traveled faster than it
does now. This conclusion dovetails perfectly with curvature-scaling
gravity that light speed has been adapting to the lowering Ricci scalar,
per \eqref{eq:scaleC}: $c\propto\left|\mathcal{R}\right|^{1/4}$,
as the universe expands. The universe is not accelerating.

\end{enumerate}
\noindent Note that we deliberately avoided the {}``light-slows-down''
misnomer since the jargon fails to reflect the right physics: in curvature-scaling
gravity, $c$ adapts to the Ricci scalar (or, equivalently, to the
cosmic scale factor); the variation in $c$ is not as a result of
the cosmic time passage.

If one had not prejudiciously subscribed to the seven-decades-old
Friedmann model and an (overreaching) belief on a universal $c$,
the ground-breaking discoveries regarding Type Ia supernovae in 1998
\cite{Riess1,Perlmutter} would have squarely been seen as compelling
evidence in favor of the variation of $c$ as the universe expands.
As such, these discoveries carry far more profound impacts than so
far thought because they help get rid of the once-sacred universality
of light speed from the scientific vocabulary. Such a conclusion is
far more significant and valuable than what ad hoc addenda such as
the {}``accelerating expansion'' and {}``dark energy'' could bring
forth. These addenda were unnecessary supplementaries trying to bring
the observational data back in line with the problematic Friedmann
model which neglects the variation of $c$ in its treatment.

\newpage{}

\section{\label{sec:9}Implications of curvature-scaling gravity in cosmology}

As we elaborated in the preceding section, the standard Friedmann
model does not take into account the variation of light speed as function
of the Ricci scalar (viz. the cosmic scale factor.) This neglect is
the root of the age problem and the deviation from the standard redshift
formulae, the deviation observed in Type Ia supernovae \cite{Riess1,Perlmutter}.
Upon incorporating the $c$-variation, we are able to resolve these
problems from first principles. Beside those two problems, the Friedmann
model encountered other serious difficulties, both observationally
and theoretically. Observationally, the Friedmann model standing alone
could not explain the near uniform horizon and the near flatness of
space. Theoretically, it suffered a series of cosmic coincidences
(also known as fine tunings), such as Dicke's instability (viz. the
oldness problem) and the budgetary shortfall. To reconcile the Friedmann
model with observations as well as to salvage the theoretical basis
of the Friedman model itself, standard cosmology has augmented the
model with supplementaries which require new physics. The supplementaries
are the dark energy and the inflationary expansion hypothesis. The
supplementaries, however, are plagued with their own fine tuning problems
and difficulties.

In this section, we shall apply curvature-scaling gravity, which allows
the variation of $c$ as function of the cosmic scale factor, to resolve
some of the most pressing problems in cosmology: (i) the horizon problem,
(ii) the flatness problem, and (ii) the oldness problem. Our results
are derived solely from the scaling rule \eqref{eq:scaleC}, $c\propto a^{-1/2}$,
and are model-independent; that is to say, they do not rely on any
model for the distribution of matter in the universe, apart from the
homogeneity and isotropy assumptions of space.

\subsection{\label{sub:evolution}The evolution of the cosmic scale}

Our treatment of cosmology is based on the modified Robertson-Walker
metric which expressly allows the dependence of $c$ on the cosmic
scale factor. The modified RW metric was given in Eq. \eqref{eq:8.2}:
\begin{eqnarray}
ds^{2} & = & c_{0}^{2}\frac{a_{0}}{a\left(t\right)}dt^{2}-a^{2}\left(t\right)\left[\frac{dr^{2}}{1-kr^{2}}+r^{2}d\Omega^{2}\right]\nonumber \\
 & = & a^{2}\left(t\right)\left[c_{0}^{2}\frac{a_{0}}{a^{3}\left(t\right)}dt^{2}-\frac{dr^{2}}{1-kr^{2}}-r^{2}d\Omega^{2}\right]\label{eq:9.1}\end{eqnarray}
in which we restore $a\left(t_{0}\right)=a_{0}$. Closed, flat, open
universe corresponds to $k>0,\ k=0,\ k<0$ respectively. Further define
the conformal time in terms of the cosmic time $t$ via\begin{equation}
d\eta=c_{0}\frac{a_{0}^{1/2}}{a^{3/2}\left(t\right)}dt.\label{eq:9.1-1}\end{equation}
The modified RW metric becomes\begin{equation}
ds^{2}=a^{2}\left(\eta\right)\left[d^{2}\eta-\frac{dr^{2}}{1-kr^{2}}-r^{2}d\Omega^{2}\right].\label{eq:9.1-2}\end{equation}
The rate of change for the cosmic scale $a\left(t\right)$ is determined
solely by the current state of the universe. At the scale factor $a$,
the rate of change should be a function of $a$ only, namely\begin{equation}
\frac{da}{dt}=f\left(a\right)\label{eq:9.2}\end{equation}
with $f$ being a functional form to be specified. With the scaling
rule for time duration \eqref{eq:scaleT}, we have\begin{eqnarray}
da & \propto & a\label{eq:9.3}\\
dt & \propto & a^{3/2}.\label{eq:9.4}\end{eqnarray}
As such

\begin{equation}
\frac{da}{dt}=\frac{\mbox{const}}{a^{1/2}}.\label{eq:9.5}\end{equation}
This equation accepts a solution\begin{equation}
a\left(t\right)=a_{0}\left(\frac{t}{t_{0}}\right)^{2/3}\label{eq:9.6}\end{equation}
with $a_{0}$ being the current scale factor. The Hubble constant
then is \begin{equation}
H\triangleq\frac{\dot{a}}{a}=\frac{2}{3t}.\label{eq:9.7}\end{equation}
The current age of the universe is related to the current value of
the Hubble constant:\begin{equation}
t_{0}=\frac{2}{3H_{0}},\label{eq:9.8}\end{equation}
from which we obtain the evolution rule for the cosmic scale factor\begin{equation}
a\left(t\right)=a_{0}\left(\frac{3}{2}H_{0}t\right)^{2/3}.\label{eq:9.9}\end{equation}
These results are entirely generic, independent of any specific details
of the matter distribution or content in universe, apart from the
homogeneity and isotropy of space.

The universe is thus expanding in the critical fashion: $a\simeq t^{\frac{2}{3}}$,
regardless of the amount of matter and/or the nature of matter content
(radiation, fermionic, etc.) and the shape of the Universe (open/closed/flat).
Unlike the Friedmann equations which typically allow $3$ different
modes %
\footnote{Plus the acceleration mode induced by the cosmological constant.%
}, in our approach, the evolution is unique and robust; the {}``critical
fashion'' of the cosmic expansion is also precisely what is seen
in observational cosmology.

Another piece of evidence in support of the evolution \eqref{eq:9.9}
can also be seen as follows. From \eqref{eq:9.1-1}\begin{equation}
\frac{d\eta}{dt}=c_{0}\frac{a_{0}^{1/2}}{a^{3/2}}=\frac{c_{0}}{a_{0}}\left(\frac{a_{0}}{a}\right)^{3/2}=\frac{c_{0}}{a_{0}}\frac{t_{0}}{t}\label{eq:9.10-1}\end{equation}
and \eqref{eq:9.8}, we get\begin{equation}
t=t_{0}\exp\left(\frac{a_{0}}{c_{0}t_{0}}\eta\right)=t_{0}\exp\left(\frac{3}{2}\frac{H_{0}a_{0}}{c_{0}}\eta\right)\label{eq:9.11}\end{equation}
with $\eta=0$ corresponding to today $t_{0}$. Combining with \eqref{eq:9.6},
we get\begin{equation}
a\left(\eta\right)=a_{0}\exp\left(\frac{H_{0}a_{0}}{c_{0}}\eta\right)\label{eq:9.12}\end{equation}
Next, for the modified RW metric, the Ricci scalar is (see Appendix
\ref{sec:E}) \begin{equation}
\mathcal{R}=-\frac{6}{a^{2}}\left(\frac{1}{a}\frac{d^{2}a}{d\eta^{2}}+\kappa\right).\label{eq:9.10}\end{equation}
and, by virtue of \eqref{eq:9.12} \begin{equation}
\mathcal{R}=-6\,\frac{H_{0}^{2}a_{0}^{2}c_{0}^{-2}+\kappa}{a^{2}}.\label{eq:9.13}\end{equation}
The Ricci length then reads\[
a_{\mathcal{R}}\triangleq\left|\mathcal{R}\right|^{-1/2}=\frac{a}{\sqrt{6\left|H_{0}^{2}a_{0}^{2}c_{0}^{-2}+\kappa\right|}}\]
which scales precisely as the cosmic scale factor. An evolution rule
away from \eqref{eq:9.6} would likely jeopardize the proportion between
the Ricci length and the cosmic scale factor.

\subsection{\label{sub:age prob}Resolution to the age problem}

\noindent Our resolution to the universe's age has been discussed
in details in Sections \ref{sec:SN1a} and \ref{sec:alternative}.
Here, we only wish to recap. The age formula \eqref{eq:9.8}\[
t_{0}=\frac{2}{3H_{0}}\]
is robust. It is derivable solely from the scaling property of time
duration. Furthermore, it is valid for all shapes of the universe.

With the corrected $H_{0}=37.4$ obtained in Section \ref{sec:SN1a},
the universe age is $t_{0}=17.4\, Gy$ comfortably accommodating its
oldest stars. The standard \textgreek{L}CDM resolution \cite{Turner}
that invokes an acceleration phase following by a deceleration phase
is no longer necessary.

\subsubsection*{The past and future of the universe:}

\noindent Regardless of its shape and/or density, the universe always
grows in accordance with a universal law \eqref{eq:9.6}: $a\propto t^{2/3}$.
The universe will not collapse nor expand supercritically. Note that
the mode of expansion in the Friedmann model is highly sensitive to
the initial state of the universe. To explain the observed critical
mode, the Friedmann model would require a fine tuning to an extraordinary
level. The anthropic principle is not needed in our approach.

\subsection{\label{sub:horizon prob}Resolution to the horizon problem}

\noindent The cosmic microwave background (CMB) observational result
shows a highly uniform distribution (to the accuracy of $10^{-5}$)
of cosmic radiation across the horizon. This uniformity presents a
serious challenge to the Friedmann model. The model needs to reconcile
the observed uniformity with the fact that the current horizon is
not causally connected. However, as we pointed out before, the Friedmann
model neglects the variability of $c$ as function of the cosmic scale
factor $a$. As the universe expands, $c$ gradually drops accordingly
in reverse proportion to $\sqrt{a}$. Conversely, as we trace back
to the origin of the cosmic time, $c$ was higher when the universe
was smaller.

More concretely, in the modified RW metric, the cosmological horizon
is given by:\begin{equation}
l_{H}\left(t_{0}\right)=\int d\eta=\int_{0}^{t_{0}}\frac{c_{0}\, a_{0}^{1/2}\, d\tau}{a^{3/2}(\tau)}\label{eq:9.14}\end{equation}
in which we utilized the conversion \eqref{eq:9.1-1}. Since $a(\tau)=a_{0}\left(\frac{\tau}{t_{0}}\right)^{2/3}$
per \eqref{eq:9.6}, we obtain\begin{equation}
l_{H}\left(t_{0}\right)=\int_{0}^{t_{0}}\frac{c_{0}\, a_{0}^{1/2}d\tau}{a_{0}^{3/2}.\left(\tau/t_{0}\right)}=\frac{c_{0}\, t_{0}}{a_{0}}\,\int_{0}^{t_{0}}\frac{d\tau}{\tau}=\infty.\label{eq:9.15}\end{equation}
The cosmological horizon is (logarithmically) divergent. Thus, the
entire universe was causally connected. This result neatly helps
explain the near uniformity in our current horizon. Our explanation
comes purely from the scaling rules and the modified RW metric. Interestingly,
the divergence in \eqref{eq:9.15} is logarithmic and is thus parsimonious
-- a rather curious result.

Our treatment allows $c$ to vary with the scale factor and, otherwise,
is in line with Moffat, and Magueijo and Albrecht's original insights
in resolving the horizon problem using variable speed of light mechanism.
In \cite{Moffat}, Moffat stipulated that if the velocity of light
in the baby universe was higher than it is now by a factor of $10^{30}$,
the horizon of the early universe would be in causal contact and thus
achieve uniformity. Beside Moffat's original proposal that $c$ underwent
a first-order phase transition from a very high value to its current
value, subsequent developments of Moffat, and of Magueijo and Albrecht
\cite{Magueijo1,Magueijo2} allowed a smooth variation in $c$. Unlike
these author's approaches, our approach does not require the need
to devise a dynamics for the variable $c$. In our work, $c$ is dependent
on the cosmic scale factor instead.

To explain the horizon problem, standard cosmology augmented the Friedmann
model with an inflationary phase in which the baby universe underwent
an exponential expansion mode \cite{Guth}. Before the inflationary
phase, the whole horizon was speculated to have been in causal contact
and have thus reached a thermal equilibrium. As the inflation kicked
in, different sections in the horizon were rapidly pulled away from
one another and, as a result, now appear separated. The inflationary
universe hypothesis in turn requires several ad hoc assumptions to
explain the mechanism of the inflation and the nature of the agent
that could cause it. These assumptions are not intrinsic to the established
physics. Furthermore, the hypothesis encounters its own fine-tuning
problems.

Curvature-scaling gravity provides a natural explanation to the horizon
problem, without any ad hoc assumptions. The underlying agent is the
adaptation of $c$ to the prevailing Ricci scalar along the course
of the universe expansion.

\subsection{\label{sub:flatness prob}Resolution to the flatness problem}

\noindent The Wilkinson Microwave Anisotropy Probe (WMAP) has confirmed
that the universe is essentially flat. The universe thus required
specific mechanisms to achieve and sustain flatness.

Our model also offers a natural explanation to the flatness problem.
As the early Universe expanded, the speed at which light travels decreased,
according the scaling rule \eqref{eq:scaleC}, $c\propto a^{-1/2}$.
As the universe grew in size several orders over, at the same time
the speed of light dropped precipitously several orders in magnitude.
The effective horizon shrank relatively to the universe's size. By
the decoupling event when light began to travel freely, only light
from a pocket around us can reach us today. This explains the flatness
problem. Note that the flatness problem is not identical to the horizon
problem; even before the decoupling event, matter still could communicate
via other channels, such as neutrinos or gravitons. The WMAP picture
is available for light at the decoupling which occurred at the decoupling
event which is believed to have occurred much later after the Big
Bang. Our approach is able to provide a unified and simultaneous explanation
for both, the flatness problem and the horizon problem.

Our explanation should be put in comparison with the inflationary
universe hypothesis which stipulated an exponential expansion in the
baby universe to help flatten out the universe regardless of its initial
state. In our approach, no new elements or assumptions are needed
to account for the flatness. Also, the universe had ample time --
from the Big Bang to the decoupling event -- to achieve flatness.

\subsection{\label{sec:Dicke prob}Resolution to Dicke's {}``runaway density
parameter'' problem or the oldness problem}

\noindent Let us first review the situation within the traditional
framework. From the traditional Friedmann equation \cite{fR1,fR2}:\begin{equation}
\frac{\dot{a}^{2}}{a^{2}}=\frac{8\pi G}{3}\rho-\frac{k\, c^{2}}{a^{2}}\label{eq:9.16}\end{equation}
and the critical density defined as $\rho_{c}=3H^{2}/(8\pi G)$, the
density parameter\begin{equation}
\Omega\triangleq\frac{\rho}{\rho_{c}}\label{eq:9.17}\end{equation}
satisfies\begin{equation}
\left(\Omega^{-1}-1\right)\rho a^{2}=-\frac{3kc^{2}}{8\pi G}\label{eq:9.18}\end{equation}
The right-hand-side is a constant, whereas $\rho a^{2}$ scales as
$a^{-1}$. To compensate the drop in $\rho a^{2}$ as $a$ grows,
$\left|\Omega^{-1}-1\right|$ must grow and $\Omega$ is driven away
from 1 if it started away from 1. This spells the instability trouble
for $\Omega$. The runaway of $\Omega$ from 1 has been known to be
rapid. In order to account for a relatively flat space at our current
era, the universe must have started out extremely flat and the value
for $\Omega$ at recombination time must have been fine-tuned to unity
at high precision. The {}``runaway'' problem posed a major challenge
to the Friedmann model.

Our approach does not suffer this problem. That is because both $\rho$
and $\rho_{c}$ scale similarly. Whereas $\rho\propto a^{-3}$, the
critical density scales as \begin{equation}
\rho_{c}\propto H^{2}=\left(\frac{2}{3t}\right)^{2}\propto a^{-3}.\label{eq:9.19}\end{equation}
As a result, $\Omega=\mbox{const}$ at all time points.

In hindsight, the reason behind the {}``runaway'' problem in the
traditional framework was the scaling mismatch in the Friedmann equation:
\begin{equation}
\frac{\dot{a}^{2}}{a^{2}}=\frac{8\pi G}{3}\rho-\frac{k\, c^{2}}{a^{2}}\label{eq:9.19-1}\end{equation}
in which the three terms scale differently. Whereas $\rho\propto a^{-3}$,
the last term scales as $a^{-2}$ and how the first term scales depends
on the shape of space. This scaling mismatch then got transferred
to Eq. \eqref{eq:9.18}, creating the dreaded instability problem.
One tentative way to fix this problem for the Friedmann equation is
to allow $c$ in the last term in \eqref{eq:9.19-1} to scale as $a^{-1/2}$,
per the scaling rule \eqref{eq:scaleC}. Note that $c$ depends on
$a$, instead of $t$. As such, the replacement $c=c_{0}a^{-1/2}$
is legitimate. In so doing, the right-hand-side of \eqref{eq:9.19-1}
scales as $a^{-3}$, and the equation subsequently yields $a\propto t^{2/3}$
in perfect agreement with the evolution rule \eqref{eq:9.6} in our
curvature-scaling gravity approach.

\subsection{\label{sub:replaceFriedmann}On the shortcomings of the Friedmann
model and its related supplementaries}

We offered a new look at cosmology in the light of curvature-scaling
gravity. The Universe evolves in a unique and robust fashion, independent
of its shape and its matter content and density. As the universe expands,
its curvature gradually drops, leading to a drop in $c$ in a precise
connection to $\mathcal{R}$ per \eqref{eq:scaleC}, $c\propto\left|\mathcal{R}\right|^{1/4}$.
With only this property and the RW metric, we are able to explain
the observational data of Type Ia supernova, the age problem, the
flatness problem, the horizon problem. The future task remains to
come up with a reasonable model for the stress-energy tensor for matter
in the universe. Any new model in this line of research based on curvature-scaling
gravity will call for a recalibration of observational data, such
as the WMAP.

Let us further comment on an important consequence of curvature-scaling
gravity: the variability of $\hbar$ as function of the universe's
Ricci scalar. In the early stage of the universe, $\mathcal{R}$ was
higher leading to smaller values for $\hbar$ according to the scaling
rule \eqref{eq:scaleH}: $\hbar\propto\left|\mathcal{R}\right|^{-1/4}$.
Quantum effects were decidedly weaker in the early universe. As a
result, any new model of cosmology will need a recalibration to observational
data, such as the WMAP data for the acoustic peaks, for example. In
another example, the decoupling event is said to have occurred at
about 380,000 years after the Big Bang. Due to the variation of $\hbar$
(which governs the coupling of matter with photons), the precise value
of this time point need recalibration %
\footnote{\label{fn:monopoles}As an aside note, absence of monopoles was first
credited as one of three motivations, the flatness problem and the
horizon problem rounding out the list, for the inflation hypothesis.
It was said that an inflation diluted the monopoles. If the field-theoretical
models for monopoles are indeed correct, the absence of monopoles
can be naturally explained in curvature-scaling gravity. The particle
physics models that predicted an abundant amount of monopoles used
the current value of $\hbar$. These models should have used smaller
value of $\hbar$.%
}.

The standard Friedmann model does not allow the variability of $c$
as a function of the cosmic scale factor $a$ (or, equivalently, the
Ricci scalar which has been falling as the universe expands.) It is
thus a problematic cosmological model to start with. The model in
its original form could not account for several observational facts
and data. It produced flawed formulae for the redshift and the Hubble
law, leading to an upward biased estimate of $H_{0}$. These problems
appear in cosmology since the cosmos is a laboratory which accumulates
the effects of non-universal $c$ over a lifespan of billions light
years. Most severe cracks show up in the age problem (via the overestimated
Hubble constant), the deviation in Type Ia supernovae, the horizon
problem, the flatness problem, and other cosmic coincidences. To resolve
these problems, standard cosmology had to resort to a series of ad
hoc solutions, compiled in the concordance \textgreek{L}CDM model.
In the light of curvature-scaling gravity, these ad hoc solutions
are unnecessary.
\begin{enumerate}
\item The lack of necessity of accelerating expansion:

As we elaborated in Sections \ref{sec:SN1a} and \ref{sec:alternative},
the data of Type Ia supernovae can be naturally explained in curvature-scaling
gravity which allows the light speed to vary with respect to the cosmic
scale factor per $c\propto a^{-1/2}$. The cosmic expansion is not
accelerating. In light of the misperceived meaning of the Michelson-Morley
experiment, the discoveries regarding Type 1a supernovae \cite{Riess1,Perlmutter}
should be seen as compelling evidence in support of the variation
of $c$, a conclusion which would profoundly deepen one's understanding
of spacetime and gravity. This is where the discoveries show their
greatest importance and impacts.

\item The resolution to the budgetary shortfall and the lack of necessity
for {}``dark energy'':

The overestimation of $H_{0}$ will help address the budgetary shortfall
problem as well. Whilst a more complete model of cosmology is required,
let us take the original Friedmann equation \eqref{eq:9.16}\[
\frac{\dot{a}^{2}}{a^{2}}=\frac{8\pi G}{3}\rho-\frac{k\, c^{2}}{a^{2}}\]
and tentatively correct it by letting $c$ depend on $a$ per $c=c_{0}\, a^{-1/2}$
with $c_{0}$ being the speed of light measure today in the outer
space (which, unlike the galaxies, is subject to cosmic expansion).
Note that $c$ does not explicitly depend on $t$; rather, $c$ adapts
to the scale factor $a$ (or, equivalently, to the Ricci scalar).
As such, the replacement $c\rightarrow c_{0}\, a^{-1/2}$ is legitimate,
yielding the following equation:\begin{equation}
\frac{\dot{a}^{2}}{a^{2}}=\frac{8\pi G}{3}\rho-\frac{k\, c_{0}^{2}}{a^{3}}\label{eq:9.20}\end{equation}
This equation adopts a unique evolution rule: $a\propto t^{2/3}$
in full agreement with \eqref{eq:9.6} in our curvature-scaling gravity
approach %
\footnote{The fact that the original Friedmann equation does admit the critical
expansion mode, $a\propto t^{2/3}$ and yields the correct age formula,
$t=2/(3H_{0})$ for $k=0$ compatible with the generic scaling rule
indicates that the Friedmann model does carry some element of truth
and should continue to be valuable upon future revisions.%
}. With the WMAP confirming a flat space, $k=0$, Eq. \eqref{eq:9.20}
admits a solution\begin{equation}
a=\left(6\pi G\rho_{c}\right)^{1/3}t^{2/3}\label{eq:9.20-1}\end{equation}
in agreement with \eqref{eq:9.6} in our curvature-scaling gravity
approach. Combined \eqref{eq:9.8} and \eqref{eq:9.20-1}, the current
critical density satisfies\begin{equation}
6\pi G\rho_{c}=\left(\frac{3}{2}H_{0}\right)^{2},\label{eq:9.20-2}\end{equation}
thus remaining to be\begin{equation}
\rho_{c}=\frac{3}{8\pi G}\, H_{0}^{2}.\label{eq:9.21}\end{equation}
Recall that $H_{0}$ has been overestimated, however. With the value
of $H_{0}$ being corrected to $37.4$ instead of $70.5$, the critical
density is reduced to $\left(\frac{37.4}{70.5}\right)^{2}\approx28\%$
of the accepted value in standard cosmology. Interestingly, this reduced
value of $\rho_{c}$ almost precisely matches the amount of baryonic
matter and dark matter in the Universe. The missing gap for the budget
thus disappears; the hypothetical {}``dark energy'' invented to
make up the shortfall is not needed. The budgetary shortfall has often
been cited as solid evidence in support of {}``dark energy''. This
is no longer the case. The overestimated $H_{0}$ was precisely the
culprit for the shortfall in the matter budget %
\footnote{Also note that although the $28\%$ leftover is said to be composed
of baryonic matter and dark matter, the dark matter component is not
necessarily some hypothetical non-luminous form of matter, such as
WIMPs. As pointed out in Section \ref{sec:4}, the linear Mannheim-Kazanas
term that is responsible for galactic rotation curves will continue
to be a surrogate for the {}``dark matter'' content for the Universe
at large. Indeed, in Mannheim's theory \cite{Mannheim1,MOB1,MOB2,MOB3},
the parameters $\gamma^{*}$ and $\gamma_{0}$ are found to acquire
universal values which act as a centripetal force toward the ordinary
mass source. We could thus stipulate that luminous matter would be
the only source of gravity, with the Mannheim-Kazanas potential playing
a surrogate to the {}``dark matter'' sector, and {}``dark energy''
being eliminated.%
}.

The cosmological constant problem and the coincidence of dark energy
density of $0.72$ at our current epoch are handily avoided.

\item The lack of necessity of inflationary expansion:

The Friedmann model cannot account for other observations -- the uniformity
of observable horizon and the flatness of space. The model furthermore
suffers from its own theoretical problem -- the oldness problem stemming
from Dicke's instability, the extraordinarily fine-tuned initial state
so that it could evolve and survive until today. In light of curvature-scaling
gravity, all these problems are non-existent. They only appeared to
be problems if they were viewed from the basis of the Friedmann model.

To reconcile these observations with the Friedmann model, the inflationary
universe hypothesis, which posited an exponential growth for the baby
universe, was proposed in the early eighties \cite{Guth} and later
became integrated into standard cosmology. There has not been any
direct evidence of inflation, to the best of our knowledge. It is,
though, often said that the WMAP data are hallmarks of inflation.
This is not necessarily so. Most of the WMAP results are generic features
of a Big Bang theory.

The inflation hypothesis introduced a new set of fine tunings and
problems on its own, the most notable of which are the slow-roll condition,
the graceful exit, and the nature of the cosmological constant\cite{Steinhardt}.
It was also particularly designed to address questions regarding the
very early stage of the universe and has little to say about the subsequent
development of the universe until our current epoch and about the
future of the universe.

\end{enumerate}
\noindent In summary, upon the reinterpretation in the light of curvature-scaling
gravity that the value of $c$ is not universal but rather a function
of the Ricci scalar as the universe expands, the issues once considered
fatal to the Friedmann model disappear. Any cosmological model to
be devised in the future must allow $c$-variability. In this direction,
curvature-scaling gravity already provides the guideline for any revision
to come: the evolution of the cosmic factor obeys the scaling rule
\eqref{eq:scaleC}. The conclusions from curvature-scaling gravity
presented in this section and the preceding one are model-independent;
they must unequivocally hold for any theory of cosmology.

\newpage{}

\section{\label{sec:10}Conclusions and outlook of curvature-scaling gravity
in quantum gravity}

Einstein's general theory of relativity was the first methodical attempt
to extrapolate the physics in man's living quarter to the entire universe.
The theory seeks guidance in the equivalence principle and the general
covariance principle that physical laws (including special relativity)
are valid in the tangent frames local to each point on the manifold.
Implicit in general relativity, though, is the assumption of a universal
absolute length scale against which all processes and phenomena --
gravitational and non-gravitational alike -- are measured. Although
general relativity offers one concrete construction of the spacetime
manifold, we provide an alternative admissible construction of the
manifold which satisfies all of Einstein's requirements with regard
to relativity and causality.

To do so, we extend Einstein's guidance further: not only are physical
laws locally valid in the tangent frames, the length scale for them
is valid only locally. We inherit fully Einstein's vision of the Riemannian
geometrical structure of the spacetime manifold and all other principles
that he establishes (the Lorentz symmetry, the relativity principle,
and the general covariance principle), whereas we further assign the
Ricci scalar $\mathcal{R}$ a new eminent role: in each local region
on the manifold, the Ricci length, defined as $a_{\mathcal{R}}\triangleq\left|\mathcal{R}\right|^{-1/2}$,
is the intrinsic length which all other lengths -- including the Bohr
radius of quantum mechanics, e.g. -- are pegged onto. Intuitively
speaking, the Ricci scalar determines the size of every object in
a local region. Mathematically speaking, the action for any physical
process is to be built from the dimensionless ratio of length and
the invariant Ricci length $a_{\mathcal{R}}$.

The spacetime manifold is thus a patchwork of local regions, obeying
two requirements (or postulates):
\begin{lyxlist}{00.00.0000}
\item [{(I)}] Special relativity and all physical laws (of non-gravitational
origin) be satisfied in each local region;
\item [{(II)}] Each region locally accept the intrinsic Ricci length as
the fundamental length scale for all processes that take place within
the region.
\end{lyxlist}
Requirement (I) is nothing but a reinforcement of the equivalence
principle. Requirement (II) is the only new feature that sets our
approach apart from Einstein's theory. From the two requirements,
we were able to construct the spacetime manifold which respects causality
globally, and obtain the Lagrangian of gravitational field coupled
with matter.

In our approach, curvature is thereby promoted: not only does it determine
the geometric structure of the underlying manifold, it is actively
involved in the dynamics of physical processes by setting the scale
for them. The intuitive rationale is that spacetime should provide
the scale (in this case, $a_{\mathcal{R}}$) for physical processes,
instead of the other way around. The ramification of this initiative
of ours is a series of conceptual departures:
\begin{enumerate}
\item The fundamental length scale $a_{\mathcal{R}}$ for physical processes
is itself a dynamical variable governed by the equations of the metric
$g_{\mu\nu}$. As such, the gravitational field is deeply involved
in the physical processes.
\item A new recipe to obtain the Lagrangian of gravity coupled with matter,
away from the standard minimal coupling procedure. The action is to
be constructed from dimensionless ratios of lengths using the Ricci
length as the common denominator. Our recipe entails the following
replacements: $dx^{\mu}\rightarrow dx^{\mu}/a_{\mathcal{R}}=\left|\mathcal{R}\right|^{1/2}dx^{\mu};\ \nabla_{\mu}\rightarrow a_{\mathcal{R}}\nabla_{\mu}=\left|\mathcal{R}\right|^{-1/2}\nabla_{\mu}$.
As such, the volume element is replaced as $d^{4}x\,\sqrt{-\det g}\rightarrow d^{4}x\,\mathcal{R}^{2}\,\sqrt{-\det g}$.
Through our recipe, the gravitational field arises organically from
the matter fields. (See Section \ref{sec:7}.)
\item Anisotropy in the scaling of time: with Requirement (I) that physical
laws retain their forms -- but not necessarily their parameters --
in every local region on the manifold, and Requirement (II) that the
physical laws adapt to the prevailing Ricci length, the scaling of
time duration is found (via the Schrödinger equation or the action
of QED) to be anisotropic with respect to the Ricci length: $dt\propto a_{\mathcal{R}}^{3/2}$
as compared to that of space, $dx\propto a_{\mathcal{R}}$. (See Section
\ref{sub:Tannisotropy}.)
\item \label{enu:C}We corrected a deep misperception that the Michelson-Morley
finding necessarily meant a universal value of light speed everywhere.
All Michelson-Morley's finding establishes is that the speed of light
at each given location is the same regardless of the direction of
the light beam and/or the motion of the observer. It has nothing to
say about the equality (or the lack thereof) in the value of $c$
at different locations. By virtue of the equivalence principle which
is local in nature, Lorentz invariance, the Michelson-Morley result,
and the relativity principle are also local in nature. The value of
$c$ is meant to be local; that is to say, $c$ acquires a new value
for each location on the manifold. In curvature-scaling gravity, $c=dx/dt\propto a_{\mathcal{R}}^{-1/2}=\left|\mathcal{R}\right|^{1/4}$
due to the anisotropic time scaling $dt\propto a_{\mathcal{R}}^{3/2}$
(whereas $dx\propto a_{\mathcal{R}}$) mentioned above. (See Section
\ref{sub:Cvariable}.) Two important points to note:

\begin{enumerate}
\item This result strictly protects causality: in each local region, $c$
is the maximum speed for all objects in the region. No objects can
surpass light at any point on the manifold. Superluminosity is strictly
forbidden. Furthermore, in each region, the Lorentz symmetry holds,
with null-geodesics strictly separating timelike trajectories from
spacelike trajectories. Causality is protected both locally and globally.
\item The speed of light is not explicitly a function of spacetime coordinates;
rather, it is intrinsically a function of the prevailing Ricci scalar.
As such, it is permissible for the Lorentz symmetry to hold locally
around every point on the manifold.
\end{enumerate}
\item We corrected another deep misperception that the all-embracing utility
of $\hbar$ and $c$ necessarily meant a universality in their value
at all locations. Requirement (I) that physical laws retain their
forms in different spacetime pocket forces their parameters -- viz.
$\hbar$ and $c$ -- to be functions of the Ricci scalar %
\footnote{For our comment regarding a line of favorite yet false critique, see
footnote \vref{fn:Ellis}.%
}. In each local region, $\hbar$ and $c$ continue to govern over
the well-established physics that take place within the region: $\hbar$
as the fundamental parameter measuring the strength of quantum effects
(such as the $SU(3)\times SU(2)\times U(1)$ model, the phonon and
specific heat in solids, the quantum Hall effect, or the nuclear shell
model) and $c$ as the fundamental parameter in the Lorentz symmetry
(being the common value of light speed regardless of the direction
of the light beam and/or observer.) Yet these parameters are pegged
onto the Ricci scalar $\mathcal{R}$ and thus adapt to the prevailing
value of $\mathcal{R}$ from one region to the next on the manifold,
per the scaling rules \eqref{eq:scaleH} and \eqref{eq:scaleC}: $\hbar\propto\left|\mathcal{R}\right|^{-1/4},\ c\propto\left|\mathcal{R}\right|^{1/4}$,
due to the anisotropic time scaling. The all-embracing principles
-- the causality principle, the relativity principle, Lorentz invariance,
the Michelson-Morley finding, the equivalence principle, and the general
covariance principle -- are fully respected. (See Section \ref{sub:Hvariable}.)
\end{enumerate}
The equivalence principle localizes physical laws (viz. special relativity
and quantum laws) to be valid in each pocket of the spacetime manifold.
The equivalence principle, in our approach, further localizes the
parameters of physical laws to be valid in each pocket only, with
the prevailing value of the Ricci scalar directly appoints their values.
The parameters that depend on the Ricci scalar are the Planck constant,
the speed of light, the length scale and the oscillatory rate of physical
processes.

The objectives of our theory are a new construction of the spacetime
manifold and the Lagrangian of gravity coupled with matter. We applied
the theory to a variety of problems regarding the foundation of relativity
and gravity, as well as astrophysics and cosmology:
\begin{itemize}
\item We addressed the misplaced fear of violation of causality and Michelson-Morley
experimental result. Being local Lorentz invariant, null geodesics
remain null geodesics in every coordinate system. Timelike paths and
spacelike paths do not mix. At any given point, the speed of light
is the upper limit for all objects. Superluminosity is strictly forbidden;
as such, causality is strictly preserved. (See Section \ref{sub:Cvariable}.)
\item In using only dimensionless ratios of lengths with the Ricci length
as denominator, gravity arises from matter in an organic, natural,
and unique manner. There is only one unified term for the action,
viz. $\mathcal{S}_{CSG}=\int d^{4}x\,\mathcal{R}^{2}\,\sqrt{g}\:\mathcal{L}_{m}$
which incorporates both matter and gravitation field. There is no
{}``free'' gravitation field that lives on its own, detached from
matter. Also, the cosmological constant is absent in $\mathcal{S}_{CSG}$.
(See Section \ref{sec:7}.)
\item In vacuo, our action $\mathcal{S}_{CSG}$ above reduces to $\mathcal{S}_{vacuo}=\int d^{4}x\,\sqrt{g}\,\mathcal{R}^{2}$
in resemblance of the $\mathcal{R}^{2}$ theory. It is a well-posed
Cauchy problem with the Cauchy data conveniently consisting of $g_{\mu\nu},\ \mathcal{R}$
and their first-order time-derivatives $\partial_{0}g_{\mu\nu},\ \partial_{0}\mathcal{R}$.
(See Section \ref{sec:3}.)
\item The action $\mathcal{S}_{vacuo}$ appears to be tractable for static
spherically symmetric setup. We provided an explicit solution and
established its connection with a solution that Mannheim-Kazanas found
in conformal gravity. Their solution and ours possess a new term,
corresponding to a linear gravitational potential $\gamma r$ which
acts on the larger scale beyond the solar system. Based on this potential
term, Mannheim has endeavored a phenomenological theory to explain
the galactic rotation curves without resorting to dark matter. Curvature-scaling
gravity thus potentially inherits and strengthens Mannheim's theory.
(See Section \ref{sec:4}.)
\item We then provided another static spherically symmetric solution which
possesses a non-constant Ricci scalar. Close to a mass source, the
Ricci scalar diverges at its event horizon, leading to an unbounded
growth of $c$ and a diminishment of $\hbar$ as one approaches the
event horizon. This is a prediction from our theory. The diminishment
of $\hbar$ and quantum effects at the event horizon could alter the
radiative behavior of Schwarzschild-type black holes. (See Section
\ref{sec:6}.)
\item For cosmology, we showed that the hypothetical dark energy is nothing
but the effects of the adaptation of light speed to the varying Ricci
scalar (and, equivalently, to the cosmic scale factor $a$.) The standard
Friedmann model neglects this important feature, thereby producing
flawed redshift formulae and resulting in theoretical predictions
irreconcilable with observational data. As a result, standard cosmology
had to invent ad hoc solutions -- dark energy, inflationary expansion,
and accelerating expansion -- to reconcile the data with the Friedmann
model. All the perceived problems and difficulties in cosmology are
non-existent, however; they arose from the deep misperception that
$c$ needs be universal on the spacetime manifold (see Point \vref{enu:C}).
By virtue of the equivalence principle, the Lorentz symmetry and its
parameter -- viz. the speed of light $c$ -- only need be valid locally
in each pocket of spacetime.

Starting only from the provision -- derivable solely from Requirements
(I) and (II) -- that $c$ adapt to the cosmic scale factor per $c\propto a^{-1/2}$
\eqref{eq:scaleC}, we provided a unified solution to all 7 most pressing
problems simultaneously: (i) the new interpretation of Type 1a supernovae,
(ii) the age problem, (iii) the horizon problem, (iv) the flatness
problem, (v) Dicke's instability problem (or the oldness problem),
(vi) the budgetary shortfall problem, and (vii) the cosmological constant
problem. Our parsimonious solution eliminates the need for the fudge
agent of dark energy and the artificial mechanisms of accelerating
expansion and inflationary expansion. We corrected the upward bias
in the estimation of the Hubble constant (with the corrected value
being $H_{0}\approx37.4$) which helps revise the universe's age to
17.4 Glys and reduce the matter budget to 28\% of what previously
thought. Our solution avoids all fine-tunings and is model-independent.
The cosmological constant and all other cosmic coincidences are avoided.
(See Sections \ref{sec:8}, \ref{sec:9}.)

With the misperceived role of causality and the Michelson-Morley experiment
corrected, the data of Type Ia supernovae should be interpreted as
concrete evidence in support of the variation of $c$ as function
of the cosmic scale factor, rather than an accelerating cosmic expansion.
Also, upon the reinterpretation of observational cosmology in the
light of curvature-scaling gravity (which supports the notion of $c$-variability),
the needs for dark energy and inflationary expansion disappear.

\end{itemize}
\noindent Regarding the quantization of gravity, there are desirable
properties of $\mathcal{R}^{2}$ gravity \cite{Sotiriou1} which could
stay valid for curvature-scaling gravity as well, at least in vacuo,
in which case curvature-scaling gravity coincides with $\mathcal{R}^{2}$
gravity:
\begin{itemize}
\item Renormalizability of $\mathcal{R}^{2}$ gravity \cite{Stelle}. One
nice feature of a renormalizable Lagrangian is that the Lagrangian
retains its original form upon the renormalization procedure.
\item As a member of the $f\left(\mathcal{R}\right)$ class of theories,
it is free of ghosts. Thus unitarity is respected \cite{ghostFree1,ghostFree2}.
\item Regarding Odstrogradsky's instability, in \cite{Woodard} Woodard
shows how the $f\left(\mathcal{R}\right)$ theories, being degenerate,
manage to avoid this fatal instability.
\item $\mathcal{R}^{2}$ gravity avoids Dolgov-Kawasaki's instability \cite{DKinstability}
since $f''\left(\mathcal{R}\right)=2>0$.
\item $\mathcal{R}^{2}$ gravity has a well-posed Cauchy problem \cite{Noakes,Teyssandier}.
The Cauchy data of $\mathcal{R}^{2}$ gravity consist of $g_{\mu\nu},\ \mathcal{R}$
and their first-order time-derivatives $\partial_{0}g_{\mu\nu},\ \partial_{0}\mathcal{R}$.
\end{itemize}
We must however note that in curvature-scaling gravity, the (fixed)
Planck scales -- Planck energy, Planck length, Planck time -- lose
their meaning. Since length scale is dependent on the Ricci scalar
($\hbar$ and $c$ being dependent on $\mathcal{R}$ at the location
they are active), the concept of a fixed Planck length is no longer
meaningful. Likewise, Planck energy and Planck length also lose their
meaning. Rather, these quantities are defined locally at each point
on the manifold and they are allowed to vary on the manifold. This
is another conceptual departure from the standard quantization paradigm.
In addition, it is a foremost task to address the conceptual question
-- which this author leaves unanswered -- if the curvature-scaling
gravity theory is to be the correct description of gravity at classical
level:\emph{ }{}``What is the actual meaning behind the quantization
of spacetime given that $\hbar$ is curvature-dependent?'' %
\footnote{An interesting excursion by Mannheim (see Section 4 of \cite{Mannheim2})
posits a quantization of gravity though its coupling with the quantized
matter fields. His thoughts could be of relevance here since in our
approach gravity directly arises from matter fields via the replacement
of lengths with their dimensionless ratios denominated by the Ricci
length. As soon as the matter fields are quantized, would this replacement
procedure automatically bring forth a quantized gravitational field
too?%
}

In summary, our theory is an alternative construction of the spacetime
manifold: the manifold is a patchwork of local pockets of spacetime,
each satisfying special relativity and adopting a local scale. The
fundamental constants $\hbar$ and $c$ continue to control the physics
in each individual local region, yet they must adapt to the prevailing
value of the Ricci scalar. (Note that they are not auxiliary fields
that live on the manifold.) We preserve all of Einstein's vision and
insights in relativity and gravity while at the same time deepening
the role of curvature. The Ricci scalar actively partakes in the dynamics
of physical processes (by setting the scale for them), thus entailing
a small step closer toward a Machian spirit which posits that physics
-- in this case, the parameters $\hbar$ and $c$ -- at each point
is determined by the distribution of matter in the universe as a whole.

Lastly, our theory is primarily a theory of spacetime and gravitation.
It is: (a) Not a theory of time anisotropy %
\footnote{In $2+1$ dimensions, time and space would scale similarly, $dt\propto dx\propto a_{\mathcal{R}}$.
Yet our Requirements (I) and (II) mentioned in the preceding page
would continue to hold. (As an aside note, due to the said isotropy,
$c$ and $\hbar$ would be constant everywhere in $2+1$ dimensions.)%
} or $c$-variation per se %
\footnote{The variability of $c$ and $\hbar$ is not an input but instead a
by-product of the theory, logically resulting from the dependence
of local scale on the Ricci curvature. It is of secondary importance
and is a logical consequence of the promotion of the Ricci scalar
as the scale-setter for physical processes. In our approach, the variability
of $c$ and $\hbar$ arises via their endogenous dependency on the
Ricci scalar. These parameters do not have a dynamics on their own
merit, apart from the dynamics of $\mathcal{R}$. Our theory employs
no terms such as $\partial_{t}c$ or $\vec{\nabla}\hbar$ which would
otherwise inflict structural damages to existing physical laws --
a major challenge that plagued the efforts of \cite{Moffat,Magueijo1}
to build a workable account for variable light speed. For example,
in those attempts, $c$ was allowed to acquire a dynamics governed
by some additional quantum mechanical field. It is neither warranted
nor necessary to do so in our approach.

We adopt the relativist's view: it is not necessary to design a mechanistic
model to account for $c$- and $\hbar$-variabilities. Here is an
apt analogy: prior to the advance of special relativity, the Lorentz-Fitzgerald
length contraction and the Lorentz transformation had been devised
to account for the Michelson-Morley experiment. Yet Lorentz attributed
the length contraction to some yet-to-discover mechanism, such as
the interaction between the electrons in the moving ruler and the
moving clock with an ether. In contrast, Einstein considered these
effects a matter of principle -- the relativity principle and the
postulate of constant $c$ -- both of which together dictate the form
of the Lorentz transformation (and of the length contraction). Indeed
Einstein insisted on using the term Relativitätsprinzip (i.e., a descriptive
approach) before finally adopting Relativitätstheorie (i.e., an explanatory
approach), a term coined by Planck (e.g., see \cite{Ohanian}).\emph{
}We follow Einstein's pioneer footstep in this regard.%
}; (b) Not a theory of cosmology alone. It was not devised to solely
resolve problems in cosmology but to address foundational issues of
gravitation; (c) Not a theory of $\mathcal{R}^{2}$ gravity, although
its Lagrangian in vacuo shares a resemblance to the that of $\mathcal{R}^{2}$
gravity (the latter not enabling the variations of the local scale
and of $c$ and $\hbar$); (d) Not related to Brans-Dicke or dilaton
theories.

Considering the wide-ranging natural outcomes that our new construction
of spacetime achieves based on such parsimonious Requirements (I)
and (II), we believe it represents an important measure toward the
extrapolation of our solar system wisdom into the realm of universe
as a whole and the interface with the quantum rules.

Although any scientific endeavor will ultimately be judged by its
success or failure to account for physical reality, our theory of
curvature-scaling gravity was decidedly originated from a philosophical
inspiration: {}``What sets the size for things around us?'' %
\footnote{To the curious-minded reader who, together with the Earth, are in
a free fall toward the Sun, the Moon, and all other heavenly bodies,
it is tantalizing to inquire what otherwise sets the size of his desk,
his office, his own body, as well as the rate of his wristwatch and
his own heartbeats. Conventional wisdom would have the omnipotent
Planck constant as the scale-setter since quantum mechanics should
decide the dimension and vibration rate of the atoms that constitute
physical things. We propose an alternative view: it is spacetime itself
-- in particular, the Ricci scalar -- that appoints the value of the
Planck constant which in turn sets the size and rate for physical
things via quantum mechanics.%
}. We were motivated to assign the scale-setter role to the scalar
curvature. In so doing, we strengthen the position of the curvature
within Einstein's grand structure of space and time while preserving
intact the inner beauty of his insights. Our undertaking is not unlike
the conceptual path Einstein adopted in the early stage of his development
of general relativity; he was initially guided by the equivalence
principle and, to a certain extent, Mach's principle. These threads
of philosophical guidance inspired him to take a leap of faith --
that gravity is a manifestation of wrapped spacetime -- to arrive
at the general covariance principle -- that physical laws be expressed
in tensorial forms. Only in the later years starting 1912-13 did he
-- in a frantic race against Hilbert, Einstein the Purist ever converting
into Einstein the Pragmatist -- abandon the philosophy-inspiring route
and adopt a practical approach toward his final gravitational field
equations which were mostly aimed at recovering Newtonian gravity
in the weak-field limit.

Philosophical enquiries can have encompassing power and scope; their
ramification can be far-reaching. It was remarkable -- to this author
when a young student -- that Einstein's imagination of a hapless falling
man eventually offered an explanation for Mercury's perihelion precession.
Throughout this report, we continue the conceptual route that Einstein
the Purist departed some time in 1912-13, extend his all-embracing
philosophical insights -- by ascribing the curvature a new privileged
status -- and hope to advance his quest for a better understanding
of space and time and gravity.

\section*{Acknowledgments}

I first and foremost wish to thank Travis W. Fisher for his critical
insights proved valuable in the progress of this work. Philip Mannheim
helpfully clarified important aspects of his theory of galactic rotation
curves. I also like to thank Richard Shurtleff, Soebur Razzaque, Alexandr
Yelnikov, V. Parameswaran Nair, Tuan A. Tran, Andrei Pokotilov, Antonio
Alfonso-Faus for their constructive feedbacks. Ngoc-Khanh Tran was
of crucial help in providing reference material for this work.\newpage{}

\appendix

\section{\label{sec:A}Derivation of the scaling rule for time duration}

Consider the Schrödinger equation of the Hydrogen atom in $d+1$ dimensions:
\begin{equation}
i\hbar\frac{\partial}{\partial t}\Psi=-\frac{\hbar^{2}}{2m_{e}}\nabla^{2}\Psi-\frac{e^{2}}{r^{d-2}}\Psi\label{eq:a.1}\end{equation}
Expressing the coordinate differentials in term of their dimensionless:\begin{equation}
\begin{cases}
\ d\vec{x} & =\ a_{\mathcal{R}}\,\vec{d\tilde{x}}\\
\ dt & =\ a_{\mathcal{R}}^{\eta}\, d\tilde{t}\end{cases}\label{eq:a.2}\end{equation}
in which $dt$ could acquire a different scale factor than $dx$ does
(hence, the exponent $\eta$). The Schrödinger equation becomes:\begin{equation}
i\frac{\hbar}{a_{\mathcal{R}}^{\eta}}\frac{\partial}{\partial\tilde{t}}\Psi=-\frac{\hbar^{2}}{2m_{e}a_{\mathcal{R}}^{2}}\tilde{\nabla}^{2}\Psi-\frac{e^{2}}{a_{\mathcal{R}}^{d-2}\tilde{r}^{d-2}}\Psi\label{eq:a.3}\end{equation}
or\begin{equation}
i\frac{\hbar}{a_{\mathcal{R}}^{\eta+2-d}}\frac{\partial}{\partial\tilde{t}}\Psi=-\frac{\hbar^{2}}{2m_{e}a_{\mathcal{R}}^{4-d}}\tilde{\nabla}^{2}\Psi-\frac{e^{2}}{\tilde{r}^{d-2}}\Psi\label{eq:a.4}\end{equation}
which requires that\begin{equation}
\begin{cases}
\ \hbar^{2} & \propto\ a_{\mathcal{R}}^{4-d}\\
\ \hbar & \propto\ a_{\mathcal{R}}^{\eta+2-d}\end{cases}\label{eq:a.5}\end{equation}
and thus\[
\eta=\frac{d}{2}.\]
In $3+1$ dimensions, \[
\eta=\frac{3}{2},\]
the anisotropy in scaling of time and space thus emerges. The anisotropy
essentially means that whilst the observer's ruler scales linearly
with the Ricci length $a_{\mathcal{R}}$, the vibrating rate of his
clocks scale as $a_{\mathcal{R}}^{3/2}$. In $2+1$ dimensions, time
duration and length are isotropic in their scaling behaviors.

\subsubsection*{Verification of the scaling rules for QED:}

In $3+1$ dimensions, the scaling rules for physical quantities and
constants are:\begin{equation}
\begin{cases}
\ d\vec{x} & \rightarrow\ a_{\mathcal{R}}\, d\vec{x}\\
\ dt & \rightarrow\ a_{\mathcal{R}}^{3/2}\, dt\end{cases}\label{eq:a.7}\end{equation}
and\begin{equation}
\begin{cases}
\ \vec{p} & \rightarrow\ a_{\mathcal{R}}^{-1/2}\,\vec{p}\\
\ E & \rightarrow\ a_{\mathcal{R}}^{-1}\, E\end{cases}\label{eq:a.8}\end{equation}
whereas\begin{equation}
\begin{cases}
\ \hbar & \rightarrow\ a_{\mathcal{R}}^{1/2}\,\hbar\\
\ c & \rightarrow\ a_{\mathcal{R}}^{-1/2}\, c\\
\ m & \rightarrow\ m\\
\ e & \rightarrow\ e\\
\ \alpha & \rightarrow\ \alpha\\
\ G & \rightarrow\ G\end{cases}\label{eq:a.9}\end{equation}
with $\vec{x},\, t,\,\vec{p},\, E,\,\hbar,\, c,\, m,\, e,\,\alpha,\, G$
being coordinates, time, momentum, energy, the Planck constant, the
speed of light, the electron mass, the elementary electric charge,
the fine coupling constant, and the gravitational constant respectively.
In addition, the electromagnetic potential $A^{\mu}=(\varphi,\,\vec{A})$
and fields $\vec{E}=-\nabla\varphi-\frac{1}{c}\partial_{t}\vec{A},\ \vec{B}=\nabla\times\vec{A}$
and fermion field $\psi$ transform as:\begin{equation}
\begin{cases}
\ \varphi & \rightarrow\ a_{\mathcal{R}}^{-1}\,\varphi\\
\ \vec{A} & \rightarrow\ a_{\mathcal{R}}^{-1}\,\vec{A}\\
\ \vec{E} & \rightarrow\ a_{\mathcal{R}}^{-2}\,\vec{E}\\
\ \vec{B} & \rightarrow\ a_{\mathcal{R}}^{-2}\,\vec{B}\\
\ \psi & \rightarrow\ a_{\mathcal{R}}^{-3/2}\,\psi\end{cases}\label{eq:a.10}\end{equation}
It is straightforward to verify that the above scaling rules leave
the action of QED (with $x^{\mu}\triangleq(c\, t,\ \vec{x})$):\begin{eqnarray}
\mathcal{S} & = & \int d^{4}x\,\mathcal{L}\nonumber \\
\mathcal{L} & = & \psi^{\dagger}\left(i\hbar\frac{\partial}{\partial t}+ic\vec{\alpha}.\hbar\vec{\nabla}-mc^{2}\beta\right)\psi+e\psi^{\dagger}\left(-\varphi+\vec{\alpha}.\vec{A}\right)\psi-\frac{1}{4}F_{\mu\nu}F^{\mu\nu}\label{eq:a.11}\\
F_{\mu\nu} & = & \partial_{\mu}A_{\nu}-\partial_{\nu}A_{\mu}\nonumber \end{eqnarray}
unchanged since $\mathcal{L}\propto a_{\mathcal{R}}^{-4}$ self-evidently.
As such, the (relativistic) Dirac equation and Maxwell equations are
unchanged.

\subsubsection*{The scaling of energy:}

The scaling rules also transform all forms of energy in the same fashion,
$E\propto a_{\mathcal{R}}^{-1}$. For example: 
\begin{itemize}
\item The electromagnetic field energy $\frac{1}{2}\int d\vec{x}(\vec{E}^{2}+\vec{B}^{2})\propto a_{\mathcal{R}}^{3}(a_{\mathcal{R}}^{-2})^{2}=a_{\mathcal{R}}^{-1}$
since $\vec{E}\propto a_{\mathcal{R}}^{-2}$ and $\vec{B}\propto a_{\mathcal{R}}^{-2}$. 
\item Photon energy $E=\hbar\nu\propto a_{\mathcal{R}}^{1/2}a_{\mathcal{R}}^{-3/2}=a_{\mathcal{R}}^{-1}$. 
\item Energy of a massive object: $E=mc^{2}\propto a_{\mathcal{R}}^{-1}$. 
\item Energy levels of the hydrogen atom $E_{jn}=-\frac{m_{e}e^{4}}{2\hbar^{2}n^{2}}\left[1+\frac{\alpha^{2}}{n^{2}}\left(\frac{n}{j+\frac{1}{2}}-\frac{3}{4}\right)\right]\propto a_{\mathcal{R}}^{-1}$. 
\end{itemize}
The requirement that all forms of energy obey a universal scaling
rule is important. Let us stress that our consideration is unconventional
in contrast to the standard cosmological paradigm which discriminates
radiation from matter and the {}``Doppler theft'' is said to only
affect photons but leave nonrelativistic matter untouched. In our
theory, all forms of energy obey a common scaling rule, $E\propto a_{\mathcal{R}}^{-1}$.

\section{\label{sec:B}Buchdahl's treatment of $\mathcal{R}^{2}$ gravity
revisited}

In this section, we shall revisit Buchdahl's formulation of $\mathcal{R}^{2}$-field
equation in vacuo. In so doing, we simplify some parts of his derivation
and correct a number of typos in his paper. Toward to end is a new
representation of his results. We also illustrate the $4$ degrees
of freedom for spherical coordinate. Following Buchdahl's notation,
the metric in spherical coordinate is written in the form: \begin{eqnarray}
ds^{2} & = & -e^{\nu}(dx^{0})^{2}+e^{\lambda}dr^{2}+e^{\mu}d\Omega^{2}\label{eq:b.1}\\
d\Omega^{2} & = & d\theta^{2}+\sin^{2}\theta d\varphi^{2}\nonumber \end{eqnarray}
The relevant components of the Ricci and metric tensors and Christoffel
symbols are:\begin{equation}
\begin{cases}
\ \ \ \mathcal{R}_{tt}e^{\lambda-\nu} & =\frac{\nu''}{2}+\frac{\nu'^{2}}{4}-\frac{\nu'\lambda'}{4}+\frac{\nu'\mu'}{2}\\
\ -\mathcal{R}_{\theta\theta}e^{\lambda-\mu} & =-e^{\lambda-\mu}+\frac{\mu''}{2}+\frac{\mu'^{2}}{2}+\frac{\nu'\mu'}{4}-\frac{\lambda'\mu'}{4}\\
\ -\mathcal{R}_{rr} & =\frac{\nu''}{2}+\frac{\nu'^{2}}{4}+\mu''+\frac{\mu'^{2}}{2}-\frac{\nu'\lambda'}{4}-\frac{\lambda'\mu'}{2}\end{cases}\label{eq:b.2}\end{equation}
\begin{equation}
\begin{cases}
\ g_{tt}e^{-\nu} & =-1\\
\ g_{\theta\theta}e^{-\mu} & =1\\
\ g_{rr}e^{-\lambda} & =1\end{cases}\label{eq:b.3}\end{equation}
\begin{equation}
\begin{cases}
\ \Gamma_{tt}^{r}e^{\lambda-\nu} & =\frac{\nu'}{2}\\
\ \Gamma_{\theta\theta}^{r}e^{\lambda-\mu} & =-\frac{\mu'}{2}\\
\ \Gamma_{rr}^{r} & =\frac{\lambda'}{2}\end{cases}\label{eq:b.4}\end{equation}
\begin{equation}
-\mathcal{R}e^{\lambda}=-2e^{\lambda-\mu}+\nu''+\frac{\nu'^{2}}{2}+2\mu''+\frac{3\mu'^{2}}{2}-\frac{\nu'\lambda'}{2}+\nu'\mu'-\lambda'\mu'\label{eq:b.5}\end{equation}
Furthermore, the Jacobian:\begin{eqnarray}
\sqrt{g} & \triangleq & \sqrt{-\det g}=e^{\frac{\nu}{2}+\frac{\lambda}{2}+\mu}\sin^{2}\theta\label{eq:b.6}\\
\sqrt{g}\, g^{rr} & = & e^{\frac{\nu}{2}-\frac{\lambda}{2}+\mu}\sin^{2}\theta\label{eq:b.7}\end{eqnarray}
The three functions $\nu,\ \lambda,\ \mu$ are subject to an arbitrary
coordinate transform. Buchdahl chooses that:\begin{equation}
\mu=\frac{1}{2}\left(\lambda-\nu\right)\label{eq:b.8}\end{equation}
so that $\sqrt{g}\, g^{rr}=\sin^{2}\theta$. As such, the equation
\begin{equation}
\square\mathcal{R}=0\label{eq:b.9}\end{equation}
 or its equivalent form \begin{equation}
\left(\sqrt{g}\, g^{rr}\,\mathcal{R}'\right)'=0\label{eq:b.10}\end{equation}
 is simplified to\begin{equation}
\mathcal{R}''=0\label{eq:b.11}\end{equation}
 or\begin{equation}
\mathcal{R}=\Lambda+kr\label{eq:b.12}\end{equation}
where $\Lambda$ and $k$ are constants of integration. Note that,
in Buchdahl's coordinate, the limit of spatial infinity corresponds
to $r\rightarrow0$. Thus $\Lambda$ corresponds to the large-distance
curvature (de Sitter parameter). 

With Buchdahl's choice, the relevant components become:\begin{equation}
\begin{cases}
\ \mathcal{R}_{tt} & =\frac{\nu''}{2}e^{\nu-\lambda}\\
\ \mathcal{R}_{\theta\theta} & =1+e^{-\frac{\nu}{2}-\frac{\lambda}{2}}\left(\frac{\nu''}{4}-\frac{\lambda''}{4}\right)\\
\ \mathcal{R}_{rr} & =-\frac{\lambda''}{2}+\frac{\lambda'^{2}}{8}-\frac{3\nu'^{2}}{8}+\frac{\nu'\lambda'}{4}\\
\ \mathcal{R} & =2e^{\frac{\nu}{2}-\frac{\lambda}{2}}-e^{-\lambda}\left(\lambda''-\frac{\lambda'^{2}}{8}-\frac{\lambda'\nu'}{4}+\frac{3\nu'^{2}}{8}\right)\end{cases}\label{eq:b.13}\end{equation}
The field equations in $\mathcal{R}^{2}$ Lagrangian are:\begin{equation}
\begin{cases}
\ \mathcal{R}_{tt}-\frac{1}{4}g_{tt}\mathcal{R} & =-\Gamma_{tt}^{r}\frac{\mathcal{R}'}{\mathcal{R}}\\
\ \mathcal{R}_{\theta\theta}-\frac{1}{4}g_{\theta\theta}\mathcal{R} & =-\Gamma_{\theta\theta}^{r}\frac{\mathcal{R}'}{\mathcal{R}}\\
\ \mathcal{R}_{rr}-\frac{1}{4}g_{rr}\mathcal{R} & =-\Gamma_{rr}^{r}\frac{\mathcal{R}'}{\mathcal{R}}+\frac{\mathcal{R}''}{\mathcal{R}}\end{cases}\label{eq:b.14}\end{equation}
The $tt-$equation:\begin{eqnarray}
\frac{\nu''}{2}e^{\nu-\lambda}+\frac{1}{4}e^{\nu}\left(\Lambda+kr\right) & = & -\frac{\nu'}{2}e^{\nu-\lambda}\frac{k}{\Lambda+kr}\label{eq:b.15}\\
\nu''+\frac{k}{\Lambda+kr}\nu'+\frac{1}{2}\left(\Lambda+kr\right)e^{\lambda} & = & 0\label{eq:b.16}\end{eqnarray}
The $\theta\theta-$equation:\begin{eqnarray}
1+e^{-\frac{\nu}{2}-\frac{\lambda}{2}}\left(\frac{\nu''}{4}-\frac{\lambda''}{4}\right)-\frac{1}{4}e^{\frac{\lambda}{2}-\frac{\nu}{2}}\left(\Lambda+kr\right) & = & \left(\frac{\lambda'}{4}-\frac{\nu'}{4}\right)e^{-\frac{\nu}{2}-\frac{\lambda}{2}}\frac{k}{\Lambda+kr}\label{eq:b.17}\\
\lambda''-\nu''+\frac{k}{\Lambda+kr}\left(\lambda'-\nu'\right)+\left(\Lambda+kr\right)e^{\lambda} & = & 4e^{\frac{\nu}{2}+\frac{\lambda}{2}}\label{eq:b.18}\end{eqnarray}
which, combined with \eqref{eq:b.16}, becomes:\begin{equation}
\lambda''+\frac{k}{\Lambda+kr}\lambda'+\frac{3}{2}\left(\Lambda+kr\right)e^{\lambda}=4e^{\frac{\nu}{2}+\frac{\lambda}{2}}\label{eq:b.19}\end{equation}
The $rr-$equation:\begin{eqnarray}
-\frac{\lambda''}{2}+\frac{\lambda'^{2}}{8}-\frac{3\nu'^{2}}{8}+\frac{\nu'\lambda'}{4}-\frac{1}{4}e^{\lambda}\left(\Lambda+kr\right) & = & -\frac{\lambda'}{2}\frac{k}{\Lambda+kr}\label{eq:b.20}\\
\lambda''-\frac{k}{\Lambda+kr}\lambda'+\frac{1}{2}\left(\Lambda+kr\right)e^{\lambda}-\frac{\lambda'^{2}}{4}+\frac{3\nu'^{2}}{4}-\frac{\nu'\lambda'}{2} & = & 0\label{eq:b.21}\end{eqnarray}
From the definition of $\mathcal{R}$:\begin{eqnarray}
2e^{\frac{\nu}{2}-\frac{\lambda}{2}}-e^{-\lambda}\left(\lambda''-\frac{\lambda'^{2}}{8}+\frac{3\nu'^{2}}{8}-\frac{\nu'\lambda'}{4}\right) & = & \Lambda+kr\label{eq:b.22}\\
\lambda''-\frac{\lambda'^{2}}{8}+\frac{3\nu'^{2}}{8}-\frac{\nu'\lambda'}{4}+\left(\Lambda+kr\right)e^{\lambda} & = & 2e^{\frac{\nu}{3}+\frac{\lambda}{2}}\label{eq:b.23}\end{eqnarray}
Note that the $\mathcal{R}-$equation is the average of the $\theta\theta-$
and $rr-$equations. Now, eliminating $\lambda''$ from the $\theta\theta-$
and $\mathcal{R}-$equations, we get:\begin{equation}
2e^{\frac{\nu}{2}+\frac{\lambda}{2}}-\frac{k}{\Lambda+kr}\lambda'-\frac{1}{2}\left(\Lambda+kr\right)e^{\lambda}-\frac{\lambda'^{2}}{8}+\frac{3\nu'^{2}}{8}-\frac{\nu'\lambda'}{4}=0\label{eq:b.24}\end{equation}
In summary, the two equations determine $\nu$ and $\lambda$ as functions
of $r$. The last equation is not an additional {}``constraint''
but simply a {}``conservation law'' among $\nu,\ \lambda,\ \nu',\ \lambda'$
at any given $r$.

Next, we make a coordinate change, which is slightly different from
Buchdahl's:\begin{equation}
\Lambda+kr=\pm\Lambda e^{kz}\label{eq:b.25}\end{equation}
\begin{eqnarray*}
\frac{d}{dr} & = & \frac{dz}{dr}\frac{d}{dz}=\pm\frac{e^{-kz}}{\Lambda}\frac{d}{dz}\\
\frac{d^{2}}{dr^{2}} & = & \frac{dz}{dr}\frac{d}{dz}\left(\pm\frac{e^{-kz}}{\Lambda}\frac{d}{dz}\right)=\pm\frac{e^{-kz}}{\Lambda}\left(\mp\frac{ke^{-kz}}{\Lambda}\frac{d}{dz}\pm\frac{e^{-kz}}{\Lambda}\frac{d^{2}}{dz^{2}}\right)=\frac{e^{-2kz}}{\Lambda^{2}}\left(\frac{d^{2}}{dz^{2}}-k\frac{d}{dz}\right)\end{eqnarray*}
The three equations become:\begin{equation}
\begin{cases}
\ \frac{e^{-2kz}}{\Lambda^{2}}\left(\nu_{zz}-k\nu_{z}\right)+\frac{ke^{-2kz}}{\Lambda^{2}}\nu_{z}\pm\frac{\Lambda}{2}e^{kz+\lambda} & =0\\
\ \frac{e^{-2kz}}{\Lambda^{2}}\left(\lambda_{zz}-k\lambda_{z}\right)+\frac{ke^{-2kz}}{\Lambda^{2}}\lambda_{z}\pm\frac{3\Lambda}{2}e^{kz+\lambda} & =4e^{\frac{\nu}{2}+\frac{\lambda}{2}}\\
\ \frac{ke^{-2kz}}{\Lambda^{2}}\lambda_{z}\pm\frac{\Lambda}{2}e^{kz+\lambda}+\frac{e^{-2kz}}{8\Lambda^{2}}\lambda_{z}^{2}-\frac{3e^{-2kz}}{8\Lambda^{2}}\nu_{z}^{2}+\frac{e^{-2kz}}{4\Lambda^{2}}\nu_{z}\lambda_{z} & =2e^{\frac{\nu}{2}+\frac{\lambda}{2}}\end{cases}\label{eq:b.26}\end{equation}
or\begin{equation}
\begin{cases}
\ \nu_{zz}\pm\frac{\Lambda^{3}}{2}e^{3kz+\lambda} & =0\\
\ \lambda_{zz}\pm\frac{3\Lambda^{3}}{2}e^{3kz+\lambda} & =4\Lambda^{2}e^{2kz+\frac{\nu}{2}+\frac{\lambda}{2}}\\
\ \lambda_{z}^{2}-3\nu_{z}^{2}+2\nu_{z}\lambda_{z}+8k\lambda_{z}\pm4\Lambda^{3}e^{3kz+\lambda} & =16\Lambda^{2}e^{2kz+\frac{\nu}{2}+\frac{\lambda}{2}}\end{cases}\label{eq:b.27}\end{equation}
Further defining\begin{equation}
\begin{cases}
\ \nu & =-u+v-kz-\ln\frac{\Lambda}{4}\\
\ \lambda & =3u+v-3kz-3\ln\frac{\Lambda}{4}\\
\ \mu & =\frac{\lambda}{2}-\frac{\nu}{2}=2u-kz-\ln\frac{\Lambda}{4}\end{cases}\label{eq:b.28}\end{equation}
we get\begin{equation}
\begin{cases}
\ u_{zz} & =16e^{u}\left(1\mp e^{2u}\right)e^{v}\\
\ v_{zz} & =16e^{u}\left(1\mp3e^{2u}\right)e^{v}\\
\ u_{z}v_{z} & =16e^{u}\left(1\mp e^{2u}\right)e^{v}+\frac{3k^{2}}{4}\end{cases}\label{eq:b.29}\end{equation}
It is also easy to check that upon taking derivative w.r.t $z,$ the
last equation is consistent with the two former equations. Therefore,
we could interpret -- \emph{à la} Buchdahl -- the first two equations
as {}``equations of motion'' (in {}``time'' $z$) and the last
equation as a {}``conservation law'' in which $k$ is a {}``constant
of motion''. The parameter $k$ also plays another role: it sets
the degree of deviation of the curvature away from the constant $\Lambda$.
A non-zero value of $k$ means a static metric with non-constant Ricci
scalar.

Buchdahl next exploited some clever analogy to Hamiltonian dynamics
to simplify these equations. However, with the benefit of hindsight,
we find a shortcut which we present in what follows:

Upon differentiating the last equation w.r.t. $z$:\begin{equation}
u_{zz}v_{z}+u_{z}v_{zz}=16\left(e^{u}\mp3e^{3u}\right)e^{v}u_{z}+16\left(e^{u}\mp e^{3u}\right)e^{v}v_{z}\label{eq:b.30}\end{equation}
and utilizing the first equation, we re-obtain the second equation.
Thus, we can ignore the second equation from now on.

Define $q$ as a function of $u$:\begin{equation}
q=u_{z}\label{eq:b.31}\end{equation}
Also, viewing $v$ as a function of $u$: \begin{eqnarray}
u_{zz} & = & q_{z}=q_{u}u_{z}=q_{u}q\label{eq:b.32}\\
v_{z} & = & v_{u}u_{z}=v_{u}q\label{eq:b.33}\end{eqnarray}
The first and last equations become:\begin{eqnarray}
qq_{u} & = & 16e^{u}\left(1\mp e^{2u}\right)e^{v}\label{eq:b.34}\\
q^{2}v_{u} & = & 16e^{u}\left(1\mp e^{2u}\right)e^{v}+\frac{3k^{2}}{4}\label{eq:b.35}\end{eqnarray}
Now, setting \begin{eqnarray}
u & = & \ln x\label{eq:b.36}\\
q_{u} & = & \frac{q_{x}}{u_{x}}=xq_{x}\label{eq:b.37}\\
v_{u} & = & \frac{v_{x}}{u_{x}}=xv_{x}\label{eq:b.38}\end{eqnarray}
we thus get:\begin{eqnarray}
qq_{x} & = & 16\left(1\mp x^{2}\right)e^{v}\label{eq:b.39}\\
q^{2}v_{x} & = & 16\left(1\mp x^{2}\right)e^{v}+\frac{3k^{2}}{4x}=qq_{x}+\frac{3k^{2}}{4x}\label{eq:b.40}\end{eqnarray}
Differentiating the first equation w.r.t. $x$:\begin{eqnarray}
q_{x}^{2}+qq_{xx} & = & 16\left(1\mp x^{2}\right)e^{v}v_{x}\mp32xe^{v}\nonumber \\
 & = & qq_{x}v_{x}\mp\frac{2xqq_{x}}{1\mp x^{2}}\nonumber \\
 & = & q_{x}^{2}+\frac{3k^{2}q_{x}}{4xq}\mp\frac{2xqq_{x}}{1\mp x^{2}}\label{eq:b.41}\end{eqnarray}
\begin{eqnarray}
q_{xx}\pm\frac{2x}{1\mp x^{2}}q_{x} & = & \frac{3k^{2}}{4xq^{2}}q_{x}\label{eq:b.42}\\
q_{xx}-\frac{2x}{x^{2}\mp1}q_{x} & = & \frac{3k^{2}}{4xq^{2}}q_{x}\label{eq:b.43}\\
\partial_{x}\left(\frac{q_{x}}{x^{2}\mp1}\right) & = & \frac{3k^{2}}{4xq}\left(\frac{q_{x}}{x^{2}\mp1}\right)\label{eq:b.44}\end{eqnarray}
This is the equation that Buchdahl obtained.

\subsubsection*{Our representation:}

Define a new function $p$ as function of $x$: \begin{equation}
p=\frac{q_{x}}{1\mp x^{2}}\label{eq:b.45}\end{equation}
We thus have:\begin{eqnarray}
q_{x} & = & \left(1\mp x^{2}\right)p\label{eq:b.46}\\
p_{x} & = & \frac{3k^{2}}{4x}\frac{p}{q}\label{eq:b.47}\end{eqnarray}
In terms of $x$:\begin{eqnarray}
e^{u} & = & x\label{eq:b.48}\\
e^{v} & = & \frac{qq_{x}}{16\left(1\mp x^{2}\right)}=\frac{qp}{16}\label{eq:b.49}\\
e^{\nu} & = & e^{-u+v-kz-\ln\frac{\Lambda}{4}}=\frac{4}{\Lambda e^{kz}}\frac{qp}{16x}\label{eq:b.50}\\
e^{\lambda} & = & e^{3u+v-3kz-3\ln\frac{\Lambda}{4}}=\frac{64}{\Lambda^{3}e^{3kz}}\frac{x^{3}qp}{16}\label{eq:b.51}\\
e^{\mu} & = & e^{2u-kz-\ln\frac{\Lambda}{4}}=\frac{4}{\Lambda e^{kz}}x^{2}\label{eq:b.52}\end{eqnarray}
With $\mathcal{R}=\Lambda+kr=\pm\Lambda e^{kz}$:\[
dr=\pm\Lambda e^{kz}dz\]
and we also know that \begin{eqnarray}
q & = & u_{z}=\frac{x_{z}}{x}=\frac{1}{x}\frac{dx}{dz}\label{eq:b.53}\\
dz & = & \frac{dx}{xq}\label{eq:b.54}\\
dr & = & \pm\Lambda e^{kz}\frac{1}{xq}dx\label{eq:b.55}\end{eqnarray}
the metric becomes:\begin{eqnarray}
ds^{2} & = & -e^{\nu}\left(dx^{0}\right)^{2}+e^{\lambda}dr^{2}+e^{\mu}d\Omega^{2}\nonumber \\
 & = & -\frac{4}{\Lambda e^{kz}}\frac{qp}{16x}\left(dx^{0}\right)^{2}+\frac{64}{\Lambda^{3}e^{3kz}}\frac{x^{3}qp}{16}\left(\Lambda^{2}e^{2kz}\frac{1}{x^{2}q^{2}}dx^{2}\right)+\frac{4}{\Lambda e^{kz}}x^{2}d\Omega^{2}\nonumber \\
 & = & \frac{4}{\Lambda e^{kz}}\left\{ \frac{p}{4}\left[-\frac{q}{4x}\left(dx^{0}\right)^{2}+\frac{4x}{q}dx^{2}\right]+x^{2}d\Omega^{2}\right\} \label{eq:b.56}\end{eqnarray}
In summary, the metric is: \begin{equation}
ds^{2}=\frac{4}{\mathcal{R}(x)}\left\{ \frac{p(x)}{4}\left[-\frac{q(x)}{4x}\left(dx^{0}\right)^{2}+\frac{4x}{q(x)}dx^{2}\right]+x^{2}d\Omega^{2}\right\} \label{eq:b.57}\end{equation}
in which\begin{equation}
\begin{cases}
\ \mathcal{R}(x) & =\pm\Lambda e^{k\int\frac{dx}{xq(x)}}\\
\ q_{x} & =\left(1\mp x^{2}\right)p\\
\ p_{x} & =\frac{3k^{2}}{4x}\frac{p}{q}\end{cases}\label{eq:b.58}\end{equation}
The $4$ parameters: $k$ (the anomalous curvature), $\Lambda$ (the
large-scale curvature), $p\left(x_{0}\right)$ and $q\left(x_{0}\right)$
at some $x_{0}$ of convenience. Note that there is a constant of
integration in $\int\frac{dx}{xq(x)}$ but it can be absorbed into
$\Lambda$.

\section{\label{sec:C}An explicit solution for spherically symmetric case}

Let us find a solution in the form:\begin{equation}
ds^{2}=e^{\alpha}\left[-\Psi(dx^{0})^{2}+\frac{dr^{2}}{\Psi}+r^{2}d\Omega^{2}\right]\label{eq:c.1}\end{equation}
The relevant quantities are computed\begin{equation}
\begin{cases}
\ \ \ \frac{\mathcal{R}_{tt}}{\Psi} & =\left(\frac{\alpha''}{2}+\frac{\alpha'^{2}}{2}+\frac{\alpha'}{r}\right)\Psi+\left(\alpha'+\frac{1}{r}\right)\Psi'+\frac{\Psi''}{2}\\
\ -\frac{\mathcal{R}_{\theta\theta}}{r^{2}} & =-\frac{1}{r^{2}}+\left(\frac{\alpha''}{2}+\frac{\alpha'^{2}}{2}+\frac{2\alpha'}{r}+\frac{1}{r^{2}}\right)\Psi+\left(\frac{\alpha'}{2}+\frac{1}{r}\right)\Psi'\\
\ \mathcal{A}\triangleq-\mathcal{R}e^{\alpha} & =-\frac{2}{r^{2}}+\left(3\alpha''+\frac{3\alpha'^{2}}{2}+\frac{6\alpha'}{r}+\frac{2}{r^{2}}\right)\Psi+\left(3\alpha'+\frac{4}{r}\right)\Psi'+\Psi''\end{cases}\label{eq:c.2}\end{equation}
\begin{equation}
\begin{cases}
\ g_{tt} & =-e^{\alpha}\Psi\\
\ g_{\theta\theta} & =e^{\alpha}r^{2}\\
\ \frac{\Gamma_{tt}^{r}}{\Psi} & =\frac{\alpha'\Psi}{2}+\frac{\Psi'}{2}\\
\ \frac{\Gamma_{\theta\theta}^{r}}{r^{2}} & =-\left(\frac{\alpha'}{2}+\frac{1}{r}\right)\Psi\end{cases}\label{eq:c.3}\end{equation}
The $tt$- and $\theta\theta$- field equations\begin{eqnarray}
\left(\mathcal{R}_{tt}-\frac{1}{4}g_{tt}\mathcal{R}\right)\mathcal{R} & = & -\Gamma_{tt}^{r}\mathcal{R}'\label{eq:c.4}\\
\left(\mathcal{R}_{\theta\theta}-\frac{1}{4}g_{\theta\theta}\mathcal{R}\right)\mathcal{R} & = & -\Gamma_{\theta\theta}^{r}\mathcal{R}'\label{eq:c.5}\end{eqnarray}
can be recast as\begin{eqnarray}
\left(\frac{\mathcal{R}_{tt}}{\Psi}+\frac{1}{4}\frac{g_{tt}e^{-\alpha}}{\Psi}\left(-\mathcal{R}e^{\alpha}\right)\right)\left(-\mathcal{R}e^{\alpha}\right) & = & \frac{\Gamma_{tt}^{r}}{\Psi}\left(\mathcal{R}'e^{\alpha}\right)\label{eq:c.6}\\
\left(-\frac{\mathcal{R}_{\theta\theta}}{r^{2}}-\frac{1}{4}\frac{g_{\theta\theta}e^{-\alpha}}{r^{2}}\left(-\mathcal{R}e^{\alpha}\right)\right)\left(-\mathcal{R}e^{\alpha}\right) & = & -\frac{\Gamma_{\theta\theta}^{r}}{r^{2}}\left(\mathcal{R}'e^{\alpha}\right)\label{eq:c.7}\end{eqnarray}
With\begin{eqnarray}
\mathcal{A}' & = & -\mathcal{R}'e^{\alpha}-\alpha'\mathcal{R}e^{\alpha}\label{eq:c.8}\\
\mathcal{R}'e^{\alpha} & = & -\mathcal{A}'+\alpha'\mathcal{A}\label{eq:c.9}\end{eqnarray}
we thus obtain three equations for the two unknowns $\alpha,\ \Psi$
and the auxiliary $\mathcal{A}$:\begin{eqnarray}
\left[\left(\frac{\alpha''}{2}+\frac{\alpha'^{2}}{2}+\frac{\alpha'}{r}\right)\Psi+\left(\alpha'+\frac{1}{r}\right)\Psi'+\frac{\Psi''}{2}-\frac{\mathcal{A}}{4}\right]\mathcal{A} & = & \left(\frac{\alpha'\Psi}{2}+\frac{\Psi'}{2}\right)\left(-\mathcal{A}'+\alpha'\mathcal{A}\right)\label{eq:c.10}\\
\left[-\frac{1}{r^{2}}+\left(\frac{\alpha''}{2}+\frac{\alpha'^{2}}{2}+\frac{2\alpha'}{r}+\frac{1}{r^{2}}\right)\Psi+\left(\frac{\alpha'}{2}+\frac{1}{r}\right)\Psi'-\frac{\mathcal{A}}{4}\right]\mathcal{A} & = & \left(\frac{\alpha'}{2}+\frac{1}{r}\right)\Psi\left(-\mathcal{A}'+\alpha'\mathcal{A}\right)\label{eq:c.11}\\
-\frac{2}{r^{2}}+\left(3\alpha''+\frac{3\alpha'^{2}}{2}+\frac{6\alpha'}{r}+\frac{2}{r^{2}}\right)\Psi+\left(3\alpha'+\frac{4}{r}\right)\Psi'+\Psi'' & = & \mathcal{A}\label{eq:c.12}\end{eqnarray}
We can simplify these equation further by directly varying the action:\begin{equation}
\mathcal{S}=\int d^{4}x\,\sqrt{g}\,\mathcal{R}^{2}=\int dx^{0}dr\, d\theta\, d\varphi\sin^{2}\theta e^{2\alpha}r^{2}\,\mathcal{R}^{2}=4\pi\int dx^{0}dr\, r^{2}\mathcal{A}^{2}\label{eq:c.13}\end{equation}
in which\begin{eqnarray}
\sqrt{g} & = & e^{2\alpha}\, r^{2}\sin^{2}\theta\label{eq:c.14}\\
\sqrt{g}\, g^{rr} & = & e^{\alpha}\,\Psi\, r^{2}\sin^{2}\theta\label{eq:c.15}\end{eqnarray}
The first equation of motion:\begin{eqnarray}
\left(\frac{\partial\left(r^{2}\mathcal{A}^{2}\right)}{\partial\alpha''}\right)''-\left(\frac{\partial\left(r^{2}\mathcal{A}^{2}\right)}{\partial\alpha'}\right)'+\frac{\partial\left(r^{2}\mathcal{A}^{2}\right)}{\partial\alpha} & = & 0\label{eq:c.16}\\
\left(\frac{\partial\left(r^{2}\mathcal{A}^{2}\right)}{\partial\alpha''}\right)'-\frac{\partial\left(r^{2}\mathcal{A}^{2}\right)}{\partial\alpha'} & = & c\nonumber \\
\left(r^{2}\mathcal{A}\right)\left[\left(\frac{\partial\mathcal{A}}{\partial\alpha''}\right)'-\frac{\partial\mathcal{A}}{\partial\alpha'}\right]+\left(r^{2}\mathcal{A}\right)'\frac{\partial\mathcal{A}}{\partial\alpha''} & = & c\nonumber \end{eqnarray}
\begin{eqnarray*}
\left(\frac{\partial\mathcal{A}}{\partial\alpha''}\right)'-\frac{\partial\mathcal{A}}{\partial\alpha'} & = & 3\left[\Psi'-\left(\alpha'\Psi+\frac{2\Psi}{r}+\Psi'\right)\right]=-3\left(\alpha'+\frac{2}{r}\right)\Psi\\
\frac{\partial\mathcal{A}}{\partial\alpha''} & = & 3\Psi\\
\left(r^{2}\mathcal{A}\right)' & = & r^{2}\mathcal{A}'+2r\mathcal{A}=r^{2}\left(\mathcal{A}'+\frac{2}{r}\mathcal{A}\right)\end{eqnarray*}
\begin{eqnarray}
-\left(\alpha'+\frac{2}{r}\right)r^{2}\Psi\mathcal{A}+r^{2}\Psi\left(\mathcal{A}'+\frac{2}{r}\mathcal{A}\right) & = & c\nonumber \\
\left(\mathcal{A}'-\alpha'\mathcal{A}\right)r^{2}\Psi & = & c\nonumber \\
\mathcal{R}'e^{\alpha}r^{2}\Psi & = & c\label{eq:c.17}\end{eqnarray}
which is equivalent to\[
\square\mathcal{R}=0.\]
Indeed\[
\left(\sqrt{g}\, g^{rr}\,\mathcal{R}'\right)'=0\]
or\[
\left(e^{\alpha}\,\Psi\, r^{2}\,\mathcal{R}'\right)'=0.\]
The second equation of motion: \begin{eqnarray}
\left(\frac{\partial\left(r^{2}\mathcal{A}^{2}\right)}{\partial\Psi''}\right)''-\left(\frac{\partial\left(r^{2}\mathcal{A}^{2}\right)}{\partial\Psi'}\right)'+\frac{\partial\left(r^{2}\mathcal{A}^{2}\right)}{\partial\Psi} & = & 0\label{eq:c.18}\\
\left(r^{2}\mathcal{A}\right)\left[\left(\frac{\partial\mathcal{A}}{\partial\Psi''}\right)''-\left(\frac{\partial\mathcal{A}}{\partial\Psi'}\right)'+\frac{\partial\mathcal{A}}{\partial\Psi}\right]+\left(r^{2}\mathcal{A}\right)'\left[2\left(\frac{\partial\mathcal{A}}{\partial\Psi''}\right)'-\frac{\partial\mathcal{A}}{\partial\Psi'}\right]+\left(r^{2}\mathcal{A}\right)''\frac{\partial\mathcal{A}}{\partial\Psi''} & = & 0\nonumber \end{eqnarray}
\begin{eqnarray*}
\left(\frac{\partial\mathcal{A}}{\partial\Psi''}\right)''-\left(\frac{\partial\mathcal{A}}{\partial\Psi'}\right)'+\frac{\partial\mathcal{A}}{\partial\Psi} & = & -\left(3\alpha''-\frac{4}{r^{2}}\right)+\left(3\alpha''+\frac{3\alpha'^{2}}{2}+\frac{6\alpha'}{r}+\frac{2}{r^{2}}\right)=\frac{3\alpha'^{2}}{2}+\frac{6\alpha'}{r}+\frac{6}{r^{2}}\\
2\left(\frac{\partial\mathcal{A}}{\partial\Psi''}\right)'-\frac{\partial\mathcal{A}}{\partial\Psi'} & = & -\left(3\alpha'+\frac{4}{r}\right)\\
\frac{\partial\mathcal{A}}{\partial\Psi''} & = & 1\\
\left(r^{2}\mathcal{A}\right)' & = & r^{2}\mathcal{A}'+2r\mathcal{A}=r^{2}\left(\mathcal{A}'+\frac{2}{r}\mathcal{A}\right)\\
\left(r^{2}\mathcal{A}\right)'' & = & r^{2}\mathcal{A}''+4r\mathcal{A}'+2\mathcal{A}=r^{2}\left(\mathcal{A}''+\frac{4}{r}\mathcal{A}'+\frac{2}{r^{2}}\mathcal{A}\right)\end{eqnarray*}
\begin{eqnarray}
\left(\frac{3\alpha'^{2}}{2}+\frac{6\alpha'}{r}+\frac{6}{r^{2}}\right)r^{2}\mathcal{A}-\left(3\alpha'+\frac{4}{r}\right)r^{2}\left(\mathcal{A}'+\frac{2}{r}\mathcal{A}\right)+r^{2}\left(\mathcal{A}''+\frac{4}{r}\mathcal{A}'+\frac{2}{r^{2}}\mathcal{A}\right) & = & 0\nonumber \\
\mathcal{A}''-3\alpha'\mathcal{A}'+\frac{3\alpha'^{2}}{2}\mathcal{A} & = & 0\nonumber \\
\mathcal{R}''-\alpha'\mathcal{R}'+\left(\alpha''-\frac{\alpha'^{2}}{2}\right)\mathcal{R} & = & 0\label{eq:c.19}\end{eqnarray}
Let us find the solution with $c=0,$ which implies $\mathcal{R}'=0$.
Then\begin{eqnarray}
\alpha''-\frac{\alpha'^{2}}{2} & = & 0\label{eq:c.20}\\
\alpha & = & -2\ln\left(1+a\, r\right)\label{eq:c.21}\\
e^{\alpha} & = & \frac{1}{\left(1+ar\right)^{2}}\label{eq:c.22}\\
\alpha' & = & -\frac{2a}{1+ar}\label{eq:c.23}\end{eqnarray}
and \begin{eqnarray}
\mathcal{A}' & = & \alpha'\mathcal{A}\label{eq:c.24}\\
\mathcal{A} & = & -\mathcal{R}_{0}e^{\alpha}=-\frac{\mathcal{R}_{0}}{\left(1+ar\right)^{2}}\label{eq:c.25}\end{eqnarray}
Plug this into Eq. \eqref{eq:c.11}:\begin{equation}
-\frac{1}{r^{2}}+\left(\frac{\alpha''}{2}+\frac{\alpha'^{2}}{2}+\frac{2\alpha'}{r}+\frac{1}{r^{2}}\right)\Psi+\left(\frac{\alpha'}{2}+\frac{1}{r}\right)\Psi'=\frac{\mathcal{A}}{4}=-\frac{\mathcal{R}_{0}}{4}e^{\alpha}\label{eq:c.26}\end{equation}
With\begin{equation}
\frac{\alpha''}{2}+\frac{\alpha'^{2}}{2}+\frac{2\alpha'}{r}+\frac{1}{r^{2}}=\frac{3a^{2}}{\left(1+ar\right)^{2}}-\frac{4a}{r\left(1+ar\right)}+\frac{1}{r^{2}}=\frac{1-2ar}{r^{2}\left(1+ar\right)^{2}}\label{eq:c.27}\end{equation}
and\begin{equation}
\frac{\alpha'}{2}+\frac{1}{r}=-\frac{a}{1+ar}+\frac{1}{r}=\frac{1}{r\left(1+ar\right)}\label{eq:c.28}\end{equation}
\begin{equation}
r\left(1+ar\right)\Psi'+\left(1-2ar\right)\Psi-\left(1+ar\right)^{2}+\frac{\mathcal{R}_{0}}{4}r^{2}=0\label{eq:c.29}\end{equation}
or\begin{equation}
\left(1+ar\right)\left(r\Psi\right)'-3a\left(r\Psi\right)-\left(1+ar\right)^{2}+\frac{\mathcal{R}_{0}}{4}r^{2}=0\label{eq:c.30}\end{equation}
Choose\begin{equation}
r\Psi=br-r_{s}-\Lambda r^{3}+\gamma r^{2}\label{eq:c.31}\end{equation}
the equation successively becomes\begin{equation}
\left(1+ar\right)\left(b-3\Lambda r^{2}+2\gamma r\right)-3a\left(br-r_{s}-\Lambda r^{3}+\gamma r^{2}\right)-1-2ar-a^{2}r^{2}+\frac{\mathcal{R}_{0}}{4}r^{2}=0\label{eq:c.32}\end{equation}
\begin{equation}
b-3\Lambda r^{2}+2\gamma r+abr-3a\Lambda r^{3}+2a\gamma r^{2}-3abr+3ar_{s}+3a\Lambda r^{3}-3a\gamma r^{2}-1-2ar-a^{2}r^{2}+\frac{\mathcal{R}_{0}}{4}r^{2}=0\label{eq:c.33}\end{equation}
\begin{equation}
\left(b+3ar_{s}-1\right)+2\left(\gamma-ab-a\right)r+\left(-3\Lambda-a\gamma-a^{2}+\frac{\mathcal{R}_{0}}{4}\right)r^{2}=0\label{eq:c.34}\end{equation}
which solves for\begin{equation}
\begin{cases}
\ b & =1-3ar_{s}\\
\ \gamma & =ab+a=a\left(2-3ar_{s}\right)\\
\ \Lambda & =\frac{\mathcal{R}_{0}}{12}-\frac{a\gamma+a^{2}}{3}=\frac{\mathcal{R}_{0}}{12}+a^{2}\left(ar_{s}-1\right)\end{cases}\label{eq:c.35}\end{equation}
It is straightforward to verify that $\Psi$ and $\alpha$ with $b,\ \gamma,\ \Lambda$
given above also satisfy Eqs. (\ref{eq:c.10}) and (\ref{eq:c.12}).
In summary, one spherically symmetric solution to $\mathcal{R}^{2}$
Lagrangian is:

\subsubsection*{Three representations:}
\begin{enumerate}
\item Representation I:

\begin{equation}
\begin{cases}
\ \Psi & =\left(1-3ar_{s}\right)-\frac{r_{s}}{r}-\Lambda r^{2}+a\left(2-3ar_{s}\right)r\\
\ e^{\alpha} & =\left(1+ar\right)^{-2}\\
\ \Lambda & =\frac{\mathcal{R}_{0}}{12}+a^{2}\left(ar_{s}-1\right)\end{cases}\label{eq:c.36}\end{equation}

\item Representation II (analogous to Mannheim-Kazanas):

If we define\begin{equation}
\begin{cases}
\ ar_{s} & \triangleq\beta\gamma\\
\ a\left(2-3ar_{s}\right) & \triangleq\gamma\end{cases}\label{eq:c.37}\end{equation}
which leads to\begin{equation}
\begin{cases}
\ a & =\frac{\gamma}{2-3ar_{s}}=\frac{\gamma}{2-3\beta\gamma}\\
\ r_{s} & =\frac{\beta\gamma}{a}=\beta\left(2-3\beta\gamma\right)\end{cases}\label{eq:c.38}\end{equation}
we thus get:\begin{equation}
\begin{cases}
\ \Psi & =\left(1-3\beta\gamma\right)-\frac{\beta\left(2-3\beta\gamma\right)}{r}-\Lambda r^{2}+\gamma r\\
\ e^{\alpha} & =\left(1+\frac{\gamma}{2-3\beta\gamma}r\right)^{-2}\end{cases}\label{eq:c.39}\end{equation}

\item Representation III:

Define a dimensionless parameter $\kappa$:\begin{equation}
\kappa\triangleq-3\gamma r_{s}\label{eq:c.40}\end{equation}
then we successively get \begin{eqnarray*}
-\frac{\kappa}{3} & = & \gamma r_{s}=2\left(ar_{s}\right)-3\left(ar_{s}\right)^{2}\\
ar_{s} & = & \frac{1-\sqrt{1+\kappa}}{3}\\
b & = & 1-3ar_{s}=\sqrt{1+\kappa}\\
a^{2}\left(ar_{s}-1\right) & = & -\frac{1}{27r_{s}^{2}}\left(1-\sqrt{1+\kappa}\right)^{2}\left(2+\sqrt{1+\kappa}\right)\end{eqnarray*}
in which the sign of the square root is chosen such that for small
$\kappa$, $a\simeq\kappa$ and $b\approx1$. We thus get:\begin{equation}
\begin{cases}
\ \Psi & =\sqrt{1+\kappa}-\frac{r_{s}}{r}-\Lambda r^{2}-\frac{\kappa}{3r_{s}}r\\
\ e^{\alpha} & =\left(1+\frac{1-\sqrt{1+\kappa}}{3r_{s}}r\right)^{-2}\\
\ \Lambda & =\frac{\mathcal{R}_{0}}{12}-\frac{1}{27r_{s}^{2}}\left(1-\sqrt{1+\kappa}\right)^{2}\left(2+\sqrt{1+\kappa}\right)\end{cases}\label{eq:c.41}\end{equation}

\end{enumerate}

\section{\label{sec:D}A perturbative solution to the vacuo field equation:
The anomalous curvature}

The metric is written in the following form with two unknown functions
$\alpha$ and $\Psi$: \begin{equation}
ds^{2}=e^{\alpha}\left[-\Psi\left(dx^{0}\right)^{2}+\frac{dr^{2}}{\Psi}+r^{2}d\Omega^{2}\right]\label{eq:d.1}\end{equation}
Taking a cue from our previous solution, we shall find the two unknown
functions in the form:\begin{eqnarray}
\Psi & = & \Psi_{0}+\gamma\Psi_{1}+\mathcal{O}\left(\gamma^{2}\right)\label{eq:d.2}\\
\alpha & = & \gamma\Phi_{1}+\mathcal{O}\left(\gamma^{2}\right)\label{eq:d.3}\end{eqnarray}
in which \begin{equation}
\Psi_{0}=1-\frac{r_{s}}{r}-\Lambda r^{2}\label{eq:d.4}\end{equation}
That is to say, we shall find the uniformly convergent solutions with
$\gamma$ as a perturbative parameter. The parameter $\gamma$ shall
play a role similar to that of the Mannheim-Kazanas parameter in our
solution in Appendix \ref{sec:C}.\begin{eqnarray}
\frac{\mathcal{R}_{tt}}{\Psi} & = & \left(\frac{\alpha''}{2}+\frac{\alpha'^{2}}{2}+\frac{\alpha'}{r}\right)\Psi+\frac{\Psi''}{2}+\left(\alpha'+\frac{1}{r}\right)\Psi'\nonumber \\
 & = & \gamma\left(\frac{\Phi_{1}''}{2}+\frac{\Phi_{1}'}{r}\right)\Psi_{0}+\frac{\Psi_{0}''}{2}+\gamma\frac{\Psi_{1}''}{2}+\frac{\Psi_{0}'}{r}+\gamma\left(\frac{\Psi_{1}'}{r}+\Phi_{1}'\Psi_{0}'\right)+\mathcal{O}\left(\gamma^{2}\right)\nonumber \\
 & = & \left(\frac{\Psi_{0}''}{2}+\frac{\Psi_{0}'}{r}\right)+\gamma\left(\frac{\Phi_{1}''\Psi_{0}}{2}+\frac{\Phi_{1}'\Psi_{0}}{r}+\frac{\Psi_{1}''}{2}+\frac{\Psi_{1}'}{r}+\Phi_{1}'\Psi_{0}'\right)+\mathcal{O}\left(\gamma^{2}\right)\nonumber \\
 & = & -3\Lambda+\gamma\left(\frac{\Phi_{1}''\Psi_{0}}{2}+\frac{\Phi_{1}'\Psi_{0}}{r}+\frac{\Psi_{1}''}{2}+\frac{\Psi_{1}'}{r}+\Phi_{1}'\Psi_{0}'\right)+\mathcal{O}\left(\gamma^{2}\right)\label{eq:d.5}\end{eqnarray}
\begin{eqnarray}
-\frac{\mathcal{R}_{\theta\theta}}{r^{2}} & = & -\frac{1}{r^{2}}+\left(\frac{\alpha''}{2}+\frac{\alpha'^{2}}{2}+\frac{2\alpha'}{r}+\frac{1}{r^{2}}\right)\Psi+\left(\frac{\alpha'}{2}+\frac{1}{r}\right)\Psi'\nonumber \\
 & = & -\frac{1}{r^{2}}+\frac{\Psi_{0}}{r^{2}}+\gamma\left(\frac{\Phi_{1}''\Psi_{0}}{2}+\frac{2\Phi_{1}'\Psi_{0}}{r}+\frac{\Psi_{1}}{r^{2}}\right)+\frac{\Psi_{0}'}{r}+\gamma\left(\frac{\Psi_{1}'}{r}+\frac{\Phi_{1}'\Psi_{0}'}{2}\right)+\mathcal{O}\left(\gamma^{2}\right)\nonumber \\
 & = & \left(-\frac{1}{r^{2}}+\frac{\Psi_{0}}{r^{2}}+\frac{\Psi_{0}'}{r}\right)+\gamma\left(\frac{\Phi_{1}''\Psi_{0}}{2}+\frac{2\Phi_{1}'\Psi_{0}}{r}+\frac{\Psi_{1}}{r^{2}}+\frac{\Psi_{1}'}{r}+\frac{\Phi_{1}'\Psi_{0}'}{2}\right)+\mathcal{O}\left(\gamma^{2}\right)\nonumber \\
 & = & -3\Lambda+\gamma\left(\frac{\Phi_{1}''\Psi_{0}}{2}+\frac{2\Phi_{1}'\Psi_{0}}{r}+\frac{\Psi_{1}}{r^{2}}+\frac{\Psi_{1}'}{r}+\frac{\Phi_{1}'\Psi_{0}'}{2}\right)+\mathcal{O}\left(\gamma^{2}\right)\label{eq:d.6}\end{eqnarray}
\begin{eqnarray}
-\mathcal{R}e^{\alpha} & = & -\frac{2}{r^{2}}+\left(3\alpha''+\frac{3\alpha'^{2}}{2}+\frac{6\alpha'}{r}+\frac{2}{r^{2}}\right)\Psi+\left(3\alpha'+\frac{4}{r}\right)\Psi'+\Psi''\nonumber \\
 & = & -\frac{2}{r^{2}}+\frac{2\Psi_{0}}{r^{2}}+\gamma\left(3\Phi_{1}''\Psi_{0}+\frac{6\Phi_{1}'\Psi_{0}}{r}+\frac{2\Psi_{1}}{r^{2}}\right)+\frac{4\Psi_{0}'}{r}+\gamma\left(\frac{4\Psi_{1}'}{r}+3\Phi_{1}'\Psi_{0}'\right)+\Psi_{0}''+\gamma\Psi_{1}''+\mathcal{O}\left(\gamma^{2}\right)\nonumber \\
 & = & \left(-\frac{2}{r^{2}}+\frac{2\Psi_{0}}{r^{2}}+\frac{4\Psi_{0}'}{r}+\Psi_{0}''\right)+\gamma\left(3\Phi_{1}''\Psi_{0}+\frac{6\Phi_{1}'\Psi_{0}}{r}+\frac{2\Psi_{1}}{r^{2}}+\frac{4\Psi_{1}'}{r}+3\Phi_{1}'\Psi_{0}'+\Psi_{1}''\right)+\mathcal{O}\left(\gamma^{2}\right)\nonumber \\
 & = & -12\Lambda+\gamma\left(3\Phi_{1}''\Psi_{0}+\frac{6\Phi_{1}'\Psi_{0}}{r}+\frac{2\Psi_{1}}{r^{2}}+\frac{4\Psi_{1}'}{r}+3\Phi_{1}'\Psi_{0}'+\Psi_{1}''\right)+\mathcal{O}\left(\gamma^{2}\right)\label{eq:d.7}\end{eqnarray}
\begin{eqnarray}
\mathcal{R}'e^{\alpha} & = & \left(\mathcal{R}e^{\alpha}\right)'-\alpha'\left(\mathcal{R}e^{\alpha}\right)\nonumber \\
 & = & -\gamma\left(3\Phi_{1}''\Psi_{0}+\frac{6\Phi_{1}'\Psi_{0}}{r}+\frac{2\Psi_{1}}{r^{2}}+\frac{4\Psi_{1}'}{r}+3\Phi_{1}'\Psi_{0}'+\Psi_{1}''\right)'-12\Lambda\gamma\Phi_{1}'+\mathcal{O}\left(\gamma^{2}\right)\label{eq:d.8}\\
\frac{\left(\mathcal{R}'e^{\alpha}\right)}{\left(-\mathcal{R}e^{\alpha}\right)} & = & \frac{\gamma}{12\Lambda}\left[\left(3\Phi_{1}''\Psi_{0}+\frac{6\Phi_{1}'\Psi_{0}}{r}+\frac{2\Psi_{1}}{r^{2}}+\frac{4\Psi_{1}'}{r}+3\Phi_{1}'\Psi_{0}'+\Psi_{1}''\right)'+12\Lambda\Phi_{1}'\right]+\mathcal{O}\left(\gamma^{2}\right)\label{eq:d.9}\end{eqnarray}
\begin{equation}
\begin{cases}
\ \frac{g_{tt}e^{-\alpha}}{\Psi} & =-1\\
\ \frac{g_{\theta\theta}e^{\alpha}}{r^{2}} & =1\\
\ \ \ \frac{\Gamma_{tt}^{r}}{\Psi} & =\frac{\alpha'\Psi}{2}+\frac{\Psi'}{2}=\frac{\Psi_{0}'}{2}+\mathcal{O}\left(\gamma\right)\\
\ -\frac{\Gamma_{\theta\theta}^{r}}{r^{2}} & =\left(\frac{\alpha'}{2}+\frac{1}{r}\right)\Psi=\frac{\Psi_{0}}{r}+\mathcal{O}\left(\gamma\right)\end{cases}\label{eq:d.10}\end{equation}
The $tt$- and $\theta\theta-$field equations:\begin{eqnarray}
\left(\mathcal{R}_{tt}-\frac{1}{4}g_{tt}\mathcal{R}\right)\mathcal{R} & = & -\Gamma_{tt}^{r}\mathcal{R}'\label{eq:d.11}\\
\left(\mathcal{R}_{\theta\theta}-\frac{1}{4}g_{\theta\theta}\mathcal{R}\right)\mathcal{R} & = & -\Gamma_{\theta\theta}^{r}\mathcal{R}'\label{eq:d.12}\end{eqnarray}
can be cast as\begin{eqnarray}
\frac{\mathcal{R}_{tt}}{\Psi}+\frac{1}{4}\frac{g_{tt}e^{-\alpha}}{\Psi}\left(-\mathcal{R}e^{\alpha}\right) & = & \frac{\Gamma_{tt}^{r}}{\Psi}\frac{\left(\mathcal{R}'e^{\alpha}\right)}{\left(-\mathcal{R}e^{\alpha}\right)}\label{eq:d.13}\\
-\frac{\mathcal{R}_{\theta\theta}}{r^{2}}-\frac{1}{4}\frac{g_{\theta\theta}e^{-\alpha}}{r^{2}}\left(-\mathcal{R}e^{\alpha}\right) & = & -\frac{\Gamma_{\theta\theta}^{r}}{r^{2}}\frac{\left(\mathcal{R}'e^{\alpha}\right)}{\left(-\mathcal{R}e^{\alpha}\right)}\label{eq:d.14}\end{eqnarray}
Up to the first-order in $\gamma$, the equations for the two unknowns
$\Psi_{1}$ and $\Phi_{1}$ are:\begin{eqnarray}
\frac{\Psi_{0}\Phi_{1}''}{2}+\Psi_{0}'\Phi_{1}'+\frac{\Psi_{0}\Phi_{1}'}{r}+\frac{\Psi_{1}''}{2}+\frac{\Psi_{1}'}{r} & - & \frac{1}{4}\left(3\Psi_{0}\Phi_{1}''+3\Psi_{0}'\Phi_{1}'+6\frac{\Psi_{0}\Phi_{1}'}{r}+\Psi_{1}''+4\frac{\Psi_{1}'}{r}+2\frac{\Psi_{1}}{r^{2}}\right)\nonumber \\
 & = & \frac{\Psi_{0}'}{24\Lambda}\left[\left(3\Psi_{0}\Phi_{1}''+3\Psi_{0}'\Phi_{1}'+6\frac{\Psi_{0}\Phi_{1}'}{r}+\Psi_{1}''+4\frac{\Psi_{1}'}{r}+2\frac{\Psi_{1}}{r^{2}}\right)'+12\Lambda\Phi_{1}'\right]\label{eq:d.15}\\
\frac{\Psi_{0}\Phi_{1}''}{2}+\frac{\Psi_{0}'\Phi_{1}'}{2}+2\frac{\Psi_{0}\Phi_{1}'}{r}+\frac{\Psi_{1}'}{r}+\frac{\Psi_{1}}{r^{2}} & - & \frac{1}{4}\left(3\Psi_{0}\Phi_{1}''+3\Psi_{0}'\Phi_{1}'+6\frac{\Psi_{0}\Phi_{1}'}{r}+\Psi_{1}''+4\frac{\Psi_{1}'}{r}+2\frac{\Psi_{1}}{r^{2}}\right)\nonumber \\
 & = & \frac{\Psi_{0}}{12\Lambda r}\left[\left(3\Psi_{0}\Phi_{1}''+3\Psi_{0}'\Phi_{1}'+6\frac{\Psi_{0}\Phi_{1}'}{r}+\Psi_{1}''+4\frac{\Psi_{1}'}{r}+2\frac{\Psi_{1}}{r^{2}}\right)'+12\Lambda\Phi_{1}'\right]\label{eq:d.16}\end{eqnarray}
First, define:\begin{eqnarray}
\mathcal{E} & \triangleq & \frac{\Psi_{0}\Phi_{1}''}{2}+\frac{\Psi_{0}'\Phi_{1}'}{2}+\frac{\Psi_{0}\Phi_{1}'}{r}\label{eq:d.17}\end{eqnarray}
which can simplified further\begin{eqnarray}
\mathcal{E} & = & \frac{1}{2}\left(\Psi_{0}\Phi_{1}'\right)'+\frac{1}{r}\left(\Psi_{0}\Phi_{1}'\right)\nonumber \\
 & = & \frac{1}{2r^{2}}\left(r^{2}\Psi_{0}\Phi_{1}'\right)'\label{eq:d.18}\end{eqnarray}
Next, define: \begin{eqnarray}
\mathcal{F} & \triangleq & \frac{1}{4}\left(3\Psi_{0}\Phi_{1}''+3\Psi_{0}'\Phi_{1}'+6\frac{\Psi_{0}\Phi_{1}'}{r}+\Psi_{1}''+4\frac{\Psi_{1}'}{r}+2\frac{\Psi_{1}}{r^{2}}\right)\label{eq:d.19}\end{eqnarray}
which can be simplified further\begin{eqnarray}
\mathcal{F} & = & \frac{3}{2}\mathcal{E}+\frac{1}{4}\left(\Psi_{1}''+4\frac{\Psi_{1}'}{r}+2\frac{\Psi_{1}}{r^{2}}\right)\label{eq:d.20}\end{eqnarray}
and also \begin{eqnarray}
\mathcal{F} & = & \frac{1}{4}\left(\frac{3}{r^{2}}\left(r^{2}\Psi_{0}\Phi_{1}'\right)'+\frac{1}{r^{2}}\left(r^{2}\Psi_{1}\right)''\right)\nonumber \\
 & = & \frac{1}{4r^{2}}\left(3r^{2}\Psi_{0}\Phi_{1}'+\left(r^{2}\Psi_{1}\right)'\right)'\label{eq:d.21}\end{eqnarray}
Eqs. \eqref{eq:d.15} and \eqref{eq:d.16} become:\begin{eqnarray}
\left(\frac{\Psi_{0}\Phi_{1}''}{2}+\Psi_{0}'\Phi_{1}'+\frac{\Psi_{0}\Phi_{1}'}{r}+\frac{\Psi_{1}''}{2}+\frac{\Psi_{1}'}{r}\right)-\frac{\Psi_{0}'\Phi_{1}'}{2} & = & \mathcal{F}+\frac{\Psi_{0}'}{6\Lambda}\mathcal{F}'\label{eq:d.22}\\
\left(\frac{\Psi_{0}\Phi_{1}''}{2}+\frac{\Psi_{0}'\Phi_{1}'}{2}+2\frac{\Psi_{0}\Phi_{1}'}{r}+\frac{\Psi_{1}'}{r}+\frac{\Psi_{1}}{r^{2}}\right)-\frac{\Psi_{0}\Phi_{1}'}{r} & = & \mathcal{F}+\frac{\Psi_{0}}{3\Lambda r}\mathcal{F}'\label{eq:d.23}\end{eqnarray}
The left-hand-sides of Eqs. \eqref{eq:d.22} and \eqref{eq:d.23}
are\begin{eqnarray}
\mbox{LHS of \eqref{eq:d.22}} & = & \mathcal{E}+\frac{\Psi_{1}''}{2}+\frac{\Psi_{1}'}{r}\label{eq:d.24}\\
\mbox{LHS of \eqref{eq:d.23}} & = & \mathcal{E}+\frac{\Psi_{1}}{r^{2}}+\frac{\Psi_{1}'}{r}\label{eq:d.25}\end{eqnarray}
Inverting \eqref{eq:d.20}\begin{eqnarray}
\mathcal{E} & = & \frac{2}{3}\mathcal{F}-\frac{1}{6}\left(\Psi_{1}''+4\frac{\Psi_{1}'}{r}+2\frac{\Psi_{1}}{r^{2}}\right)\label{eq:d.26}\end{eqnarray}
the LHS's become\begin{eqnarray}
\mbox{LHS of \eqref{eq:d.22}} & = & \frac{2}{3}\mathcal{F}+\frac{\Psi_{1}''}{3}+\frac{\Psi_{1}'}{3r}-\frac{\Psi_{1}}{3r^{2}}\label{eq:d.27}\\
\mbox{LHS of \eqref{eq:d.23}} & = & \frac{2}{3}\mathcal{F}-\frac{\Psi_{1}''}{6}+\frac{\Psi_{1}'}{3r}+2\frac{\Psi_{1}}{3r^{2}}\label{eq:d.28}\end{eqnarray}
Eqs. \eqref{eq:d.22} and \eqref{eq:d.23} can be simplified to be\begin{eqnarray}
\frac{\Psi_{0}'}{2\Lambda}\mathcal{F}'+\mathcal{F} & = & \ \ \,\Psi_{1}''+\frac{\Psi_{1}'}{r}-\frac{\Psi_{1}}{r^{2}}\label{eq:d.29}\\
\frac{\Psi_{0}}{\Lambda r}\mathcal{F}'+\mathcal{F} & = & -\frac{\Psi_{1}''}{2}+\frac{\Psi_{1}'}{r}+2\frac{\Psi_{1}}{r^{2}}\label{eq:d.30}\end{eqnarray}
which, as algebraic equations, can be solved for $\mathcal{F}$ and
$\mathcal{F}'$ separately:\begin{eqnarray}
\mathcal{F}' & = & 3\Lambda\frac{\Psi_{1}''-2\frac{\Psi_{1}}{r^{2}}}{\Psi_{0}'-2\frac{\Psi_{0}}{r}}=3\Lambda\frac{\Psi_{1}''-2\frac{\Psi_{1}}{r^{2}}}{r^{2}\left(r^{-2}\Psi_{0}\right)'}\label{eq:d.31}\\
\mathcal{F} & = & -\frac{\Psi_{1}''}{2}+\frac{\Psi_{1}'}{r}+2\frac{\Psi_{1}}{r^{2}}-\frac{\Psi_{0}}{\Lambda r}\mathcal{F}'\label{eq:d.32}\end{eqnarray}
Next, define:\begin{eqnarray}
\mathcal{G} & \triangleq & \Psi_{1}''-2\frac{\Psi_{1}}{r^{2}}\label{eq:d.33}\end{eqnarray}
which leads to\begin{eqnarray}
\mathcal{G} & = & r\left(r^{-2}(r\Psi_{1})'\right)'\label{eq:d.34}\\
\mathcal{G}' & = & \Psi_{1}'''-2\frac{\Psi_{1}'}{r^{2}}+4\frac{\Psi_{1}}{r^{3}}\label{eq:d.35}\end{eqnarray}
We then have, from Eqs. \eqref{eq:d.31} and \eqref{eq:d.32}: \begin{eqnarray}
\mathcal{F}' & = & 3\Lambda\frac{\mathcal{G}}{r^{2}\left(r^{-2}\Psi_{0}\right)'}\label{eq:d.36}\\
\mathcal{F} & = & -\frac{\Psi_{1}''}{2}+\frac{\Psi_{1}'}{r}+2\frac{\Psi_{1}}{r^{2}}-3\frac{\Psi_{0}\mathcal{G}}{r^{3}\left(r^{-2}\Psi_{0}\right)'}\label{eq:d.37}\end{eqnarray}
Differentiating Eq. \eqref{eq:d.37} and invoking \eqref{eq:d.33}
and \eqref{eq:d.35}:\begin{eqnarray}
\mathcal{F}' & = & -\frac{\Psi_{1}'''}{2}+\frac{\Psi_{1}''}{r}+\frac{\Psi_{1}'}{r^{2}}-4\frac{\Psi_{1}}{r^{3}}-3\frac{\Psi_{0}}{r^{3}\left(r^{-2}\Psi_{0}\right)'}\mathcal{G}'-3\left(\frac{\Psi_{0}}{r^{3}\left(r^{-2}\Psi_{0}\right)'}\right)'\mathcal{G}\nonumber \\
 & = & -\frac{1}{2}\left(\Psi_{1}'''-2\frac{\Psi_{1}'}{r^{2}}+4\frac{\Psi_{1}}{r^{3}}\right)+\frac{1}{r}\left(\Psi_{1}''-2\frac{\Psi_{1}}{r^{2}}\right)-3\frac{\Psi_{0}}{r^{3}\left(r^{-2}\Psi_{0}\right)'}\mathcal{G}'-3\left(\frac{\Psi_{0}}{r^{3}\left(r^{-2}\Psi_{0}\right)'}\right)'\mathcal{G}\nonumber \\
 & = & -\frac{1}{2}\mathcal{G}'+\frac{1}{r}\mathcal{G}-3\frac{\Psi_{0}}{r^{3}\left(r^{-2}\Psi_{0}\right)'}\mathcal{G}'-3\left(\frac{\Psi_{0}}{r^{3}\left(r^{-2}\Psi_{0}\right)'}\right)'\mathcal{G}\nonumber \\
 & = & -\left[3\frac{\Psi_{0}}{r^{3}\left(r^{-2}\Psi_{0}\right)'}+\frac{1}{2}\right]\mathcal{G}'+\left[-3\left(\frac{\Psi_{0}}{r^{3}\left(r^{-2}\Psi_{0}\right)'}\right)'+\frac{1}{r}\right]\mathcal{G}\label{eq:d.38}\end{eqnarray}
Equating Eqs. \eqref{eq:d.36} and \eqref{eq:d.38}:\begin{equation}
\left[3\frac{\Psi_{0}}{r^{3}\left(r^{-2}\Psi_{0}\right)'}+\frac{1}{2}\right]\mathcal{G}'=\left[-3\left(\frac{\Psi_{0}}{r^{3}\left(r^{-2}\Psi_{0}\right)'}\right)'+\frac{1}{r}-3\Lambda\frac{1}{r^{2}\left(r^{-2}\Psi_{0}\right)'}\right]\mathcal{G}\label{eq:d.39}\end{equation}
The bracketed terms in Eq. \eqref{eq:d.39} are explicitly computed
below:\begin{eqnarray*}
\Psi_{0} & = & 1-\frac{r_{s}}{r}-\Lambda r^{2}\\
r^{-2}\Psi_{0} & = & \frac{1}{r^{2}}-\frac{r_{s}}{r^{3}}-\Lambda\\
\left(r^{-2}\Psi_{0}\right)' & = & -\frac{2}{r^{3}}+\frac{3r_{s}}{r^{4}}=-\frac{2}{r^{4}}\left(r-\frac{3}{2}r_{s}\right)\\
\frac{1}{r^{2}\left(r^{-2}\Psi_{0}\right)'} & = & -\frac{1}{2}\frac{r^{2}}{r-\frac{3}{2}r_{s}}\\
\frac{\Psi_{0}}{r^{3}\left(r^{-2}\Psi_{0}\right)'} & = & -\frac{1}{2}\frac{r-r_{s}-\Lambda r^{3}}{r-\frac{3}{2}r_{s}}\\
\left(\frac{\Psi_{0}}{r^{3}\left(r^{-2}\Psi_{0}\right)'}\right)' & = & -\frac{1}{2}\frac{\left(1-3\Lambda r^{2}\right)\left(r-\frac{3}{2}r_{s}\right)-\left(r-r_{s}-\Lambda r^{3}\right)}{\left(r-\frac{3}{2}r_{s}\right)^{2}}=\frac{\frac{1}{4}r_{s}+\Lambda r^{3}-\frac{9}{4}r_{s}\Lambda r^{2}}{\left(r-\frac{3}{2}r_{s}\right)^{2}}\end{eqnarray*}
and\begin{equation}
3\frac{\Psi_{0}}{r^{3}\left(r^{-2}\Psi_{0}\right)'}+\frac{1}{2}=-\frac{3}{2}\frac{r-r_{s}-\Lambda r^{3}}{r-\frac{3}{2}r_{s}}+\frac{1}{2}=\frac{-r+\frac{3}{4}r_{s}+\frac{3}{2}\Lambda r^{3}}{r-\frac{3}{2}r_{s}}\label{eq:d.40}\end{equation}
and\begin{equation}
-3\left(\frac{\Psi_{0}}{r^{3}\left(r^{-2}\Psi_{0}\right)'}\right)'+\frac{1}{r}-3\Lambda\frac{1}{r^{2}\left(r^{-2}\Psi_{0}\right)'}=\frac{-\frac{3}{4}r_{s}-3\Lambda r^{3}+\frac{27}{4}r_{s}\Lambda r^{2}}{\left(r-\frac{3}{2}r_{s}\right)^{2}}+\frac{1}{r}+\frac{\frac{3}{2}\Lambda r^{2}}{r-\frac{3}{2}r_{s}}=\frac{\left(-r+3r_{s}\right)\left(-r+\frac{3}{4}r_{s}+\frac{3}{2}\Lambda r^{3}\right)}{r\left(r-\frac{3}{2}r_{s}\right)^{2}}\label{eq:d.41}\end{equation}
Plugging \eqref{eq:d.40} and \eqref{eq:d.41} to Eq. \eqref{eq:d.39},
we obtain:\begin{equation}
\frac{\mathcal{G}'}{\mathcal{G}}=\frac{-r+3r_{s}}{r\left(r-\frac{3}{2}r_{s}\right)}=\frac{1}{r-\frac{3}{2}r_{s}}-\frac{2}{r}\label{eq:d.42}\end{equation}
from which\begin{equation}
\mathcal{G}=-\frac{a}{r^{2}}\left(r-\frac{3}{2}r_{s}\right)\label{eq:d.43}\end{equation}
with $a$ being a constant of integration. Combined with \eqref{eq:d.34}
we then get\begin{equation}
\Psi_{1}=\frac{a}{2}\left(r-\frac{3}{2}r_{s}\right)+br^{2}+\frac{c}{r}\label{eq:d.44}\end{equation}
with $b$ and $c$ being two additional constants of integration.
However, $a,\ b$ and $c$ can be absorbed into the definition of
$\Lambda,\ r_{s}$ and $\gamma$ respectively in $\Psi_{0}$. We thus
set $a=2,\ b=c=0$ and obtain\begin{equation}
\Psi_{1}=r-\frac{3}{2}r_{s}\label{eq:d.45}\end{equation}
which, via \eqref{eq:d.37}, neatly leads to\begin{eqnarray}
\mathcal{F} & = & -\frac{\Psi_{1}''}{2}+\frac{\Psi_{1}'}{r}+2\frac{\Psi_{1}}{r^{2}}-3\frac{\Psi_{0}}{r^{3}\left(r^{-2}\Psi_{0}\right)'}\mathcal{G}\nonumber \\
 & = & \frac{1}{r}+\frac{2}{r}-\frac{3r_{s}}{r^{2}}+3\left(1-\frac{r_{s}}{r}-\Lambda r^{2}\right)\frac{1}{r^{3}}\frac{1}{-\frac{2}{r^{3}}+\frac{3r_{s}}{r^{4}}}\frac{2}{r^{2}}\left(r-\frac{3}{2}r_{s}\right)\nonumber \\
 & = & \frac{3}{r}-\frac{3r_{s}}{r^{2}}-3\left(1-\frac{r_{s}}{r}-\Lambda r^{2}\right)\frac{1}{r}\nonumber \\
 & = & 3\Lambda r\label{eq:d.46}\end{eqnarray}
Using \eqref{eq:d.21}, we successively get \begin{eqnarray*}
\left(3r^{2}\Psi_{0}\Phi_{1}'+\left(r^{2}\Psi_{1}\right)'\right)' & = & 12\Lambda r^{3}\\
3r^{2}\Psi_{0}\Phi_{1}'+3r^{2}-3r_{s}r & = & 3\Lambda r^{4}+3\epsilon\\
r^{2}\Psi_{0}\Phi_{1}' & = & -\left(r^{2}-r_{s}r-\Lambda r^{4}\right)+\epsilon\\
\Phi_{1}' & = & -1+\frac{\epsilon}{r^{2}\Psi_{0}}\end{eqnarray*}
or, finally:\begin{equation}
\Phi_{1}=-r+\epsilon\int\frac{dr}{r^{2}\left(1-\frac{r_{s}}{r}-\Lambda r^{2}\right)}\label{eq:d.47}\end{equation}
The Ricci scalar can also be calculated from \eqref{eq:d.7} and \eqref{eq:d.19}:\[
\mathcal{R}e^{\alpha}=12\Lambda-4\gamma\mathcal{F}+\mathcal{O}\left(\gamma^{2}\right)\]
or, using \eqref{eq:d.3}, \eqref{eq:d.46}, and \eqref{eq:d.47}:
\begin{eqnarray}
\mathcal{R} & = & 12\Lambda\left(1-\alpha\right)-4\gamma\mathcal{F}+\mathcal{O}\left(\gamma^{2}\right)\nonumber \\
 & = & 12\Lambda-\gamma\left(4\mathcal{F}+12\Lambda\Phi_{1}\right)+\mathcal{O}\left(\gamma^{2}\right)\nonumber \\
 & = & 12\Lambda\left[1-\gamma\epsilon\int\frac{dr}{r^{2}\left(1-\frac{r_{s}}{r}-\Lambda r^{2}\right)}\right]+\mathcal{O}\left(\gamma^{2}\right)\label{eq:d.48}\end{eqnarray}
In perfect agreement with Buchdahl's study in Appendix \ref{sec:B},
the solution (\ref{eq:d.45}, \ref{eq:d.47}) contains $4$ parameters:
(1) $\Lambda$, the large-distance curvature (the de Sitter term);
(2) $r_{s}$ as a free parameter (the Newton term); (3) $\gamma$,
specifying the linear term as in Mannheim-Kazanas's potential in conformal
gravity (the Mannheim-Kazanas term); and (4) $\epsilon$, charactering
the anomalous curvature -- it allows the curvature to be non-constant
as evident in \eqref{eq:d.48}. The parameter $\epsilon$ is the new
feature in our solution. There is a constant of integration in $\Phi_{1}$
but it is an overall scale factor.

In summary, the metric is:\begin{equation}
ds^{2}=e^{\gamma\left[-r+\epsilon\int\frac{dr}{r^{2}\left(1-\frac{r_{s}}{r}-\Lambda r^{2}\right)}\right]}\left[-\left(1-\frac{r_{s}}{r}-\Lambda r^{2}+\gamma\left(r-\frac{3}{2}r_{s}\right)\right)\left(dx^{0}\right)^{2}+\frac{dr^{2}}{1-\frac{r_{s}}{r}-\Lambda r^{2}+\gamma\left(r-\frac{3}{2}r_{s}\right)}+r^{2}d\Omega^{2}\right]+\mathcal{O}\left(\gamma^{2}\right)\label{eq:d.49}\end{equation}

\section{\label{sec:E}The Ricci scalar in the Robertson-Walker metric and
the modified Robertson-Walker metric}

Let us first consider the closed universe case, $\kappa=1$. The RW
metric reads (with $d\Omega^{2}=d\theta^{2}+\sin^{2}\theta\, d\phi^{2}$):
\begin{equation}
ds^{2}=c^{2}dt^{2}-a^{2}(t)\left(d\chi^{2}+\sin^{2}\chi d\Omega^{2}\right)\label{eq:e.1}\end{equation}
whereas the modified RW metric reads\begin{equation}
ds^{2}=\left(c_{0}\sqrt{\frac{a_{0}}{a(t)}}\right)^{2}dt^{2}-a^{2}(t)\left(d\chi^{2}+\sin^{2}\chi d\Omega^{2}\right)=c_{0}^{2}\frac{a_{0}}{a(t)}dt^{2}-a^{2}(t)\left(d\chi^{2}+\sin^{2}\chi d\Omega^{2}\right)\label{eq:e.2}\end{equation}
Both metrics can be commonly recast in term of the conformal time
$\eta$:\begin{equation}
ds^{2}=a^{2}(\eta)\left[d\eta^{2}-\left(d\chi^{2}+\sin^{2}\chi d\Omega^{2}\right)\right]\label{eq:e.3}\end{equation}
with $\eta$ being defined as\begin{equation}
d\eta=\begin{cases}
\ c\, a^{-1}(t)\, dt & \mbox{ for RW metric}\\
\ c_{0}\, a_{0}^{1/2}\, a^{-3/2}(t)\, dt & \mbox{ for modified RW metric}\end{cases}\label{eq:e.4}\end{equation}
The variables $(x^{0},x^{1},x^{2},x^{3})\triangleq(\eta,\chi,\theta,\phi)$
are of dimensionless unit. The metric tensor is diagonal:\begin{align}
g_{00} & =a^{2}(\eta),\ \ g_{11}=-a^{2}(\eta),\ \ g_{22}=-a^{2}(\eta)\sin^{2}\chi,\ \ g_{33}=-a^{2}(\eta)\sin^{2}\chi\sin^{2}\theta.\label{eq:e.5}\end{align}
Of the 64 Christoffel symbols $\Gamma_{\mu\nu}^{\alpha}=\frac{1}{2}g^{\alpha\beta}\left(\partial_{\mu}g_{\nu\beta}+\partial_{\nu}g_{\mu\beta}-\partial_{\beta}g_{\mu\nu}\right)$,
the 19 non-vanishing components are (with the prime indicating the
derivative with respect to $\eta$):\begin{align}
\Gamma_{00}^{0} & =\Gamma_{11}^{0}=\Gamma_{01}^{1}=\Gamma_{10}^{1}=\Gamma_{02}^{2}=\Gamma_{20}^{2}=\Gamma_{03}^{3}=\Gamma_{30}^{3}=\frac{a'}{a},\nonumber \\
\Gamma_{22}^{0} & =\frac{a'}{a}\sin^{2}\chi,\ \ \Gamma_{33}^{0}=\frac{a'}{a}\sin^{2}\chi\sin^{2}\theta,\ \ \Gamma_{22}^{1}=-\sin\chi\cos\chi,\ \Gamma_{33}^{1}=-\sin\chi\cos\chi\sin^{2}\theta,\label{eq:e.6}\\
\Gamma_{12}^{2} & =\Gamma_{21}^{2}=\Gamma_{13}^{3}=\Gamma_{31}^{3}=\cot\chi,\ \ \Gamma_{33}^{2}=-\sin\theta\cos\theta,\ \ \Gamma_{23}^{3}=\Gamma_{32}^{3}=\cot\theta\nonumber \end{align}
The Ricci tensor $\mathcal{R}_{\beta}^{\alpha}=g^{\alpha\gamma}\left(\partial_{\delta}\Gamma_{\gamma\beta}^{\delta}-\partial_{\beta}\Gamma_{\gamma\delta}^{\delta}+\Gamma_{\gamma\beta}^{\delta}\Gamma_{\delta\sigma}^{\sigma}-\Gamma_{\gamma\delta}^{\sigma}\Gamma_{\beta\sigma}^{\delta}\right)$
is diagonal:\begin{equation}
\mathcal{R}_{0}^{0}=-\frac{3}{a^{2}}\left(\frac{a''}{a}-\frac{a'^{2}}{a^{2}}\right),\ \ \ \mathcal{R}_{k}^{k}=-\frac{1}{a^{2}}\left(\frac{a''}{a}+\frac{a'^{2}}{a^{2}}+2\right)\label{eq:e.7}\end{equation}
The Ricci scalar is\begin{equation}
\mathcal{R}=\mathcal{R}_{\alpha}^{\alpha}=-\frac{6}{a^{2}}\left(\frac{a''}{a}+1\right)\label{eq:e.8}\end{equation}
The case of open universe, $\kappa=-1$, can be obtained by replacing
$\sin\chi\rightarrow\sinh\chi$ in the modified RW metric. This amounts
to replacing $\chi\rightarrow i\chi$. The case of flat universe,
$\kappa=0$, can be obtained by replacing $\sin\chi\rightarrow\chi$
in the modified RW metric. The previous calculations sail through
with the Ricci scalar being generalized to\begin{equation}
\mathcal{R}=-\frac{6}{a^{2}}\left(\frac{a''}{a}+\kappa\right)\label{eq:e.9}\end{equation}
This result is applicable to both the RW metric and the modified RW
metric.

\section{\label{sec:F}Three forms of the time duration paradox}

The time duration paradox (also known as the twin paradox in popular
science) is probably the most perplexing one in relativity. The most
classic version of it started in the early day of special relativity
and is usually cast in a fanciful form which involves a pair of identical
twins. One twin stays on the Earth while the other travels at high
speed to a distant star light years away then makes a U-turn to come
back. At their rendezvouz decades later, the twin that stays on Earth
will have aged more than the other one. The twin paradox has been
resolved. There are two other versions of it, though, one in general
relativity and one in our theory of curvature-scaling gravity. To
bring forth our version, it is necessary that we review the other
two for the purpose of comparison. 
\begin{enumerate}
\item The classic and original version, valid for special relativity:

A pair of synchronized clocks, starting at one location $A$, are
made to trace out two timelike trajectories then brought back together
at a later time at $B$. For each clock, the time count (i.e., the
number of revolutions the clock will have completed, or the number
of {}``beats'' the clock will have clicked) between the two events
is $\Delta\tau=\Delta s/c$ with $\Delta s\triangleq\int_{A}^{B}ds$
and $ds^{2}=c^{2}dt^{2}-d\vec{x}^{2}$ being the infinitesimal proper
distance in the Minkowski metric. If the cumulative proper distances
of the two paths are different, say, $\Delta s_{1}>\Delta s_{2}$,
the clocks will register different time counts, $\Delta\tau_{1}>\Delta\tau_{2}$.
The metric (whether it is Minkowski or not) is not important; the
only requirement is that the two path accumulate different total proper
distances. This effect is related to the time dilation effect, observed
in the life time of muons which are created in the stratosphere yet
could reach the sea level.

\item The gravitational redshift, which takes place in general relativity:

This effect was one of the early predictions of the equivalence principle.
It has been verified in the Pound-Rebka? experiment which measures
the redshift of photons free falling in the Earth's gravitation field.
We shall recast this paradox in a form more aligned with the first
version. Two synchronized clocks, starting at one location $A$, are
quickly sent to two different regions, one in a strong gravitational
field and the other in a weak field. After a long exposure to the
fields, they are quickly brought back together. Again, the time count
for each clock is $\Delta\tau=\Delta s/c$ with $\Delta s=\int_{A}^{B}ds$
as before but $ds^{2}=g_{00}\, c^{2}dt^{2}+\dots$ with the dots denotes
the remaining 15 terms in the metric. These 15 terms vanish along
the two trajectories since the two clocks sit still at their respective
regions for the whole long exposure. Take the Schwarzschild metric:
$g_{00}=1-\frac{r_{s}}{r}$ in which $r_{s}=\frac{GM}{2c^{2}}$ where
$M$ is the mass of the field source. Therefore, although $\Delta t_{1}=\Delta t_{2}$,
the cumulative proper distances are different due to the difference
in their $g_{00}$. Take $M_{1}>M_{2}$, then $g_{00}(1)<g_{00}(2)$
and so, $\Delta s_{1}<\Delta s_{2}$. As such, the clocks register
different time counts: $\Delta\tau_{1}<\Delta\tau_{2}.$ In this case,
the clock exposed to the stronger gravitational field runs slower
(i.e., registers less time counts) then the clock exposed to the weaker
field, as expected from the equivalence principle.

\item The third version -- Time counts in curvature-scaling gravity:

Consider again the two synchronized clocks. Send them to two regions
with different Ricci scalars, say, $\mathcal{R}_{1}>\mathcal{R}_{2}$,
which means their Ricci lengths are related as $a_{\mathcal{R}}(1)<a_{\mathcal{R}}(2)$.
Note that the time count is no longer $\Delta\tau=\Delta s/c$ but
is the ratio of $\Delta\tau$ with the oscillatory period of the atoms
that make up the clock. With $a_{\mathcal{R}}(1)<a_{\mathcal{R}}(2)$,
the atoms in Region 1 require less time to vibrate than the atoms
in Region 2. The time scaling rule \eqref{eq:5.5} dictates that the
period for each vibration scale as $a_{\mathcal{R}}^{3/2}$. Therefore,
even if the total proper distances of the two paths are deliberately
prepared to be equal, Clock 1 registers more time counts than Clock
2: $\Delta\tilde{\tau}_{1}>\Delta\tilde{\tau}_{2}$. This effect works
somewhat in reverse of the gravitational redshift, since $\mathcal{R}_{1}>\mathcal{R}_{2}$
would usually require a stronger field for Region 1 than the field
in Region 2. We do not expect this effect to be material for the solar
system however since the solar system is almost Ricci flat.

\end{enumerate}

\section{\label{sec:G}On the redshift of photons from distant galaxies}

Although the Universe has been in the expansion mode since the Big
Bang, galaxies are not subject to cosmic expansion. Otherwise, the
redshift of light emitted from distant galaxies would not be detectable
on Earth because Earth-based apparatus would expand accordingly and
thus the traveling photon's wavelength and/or frequency would appear
exactly the same as the one on Earth.

In this Appendix, we establish the relationship between redshift parameter
$z$ to the cosmic scale expansion within our curvature-scaling gravity.
Note that a new element in our theory is the variability of light
speed, in which $c\propto a^{-1/2}$ with $a$ being the cosmic scale
factor at the location the photon resides.

We model the Milky Way and a distant galaxy as non-expanding objects,
devoid of the cosmic expansion. The scale factor for them are set
as a fixed value, denoted as $a_{0}=1$. Nonetheless, we do allow
the space enclosing the galaxies to expand. We even allow the space
to have been expanding well before the galaxies were formed. With
such allowance, each of the galaxies and the space enclosing it can
have different scale factor.

Consider a photon emitted from the distant galaxy at wavelength $\lambda_{0}$
and travel at speed $c_{0}$. The outskirt of the distant galaxy had
been expanding to a scale factor $a_{1}>1$, thus corresponding to
a lower light speed $c_{1}=\frac{c_{0}}{\sqrt{a_{1}}}$ . Once reaching
the outskirt of its galaxy, the light wave gets {}``compressed''
since its front end {}``slows'' down. By the time its back end hits
the outskirt, its wavelength is shortened to\begin{equation}
\begin{cases}
\ \lambda_{1}=\lambda_{0}\frac{c_{1}}{c_{0}}=\frac{\lambda_{0}}{\sqrt{a_{1}}}\\
\ c_{1}\ =\frac{c_{0}}{\sqrt{a_{1}}}\end{cases}\label{eq:g.1}\end{equation}
The light wave then began to expand together with the cosmos as it
was on its transit toward Earth. By the time it reaches the outskirt
of the Milky Way at which location the scale factor is $a_{2}$ and
the light speed is $c_{2}=\frac{c_{0}}{\sqrt{a_{2}}}$, it would have
undergone a cosmic expansion of $\frac{a_{2}}{a_{1}}$. Thus its wavelength
is expanded to\begin{equation}
\begin{cases}
\ \lambda_{2}=\frac{a_{2}}{a_{1}}\lambda_{1}=\frac{a_{2}}{a_{1}^{3/2}}\lambda_{0}\\
\ c_{2}\ =\frac{c_{0}}{\sqrt{a_{2}}}\end{cases}\label{eq:g.2}\end{equation}
As the photon enters the Milky Way in which the light speed remains
$c_{0}$ (because the Milky Way has resisted expansion), the front
end of the light wave races toward Earth, thus {}``stretching''
out the wavelength to\begin{equation}
\lambda_{3}=\lambda_{2}\frac{c_{0}}{c_{2}}=\frac{a_{2}}{a_{1}^{3/2}}\lambda_{0}\sqrt{a_{2}},\label{eq:g.3}\end{equation}
or\begin{equation}
\frac{\lambda_{3}}{\lambda_{0}}=\left(\frac{a_{2}}{a_{1}}\right)^{3/2}.\label{eq:g.4}\end{equation}
The light wave, having been {}``stretched'' out, can no longer be
absorbed by the respective atom on Earth. The light wave is said to
have experienced a redshift. The redshift defined as $z\triangleq\frac{\lambda_{observed}-\lambda_{emit}}{\lambda_{emit}}=\frac{\lambda_{3}-\lambda_{0}}{\lambda_{0}}$
thus satisfies\begin{equation}
1+z=\left(\frac{a_{2}}{a_{1}}\right)^{3/2}.\label{eq:g.5}\end{equation}
What is new in this context, as compared with standard cosmology,
is the exponent of $\frac{3}{2}$ in the redshift formula instead
of unity. Because $a_{2}>a_{1}$, $\left(\frac{a_{2}}{a_{1}}\right)^{3/2}>\frac{a_{2}}{a_{1}}$,
meaning that the redshift as detected on Earth gets enhanced as compared
with the raw ratio of the scale factors $\frac{a_{2}}{a_{1}}$. To
deduce the actual ratio $\frac{a_{2}}{a_{1}}$, we must reduce the
detected redshift value accordingly. Without this knowledge, one would
mistakenly overestimate the redshift as has been an issue in the Hubble
law and the calibration of the Hubble constant.

Note that we allow\emph{ }the expansion of the outer space to freely
take place even before the galaxies were formed (i.e., we did not
presume $a_{1}=1$). Only the ratio $\frac{a_{2}}{a_{1}}$ between
the cosmic scale factors in the outer space appears in the redshift
formula, however. We also do not make any specific assumptions about
the actual cosmic expansion process during the transit of the light
wave. %
\footnote{Our final comment is on the galactic gravitational resistance to cosmic
expansion. It is agreed that galaxies resist the cosmic expansion;
otherwise, the redshift of light from distant galaxies would not be
detectable. Yet the mechanism for the gravitational confinement is
not well addressed in standard cosmology. It is often a hand-waving
notion that the presence of matter holds the {}``fabric'' of space
inside a galaxy stationary, thus enabling the galaxy to resist the
cosmic expansion. What constitutes the {}``fabric'' is often left
unanswered. Curvature-scaling gravity answers this question. The {}``fabric''
is nothing but the Ricci scalar. It is reasonable to expect that due
to the presence of matter, $\mathcal{R}$ within galaxies is higher
than that in the void of the outer space, and that with galaxies stabilizing
due to their own rotations, $\mathcal{R}$ inside galaxies tends to
be stationary as well.%
}

\end{document}